\documentclass[twocolumn]{aastex7}

\newcommand{\avghna}{\ensuremath{\langle\mathrm{H}n\alpha \rangle} }
\newcommand{\hna}{\ensuremath{\mathrm{H}n\alpha} }
\newcommand{\HII}{H\,\textsc{ii} }

\newcommand{\HI}{H\,\textsc{i} }
\newcommand{\hi}{H\,\textsc{i} }
\newcommand{\VLSR}{V$_{\text{LSR}}$ }
\usepackage{comment}
\usepackage{longtable}
\usepackage{afterpage}
\usepackage{amsmath}
\usepackage{multirow}
\usepackage{natbib}
\defcitealias{HafnerFlipPaper}{HDW20}

\begin{document}
\title{Tracing Gas Kinematics and Interactions between H II Regions and Molecular Clouds using VLA Observations of Recombination Lines and Hydroxyl}

\correspondingauthor{E. Cappellazzo}
\email{elizabeth.cappellazzo@hdr.mq.edu.au, beth.cappellazzo@csiro.au}

\author[0009-0004-2905-6515]{E. Cappellazzo}
\affiliation{School of Mathematical and Physical Sciences, Macquarie University, Sydney, NSW 2109, Australia}
\affiliation{Astrophysics and Space Technologies Research Centre, Macquarie University, Sydney, NSW 2109, Australia}
\affiliation{ATNF, CSIRO, Space and Astronomy, PO Box 76, Epping, NSW 1710, Australia}
\email{elizabeth.cappellazzo@hdr.mq.edu.au}

\author[0000-0003-0235-3347]{J. R. Dawson}
\affiliation{School of Mathematical and Physical Sciences, Macquarie University, Sydney, NSW 2109, Australia}
\affiliation{Astrophysics and Space Technologies Research Centre, Macquarie University, Sydney, NSW 2109, Australia}
\affiliation{ATNF, CSIRO, Space and Astronomy, PO Box 76, Epping, NSW 1710, Australia}
\email{joanne.dawson@mq.edu.au}

\author[0000-0002-1737-0871]{Mark Wardle}
\affiliation{School of Mathematical and Physical Sciences, Macquarie University, Sydney, NSW 2109, Australia}
\affiliation{Astrophysics and Space Technologies Research Centre, Macquarie University, Sydney, NSW 2109, Australia}
\email{mark.wardle@mq.edu.au}

\author[0000-0003-3351-6831]{Trey V. Wenger}
\affiliation{California State University, Chico, 400 W 1st St, Chico, CA 95929, USA}
\affiliation{NSF Astronomy \& Astrophysics Postdoctoral Fellow, Department of Astronomy, University of Wisconsin--Madison, 475 N Charter St, Madison, WI 53703, USA}
\email{tvwenger@csuchico.edu}

\author[0000-0001-6179-0606]{Anita Hafner}
\affiliation{Sydney Institute for Astronomy, School of Physics, University of Sydney NSW 2006, Australia}
\email{anita.hafner@sydney.edu.au}

\author[0000-0002-2465-7803]{Dana S. Balser}
\affiliation{National Radio Astronomy Observatory, 520 Edgemont Road,
Charlottesville, VA 22903, USA}
\email{dbalser@nrao.edu}

\author[0000-0001-8800-1793]{L. D. Anderson}
\affiliation{Department of Physics and Astronomy, West Virginia University, Morgantown, WV 26506, USA}
\affiliation{Center for Gravitational Waves and Cosmology, West Virginia University, Chestnut Ridge Research Building, Morgantown, WV 26506, USA}
\affiliation{Green Bank Observatory, P.O. Box 2, Green Bank, WV 24944, USA}
\email{loren.anderson@mail.wvu.edu}

\author[0000-0002-5053-2828]{Elizabeth K. Mahony}
\affiliation{ATNF, CSIRO, Space and Astronomy, PO Box 76, Epping, NSW 1710, Australia}
\email{elizabeth.mahony@csiro.au}

\author{M. R. Rugel}
\affiliation{Deutsches Zentrum f{\"u}r Astrophysik, Postplatz 1, 02826 G{\"o}rlitz, Germany}
\affiliation{National Radio Astronomy Observatory, PO Box O, Socorro, NM 87801, USA}
\email{michael.rugel@dzastro.de}

\author[0000-0002-6300-7459]{John M. Dickey}
\affiliation{School of Natural Sciences, University of Tasmania, Hobart, TAS, Australia}
\email{john.dickey@utas.edu.au}

\begin{abstract}

Observational studies of \HII region-molecular cloud interactions constrain models of feedback and quantify its impact on the surrounding environment.
A recent hypothesis proposes that a characteristic spectral signature in ground state hyperfine lines of hydroxyl (OH)---the OH flip---may trace gas that is dynamically interacting with an expanding \HII region, offering a new means of probing such interactions. 
We explore this hypothesis using dedicated Jansky Very Large Array (VLA) observations of three Galactic \HII regions, G049.205$-$0.343, G034.256+0.145 and G024.471+0.492, in 1--2 GHz continuum emission, all four 18-cm ground-state OH lines, and multiple hydrogen radio recombination lines. 
A Gaussian decomposition of the molecular gas data reveals complex OH emission and absorption across our targets. 
We detect the OH flip towards two of our sources, G049.205$-$0.343 and G034.256+0.145, finding agreement between key predictions of flip hypothesis and the observed multi-wavelength spectra, kinematics and morphology. Specifically, we demonstrate a strong spatial and kinematic association between the OH flip and the ionized gas of the \HII regions---the first time this has been demonstrated for resolved sources---and evidence from $^{13}$CO(1--0) data that the expected OH component originates from the non-disturbed gas of the parent cloud. 
While we detect no flip in G024.471+0.492, we do find evidence of interacting molecular gas traced by OH, providing further support for OH's ability to trace \HII region-molecular cloud interactions. 

\end{abstract}

\section{Introduction}

\label{sec:intro}
Stellar feedback from high-mass stars plays a fundamental role in the evolution of galaxies.
Young, massive stars are responsible for a complex series of feedback mechanisms which add momentum and energy back into the interstellar medium \citep[ISM, e.g.][]{Krumholz2014PhR...539...49K}. This feedback occurs through photoionization, stellar winds, supernovae, thermal expansion and radiation pressure. The feedback from these stars can result in complex motion in their surrounding gas, such as driven turbulence, shocks and expansion from thermal pressure \citep[e.g.][]{VinkOBWind2000A&A...362..295V, Mellema2006ApJ...647..397M, WalchDispersal2012MNRAS.427..625W, ZhuM172023MNRAS.522..503Z}. These processes may either trigger further local star formation \citep[][]{Rahner2018MNRAS.473L..11R, RugelW49A2019A&A...622A..48R, Panwar2020ApJ...905...61P} or prevent further local star formation by injecting turbulence and impeding collapse in molecular clouds \citep[e.g.,][]{Federrath2013ApJ...763...51F}, or completely destroying the parent molecular cloud \citep[e.g.][]{Rahner2017MNRAS.470.4453R, Kim2018ApJ...859...68K, Kruijssen2019Natur.569..519K, Chevance2022MNRAS.509..272C, Keller2022MNRAS.514.5355K}.

The dynamic interactions between massive stars and their environments are, however, poorly understood, 
and the effect of stellar feedback on galaxy evolution is not well constrained in models and simulations \citep[e.g.][]{Krumholz2014PhR...539...49K, Grudic2022MNRAS.512..216G}. In particular, pre-supernova feedback might have a greater influence on the regulation of star formation in galaxies than supernova explosions \citep{Kruijssen2019Natur.569..519K, Chevance2022MNRAS.509..272C}, underscoring the importance of studying the \HII region phase of high-mass star forming regions, before the first supernovae have occurred. 

\HII regions are pockets of ionized gas surrounding massive OB-type stars. They are short-lived ($<10$ Myr) nebulae which are produced by the ultra-violet radiation from OB stars ionizing the surrounding gas within molecular clouds. The high pressure, density and temperature ($10^4$ K) of the ionized gas leads to expansion into their surrounding parent molecular cloud. Their kinematics, morphology and physical properties can provide much needed constraints on the pre-supernova feedback driven by young, high mass stars. Hydrogen radio recombination lines (RRLs) are commonly used to study the kinematics, dynamics and properties of the interior ionized gas in \HII regions \citep[e.g.,][]{Balser2021ApJ...921..176B, Khan2022A&A...664A.140K, Pabst2024arXiv240417963P, BalserWenger2024ApJ...964...47B}. However, while there are many observational studies of pre-supernova feedback through \HII regions \citep[e.g.,][]{Rahner2018MNRAS.473L..11R, RugelW49A2019A&A...622A..48R, Barnes2021MNRAS.508.5362B, Olivier2021ApJ...908...68O, Rowland2024A&A...685A..46R, Pathak2025ApJ...982..140P}, it remains challenging to map and model the impact of this feedback on the dynamics and energetics of the surrounding gas. High-resolution spectral-line observations of the interior ionized gas and surrounding molecular environment of \HII regions and their parent clouds are required to address this challenge.

Hydroxyl (OH) is a powerful tracer for probing molecular gas surrounding \HII regions \citep[e.g.][]{RugelTHOR2018A&A...618A.159R}. OH has four 18-cm ground-rotational state transitions, two main lines at 1665.402 and 1667.359-MHz and two satellite lines at 1612.231 and 1720.530-MHz. Key to the present work is that the ground-state OH lines are anomalously excited in most environments in which they are observed \citep[e.g.][]{DawsonSPLASH2014MNRAS.439.1596D, DawsonSPLASH2022MNRAS.512.3345D, HafnerGNOMES2023}, and that this excitation may be modeled to constrain the physical state of the gas. In particular, the satellite lines are highly sensitive to environmental conditions, and are often weakly inverted or sub-thermally excited \citep{Elitzur16121976ApJ...205..384E, Elitzur17201976ApJ...203..124E, Guibert1978A&A....66..395G, vanLangevelde1995, Ebisawa2019ApJ...871...89E, Ebisawa2020ApJ...904..136E}, providing a sensitive barometer of ISM conditions.

Recently, \citet[][\citetalias{HafnerFlipPaper} hereafter]{HafnerFlipPaper} have argued that a particular class of OH satellite-line conjugate profile, known as the satellite-line flip, might be a signature of dynamically-interacting (shocked) and undisturbed (unshocked) gas components in the molecular gas surrounding \HII regions. The OH flip is a blended double spectral feature where the blueshifted component shows the 1612-MHz line in absorption and the 1720-MHz line in stimulated emission, and the redshifted component has the satellite lines `flipped', with 1612-MHz in stimulated emission and 1720-MHz in absorption. The main lines are generally seen in absorption across the two components. \citetalias{HafnerFlipPaper} propose (partly on the basis of non-local thermodynamic equilibrium excitation modeling) that the redshifted component traces the ambient molecular gas surrounding an \HII region, and the blueshifted component traces dynamically-interacting gas shocked by its expansion into the surrounding molecular cloud. We refer to this as the `\citetalias{HafnerFlipPaper} hypothesis' hereafter.

If \citetalias{HafnerFlipPaper} is correct, the OH flip may be a new tool to disentangle dynamically-interacting and non-interacting molecular gas around \HII regions. Furthermore, it may provide a means of quantifying the impact of high-mass stellar feedback on molecular gas dynamics and energetics. However, there have to-date been no spatially-resolved studies focusing on the behavior of the OH satellite lines towards \HII regions.
This work therefore presents new radio interferometric observations with the VLA of the four OH lines towards three Galactic \HII regions, together with complementary observations of multiple hydrogen RRLs (to map the interior ionized gas). We aim to produce the first resolved maps of the OH flip, explore the morphology and kinematics of the flip components, and robustly confirm their spatio-kinematic association with the \HII regions and broader molecular environment. We hence aim to confirm and constrain some of the basic features of the \citetalias{HafnerFlipPaper} hypothesis. 

This paper is organized as follows. We describe our chosen \HII regions and observed data in Section \ref{sec:data}. We outline our data reduction and Gaussian decomposition procedures in Section \ref{sec:methods}. We present the continuum and kinematic results from the RRLs and OH in Section \ref{sec:results}. We assess whether the observational results are in agreement with the predictions of \citetalias{HafnerFlipPaper} in Section \ref{sec:disc}. We provide our conclusions in Section \ref{sec:conc}.

\section{Data}
\label{sec:data}
\subsection{Target Selection}
\label{sec:targets}

\begin{table*}[]
\centering
\resizebox{\textwidth}{!}{
\begin{tabular}{ccccccc}
\hline
Field & Field R.A. & Field Dec. & Target & Target R.A. & Target Dec. & Distance \\
 & J2000 & J2000 & & J2000 & J2000 \\
(Name) & (hh:mm:ss) & (dd:mm:ss) & (Name) & (hh:mm:ss) & (dd:mm:ss) & (kpc) \\ 
\hline
G049.205$-$0.343 & 19:23:00.86 & +14:16:46.7 & G049.205$-$0.343 & 19:23:00.88 & +14:16:46.75 & 5.31$^{+0.18}_{-0.22}$ \\
G034.256+0.145 & 18:53:20.22 & +01:14:40.3 & G034.256+0.136 & 18:53:19.89 & +01:14:40.17 & 3.29$^{+0.30}_{-0.38}$ \\
 &  &  & G034.256+0.155 & 18:53:18.11 & +01:14:58.4 & 3.29$^{+0.30}_{-0.38}$ \\
G024.471+0.492 & 18:34:09.28 & -07:17:52.4 & G024.471+0.492 & 18:34:09.20 & -07:17:57.18 & 7.56$^{+0.23}_{-0.23}$\\ \hline
\end{tabular}%
}
\caption{\HII Region Targets in this work. The first three columns refer to the observed field, the last four columns refer to the \HII region target(s) within those fields. G049's distance is from parallax measurements \citep{Wu2014}, G034 and G024's distances are from kinematic measurements \citep{WengerMetallicity2019ApJ...887..114W}.}
\label{tab:sources}
\end{table*}

\begin{figure*}
    \centering
    \includegraphics[width=\columnwidth]{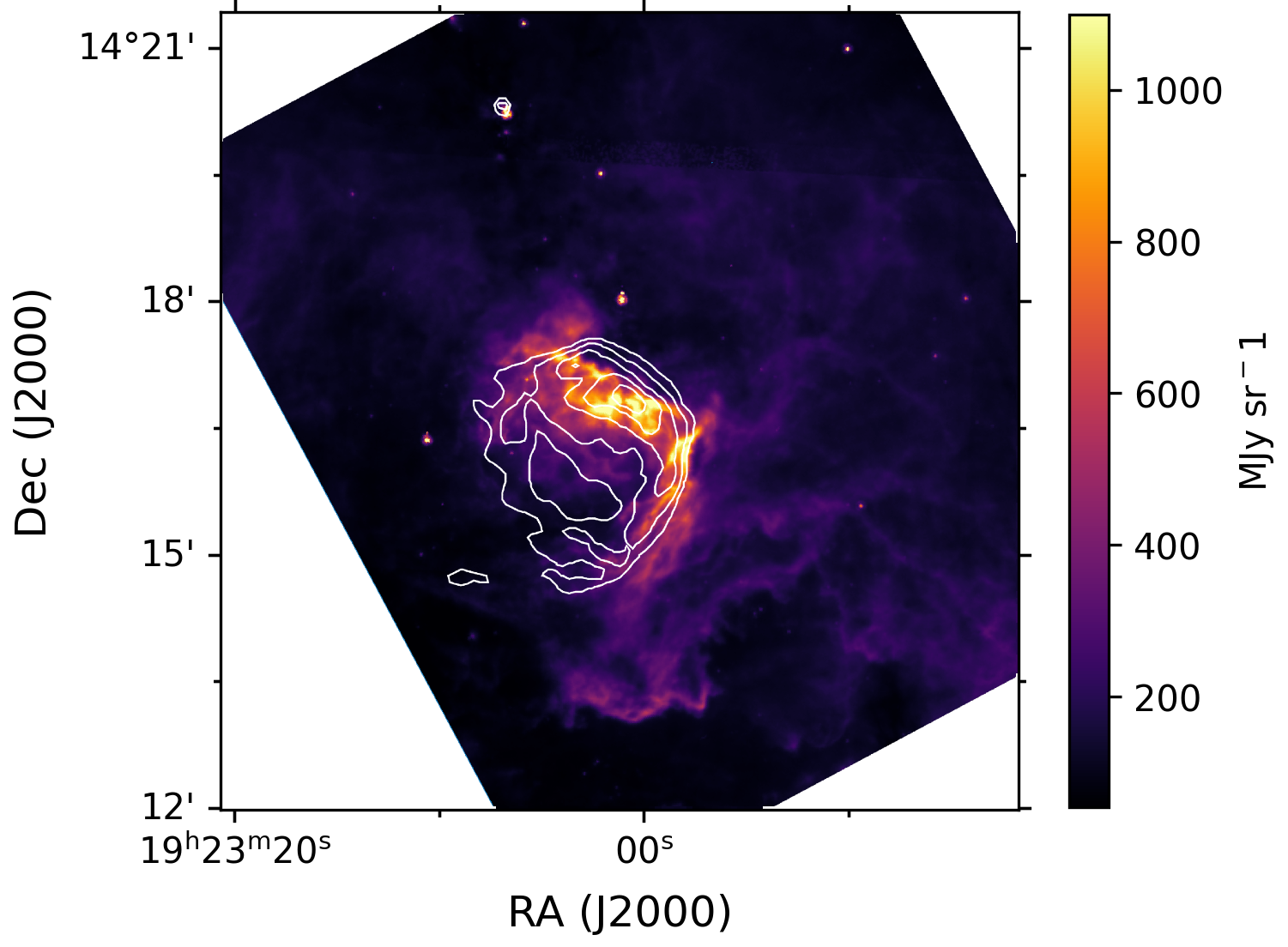}
    \includegraphics[width=\columnwidth]{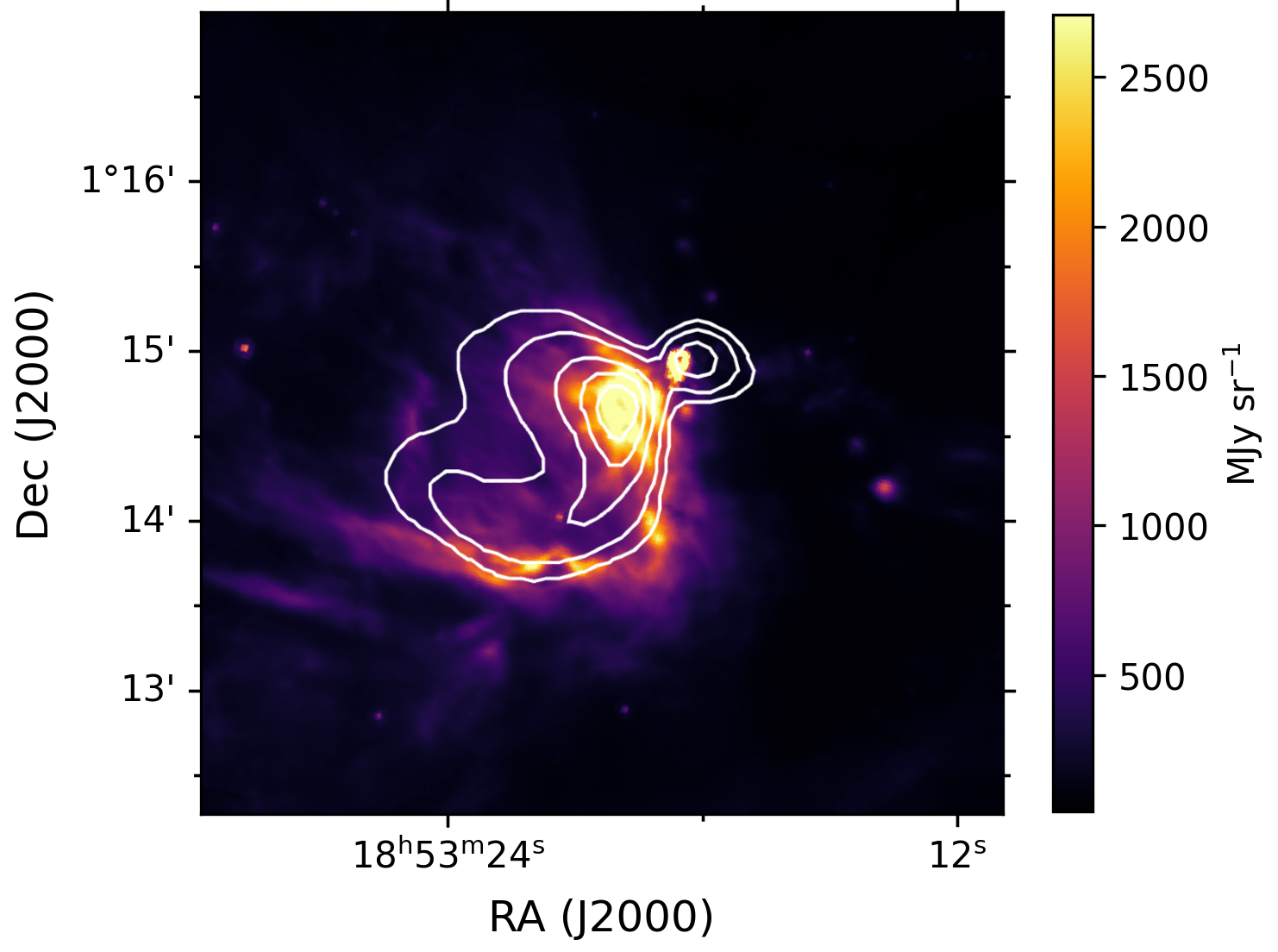}
    \includegraphics[width=\columnwidth]{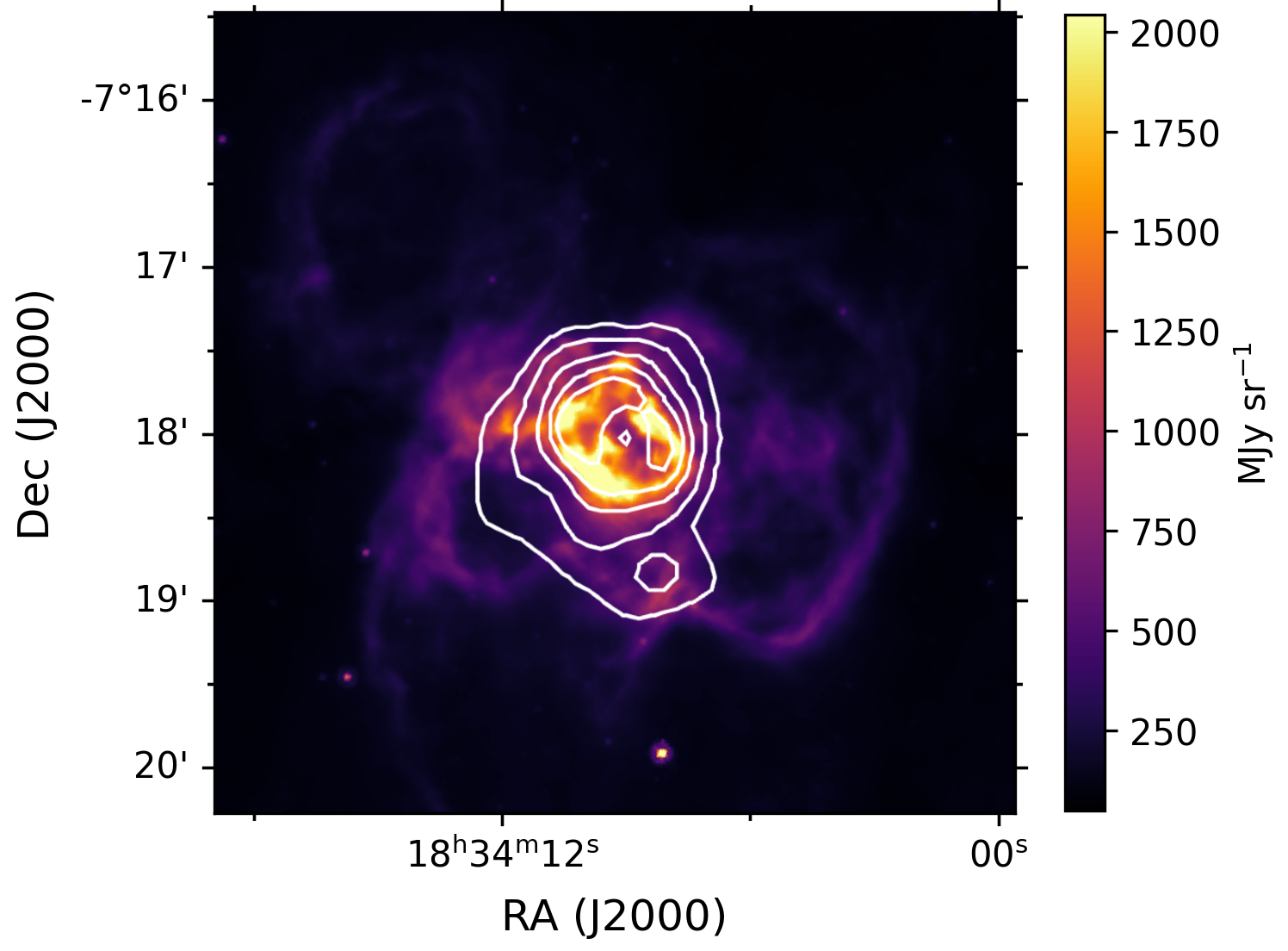}
    \caption{8$\mu$m emission towards Galactic \HII regions G049.205$-$0.343 (top left), G034.256+0.145 (top right) and G024.471+0.492 (bottom middle) from GLIMPSE survey, overlaid with VLA continuum contours from this work. The contours are 1\%, 5\%, 10\%, 20\%, 40\% 60\% and 80\% of the brightest 1--2 GHz radio continuum (T$_c$=1460 K, 2310 K, 810 K for G049.205$-$0.343, G034.256+0.145 and G024.471+0.492, respectively)} 
    \label{fig:GLIMPSE}
\end{figure*}

We select three fields for this work, containing the Galactic \HII regions G049.205$-$0.343 (`G049' hereafter), G034.256+0.145 (`G034' hereafter), and G024.471+0.492 (`G024' hereafter), as listed in Table \ref{tab:sources} and shown in Figure \ref{fig:GLIMPSE}. G049 has a champagne flow morphology \citep{Koo1999ApJ...518..760K} and is the largest and brightest \HII region within the W51B molecular cloud \citep{Kang2010ApJS..190...58K}. G034 contains multiple \HII regions, including the extended \HII region G034.256+0.136 (`G034 Extended' hereafter) and the ultra-compact (UC) `cometary \HII region' G034.256+0.155 \citep[`G034 Compact' hereafter, e.g.,][]{Reid1985ApJ...288L..17R, Mookerjea2007ApJ...659..447M, Tan2023}. G024 has a ring morphology containing several hyper-compact (HC) and UC \HII regions embedded in an expanding molecular ring \citep{Saha2024ApJ...970L..40S}.

All three sources have detections of the OH satellite-line flip, albeit sometimes marginal, from archival observations. G049 was selected from the conjugated satellite line pairs identified in \citet{RugelTHOR2018A&A...618A.159R}. G034 was identified as a satellite line flip candidate by \citet{BeutherTHOR2016A&A...595A..32B}. G024 was identified from direct inspection of the HI/OH/Recombination line survey of the Milky Way (THOR) data \citep{BeutherTHOR2016A&A...595A..32B, WangTHOR2020A&A...634A..83W}. Each source is also spatially and kinematically distinct from other nearby radio sources, and has extended radio continuum flux $>$200 mJy/beam at 1.6 GHz, which corresponds to an OH optical depth sensitivity of 0.02 (at $\sim$5.5 mJy/beam), from a 3.5-hour observation.

\subsection{VLA Observations}
\label{sec:vlaobs}

The three sources were observed with the Karl G. Jansky Very Large Array (VLA) with the L-band receiver (1--2 GHz) configured to simultaneously observe 38 H RRLs (H149$\alpha$ to H186$\alpha$), the four OH hyperfine ground-rotational state transitions (1612, 1665, 1667, 1720-MHz), the \HI 21-cm line, and broadband radio continuum (929--2029 MHz) under the project code 21A-153 (PI Dawson). 
The sources are all large enough on the sky ($\sim$2--3') to be spatially resolved by the synthesized beam ($\sim$20''), but smaller than the largest detectable angular scale ($\sim$970''). 
The observations took place between May 2021 and July 2021 over 12 observing sessions.
Each source was observed for a total of 14 hours with one 3.5 hour session in the compact D-configuration (maximum baseline: 1 km) and three 3.5 hour sessions in the more extended C-configuration (maximum baseline: 3.6 km). 
At the beginning of each session, a primary calibrator source, 3C286, was observed for $\sim$15 minutes for bandpass and flux calibration, and a secondary calibrator with low polarization, J1407+2827, was observed for $\sim$10 minutes for polarization calibration. For each source, another secondary calibrator, close to the science target on sky, was observed for 2 minute intervals every 20 minutes for complex gain calibration (J1925+2106 for G049, J1812-0648 for G034 and J1822-0938 for G024). 

The correlator set-up is detailed in Appendix \ref{sec:appendixA}. The radio continuum bandpass was split into 16 low spectral resolution windows (hereafter, continuum windows). For each spectral window, the center frequency, channel width and bandwidth are listed in Table \ref{tab:corr}. The spectral line and rest frequency for each spectral window are also listed. The H RRLs, four OH lines and \HI 21-cm line have velocity resolutions of $\sim$0.9 km s$^{-1}$, $\sim$0.1 km s$^{-1}$ and $\sim$0.4 km s$^{-1}$, respectively. \HI absorption is complex and it is challenging to isolate the components associated with \HII regions, so it is not analyzed in this work.

\subsection{Archival CO}
To complement the VLA observations and to trace the wider molecular environment of the sources we examine $^{13}$CO(1--0) from the FOREST Unbiased Galactic plane Imaging (FUGIN) survey \citep{FUGIN}. We choose $^{13}$CO instead of $^{12}$CO because $^{12}$CO becomes optically thick easily, and hence may not adequately represent the spatial morphology of the bulk molecular gas. The spectral features observed in our OH data cannot be matched identically to features in the FUGIN data due to its high and varying noise level (0.6--3.1 K) and coarser spectral resolution (0.65 km s$^{-1}$). We therefore do not attempt to fit or decompose the CO spectra. The data have an angular resolution of 21" and are fully spatially sampled. 

\subsection{Archival 8-micron}
We examine 8$\mu$m emission from the GLIMPSE survey \citep{GLIMPSE2009PASP..121..213C}. This emission is primarily from polycyclic aromatic hydrocarbons (PAHs) and traces excited dust around the \HII regions. As the sources in this work are away from other bright sources, the 8$\mu$m emission is unambiguously associated with the \HII regions. 

\section{Data Reduction \& Analysis Methods}
\label{sec:methods}

\subsection{VLA Data Reduction}
\label{sec:vla-reduct}

We calibrate, image and clean the VLA data using the \texttt{Wenger Interferometry Software Package} \citep[WISP,][]{WISP2018ascl.soft12001W}. \texttt{WISP} is a wrapper for the \texttt{Common Astronomy Software Applications} \citep[\texttt{CASA},][]{CASA} in \texttt{Python} which automates calibration, imaging and cleaning processes. The data reduction process follows a similar method to \citet{WengerSHRDS2019ApJS..240...24W}, which we outline below.

\subsubsection{Calibration}
We calibrate the data with \texttt{WISP}, following the standard VLA calibration procedure. Initially, radio frequency interference (RFI) are flagged using the \texttt{TFCROP} algorithm within \texttt{CASA}. 
The primary calibrator is used for bandpass, delay and flux density calibration. The secondary calibrators are used to calculate complex gain solutions. The polarization calibration.

We apply the calibration solutions for the bandpass, phase, gain, flux and polarization to the science data. RFI in the science data was first flagged automatically using the \texttt{RFLAG} algorithm within \texttt{CASA}. We then manually flag further RFI after visual inspection. Ten continuum windows and 13 H RRL spectral windows are dominated by RFI ($>$ 50\% of channels) so we completely flag them. These windows are labeled with an ``*'' next to their number in Appendix \ref{sec:appendixA}.

\subsubsection{Imaging \& Cleaning}
We perform the entire cleaning and imaging process using \texttt{WISP}.
Since the synthesized beam of the highest spectral window is $\sim$14'', we generate images with a pixel size of 4.5'' to ensure that there are at least 3 pixels across the full width at half maximum (FWHM) of the synthesized beam. We use Briggs weighting with \texttt{robust}$=$0.5 \citep{Briggs1995AAS...18711202B}.  
The emission morphology of our sources is extended and complex. Therefore we use the multiscale CLEAN algorithm \citep{multiscaleCornwell} on the three-dimensional data cubes, with scales of 0, 6, 12 and 18 pixels and the Multi-scale Multi-Frequency Synthesis (MSMFS) algorithm \citep{MSMFS2011A&A...532A..71R} to produce the two-dimensional continuum images. The common physical origin of the RRLs and continuum means we can use continuum CLEAN masks for RRL cubes. While OH does not have the same physical origin as the continuum, we found that all OH detections indeed lay within the generated CLEAN masks. Therefore we use the \texttt{auto-multithresh} algorithm in \texttt{CASA} to generate the CLEAN masks for all data products (see \citealt{WengerSHRDS2019ApJS..240...24W}). 
For each field, we produce four OH data cubes  (1612, 1665, 1667 and 1720-MHz), 25 RRL data cubes and one broadband continuum MSMFS image. The continuum is not subtracted from either the OH or RRL spectral line cubes. 

In order to increase the signal-to-noise of the detected RRLs, for each \HII region, we smooth the remaining higher frequency RRL cubes (H149$\alpha$ to H170$\alpha$) to a common beam size (Table \ref{tab:beam}) and spectral resolution (0.88 km s$^{-1}$) and stack them to produce a \avghna cube, following the method of \citep{WengerSHRDS2019ApJS..240...24W}. 

To prepare the OH data cubes for fitting, we also smooth the four OH data cubes to a common beam size (Table \ref{tab:beam}) and spectral resolution (0.09 km s$^{-1}$). G034 and G049 contain multiple high-gain OH masers in the main lines that are both spatially and spectrally coincident with the \HII regions. High gain masers do not appear as simultaneous four-line features and their sidelobes contaminate the spectra, therefore we mask out the velocity channels corresponding to maser emission. 

\begin{table}
\centering
\caption{Parameters of the final effective beam for RRL data cubes (top) and OH data cubes (bottom) from each \HII region.}
\begin{tabular}{c}
     \hna  \\ 
\end{tabular}

\begin{tabular}{cccc}
\hline
Source & Major Axis & Minor Axis & Position angle \\
(Name) & (arcsecond) & (arcsecond) & (degrees) \\
\hline
G049.205$-$0.343 & 18.5 & 18.2 & 1.9 \\
G034.256+0.145 & 24.9 & 21.3 & 0.3 \\
G024.471+0.492 & 24.1 & 22.4 & 2.2 \\ \hline
\end{tabular}

\begin{tabular}{c}
     Hydroxyl (OH)  \\ 
\end{tabular}

\begin{tabular}{cccc}
\hline
Source & Major Axis & Minor Axis & Position angle \\
(Name) & (arcsecond) & (arcsecond) & (degrees) \\
\hline
G049.205$-$0.343 & 14.4 & 14.4 & 1.3 \\
G034.256+0.145 & 16.7 & 15.9 & 1.8 \\
G024.471+0.492 & 18.0 & 17.2 & 1.3 \\ \hline
\end{tabular}

\label{tab:beam}
\end{table}

\subsection{RRL Gaussian Decomposition}
\label{sec:RRLfit}

We fit the \avghna spectra to identify and map the kinematics and morphology of the ionised gas towards each source.
The thermal and non-thermal motions of ionised gas within \HII regions can be assumed to form a Gaussian line profile \citep[e.g.][]{Balser2021ApJ...921..176B}. From inspection, we found the \avghna features to be well-described by single Gaussian line profiles. We fit the features with a single Gaussian profile using the \texttt{scipy.optimize} package \texttt{curve\_fit} \citep{scipy}. We discard fits with signal-to-noise S/N $<10$, following \citet{WengerSHRDS2019ApJS..240...24W}.

\subsection{OH Gaussian Decomposition}
\label{sec:OHfit}

We fit the blended OH spectra towards our sources using a newly developed approach for spatially coherent Gaussian decomposition. A spectral feature should be present in all four OH 18-cm transitions, so we assume that the centroid velocities and FWHM of a given Gaussian component are the same in each of the four transitions, such that the lines can be fit simultaneously. In addition, the OH emission and absorption should be spatially coherent across the source: each spectral feature should appear in the data across multiple pixels, although the degree of blending and S/N of each feature may vary.

Existing algorithms cannot simultaneously fit the four OH lines in a spatially coherent way. For example, \texttt{ROHSA}, which is designed for \hi 21-cm emission line decomposition, incorporates spatial regularization, but cannot simultaneously fit the four OH transitions \citep{ROHSA2019A&A...626A.101M}. The \texttt{AMOEBA} algorithm aims to identify the best number of Gaussian components for single sightline OH spectra, but cannot deal with spatial regularization across neighboring sightlines \citep{HafnerAMOEBA2021}. Therefore, we develop a new approach for this work. 

\begin{figure}
    \centering
    \includegraphics[width=\columnwidth]{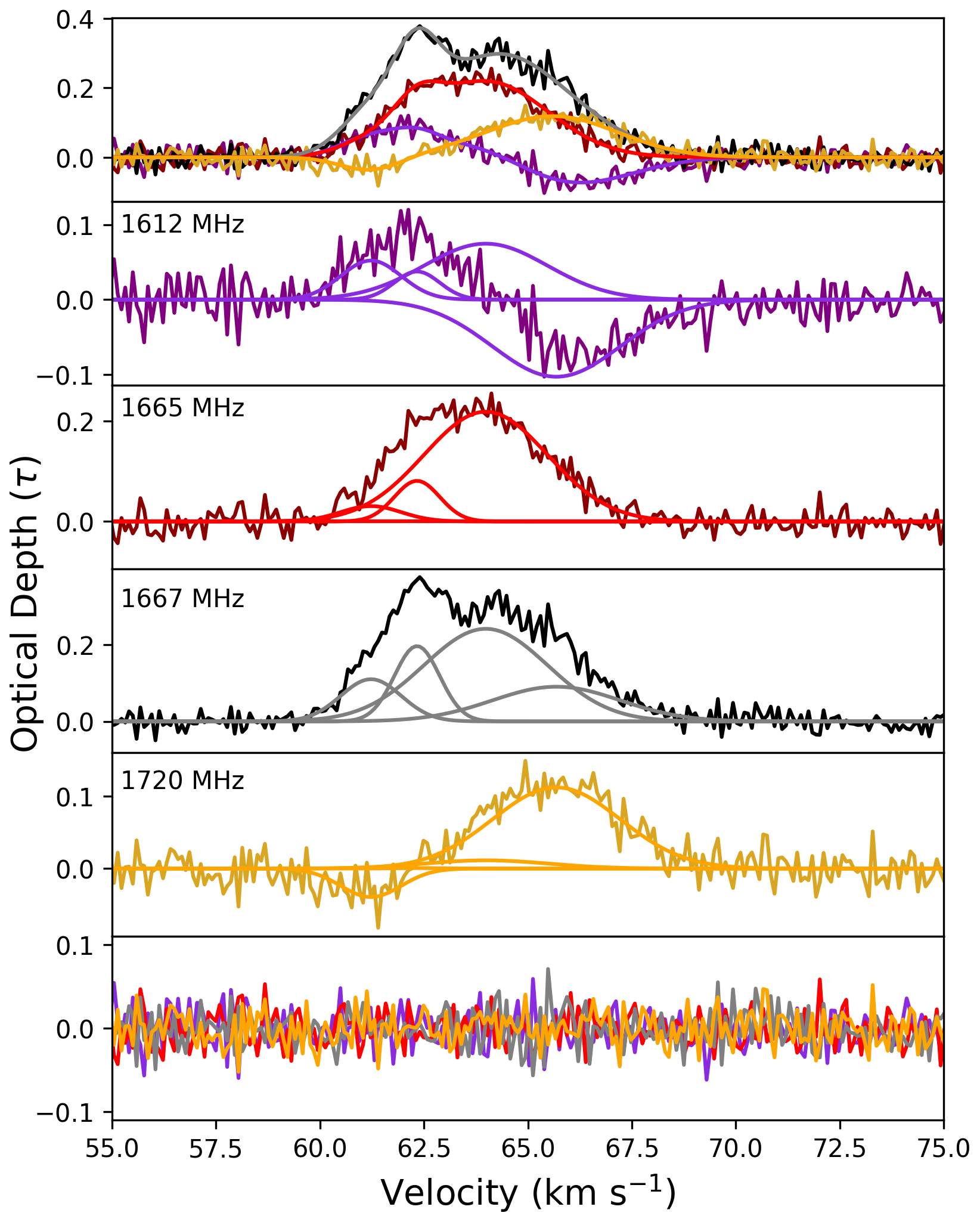}
    \caption{Example of Gaussian Decomposition of OH optical depth spectra from a single sightline towards G049.205$-$0.343. The top panel shows the total model for each OH line (1612-MHz in purple, 1665-MHz in red, 1667-MHz in gray and 1720-MHz in gold). The following panels show the individual fitted Gaussian components for the 1612, 1665, 1667 and 1720-MHz lines. The bottom panel shows the fit residuals. See Section \ref{sec:OHfit} for details.}
    \label{fig:G049MIMOSA}
\end{figure}

\begin{figure}
    \centering
    \includegraphics[width=\columnwidth]{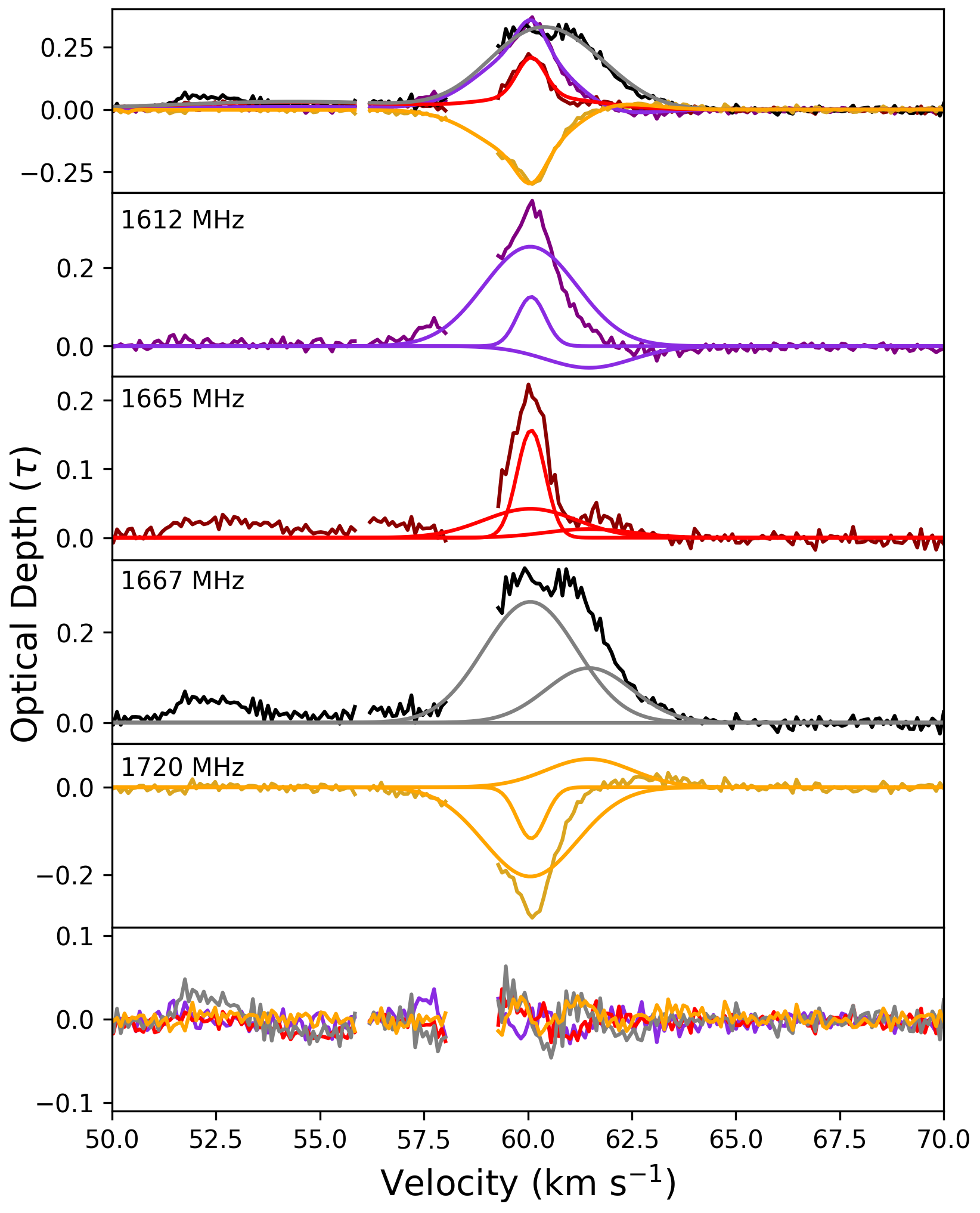}
    \caption{Same as Figure \ref{fig:G049MIMOSA} for a single sightline towards G034.256+0.145. Note that the features at $\sim$52.5 km s$^{-1}$ and $\sim$56 km s$^{-1}$ are fit during the Gaussian decomposition process described in Section \ref{sec:OHfit}. These components were excluded as they corresponded to diffuse foreground clouds which are not related to the \HII region.}
    \label{fig:G034MIMOSA}
\end{figure}

\begin{figure}
    \centering
    \includegraphics[width=\columnwidth]{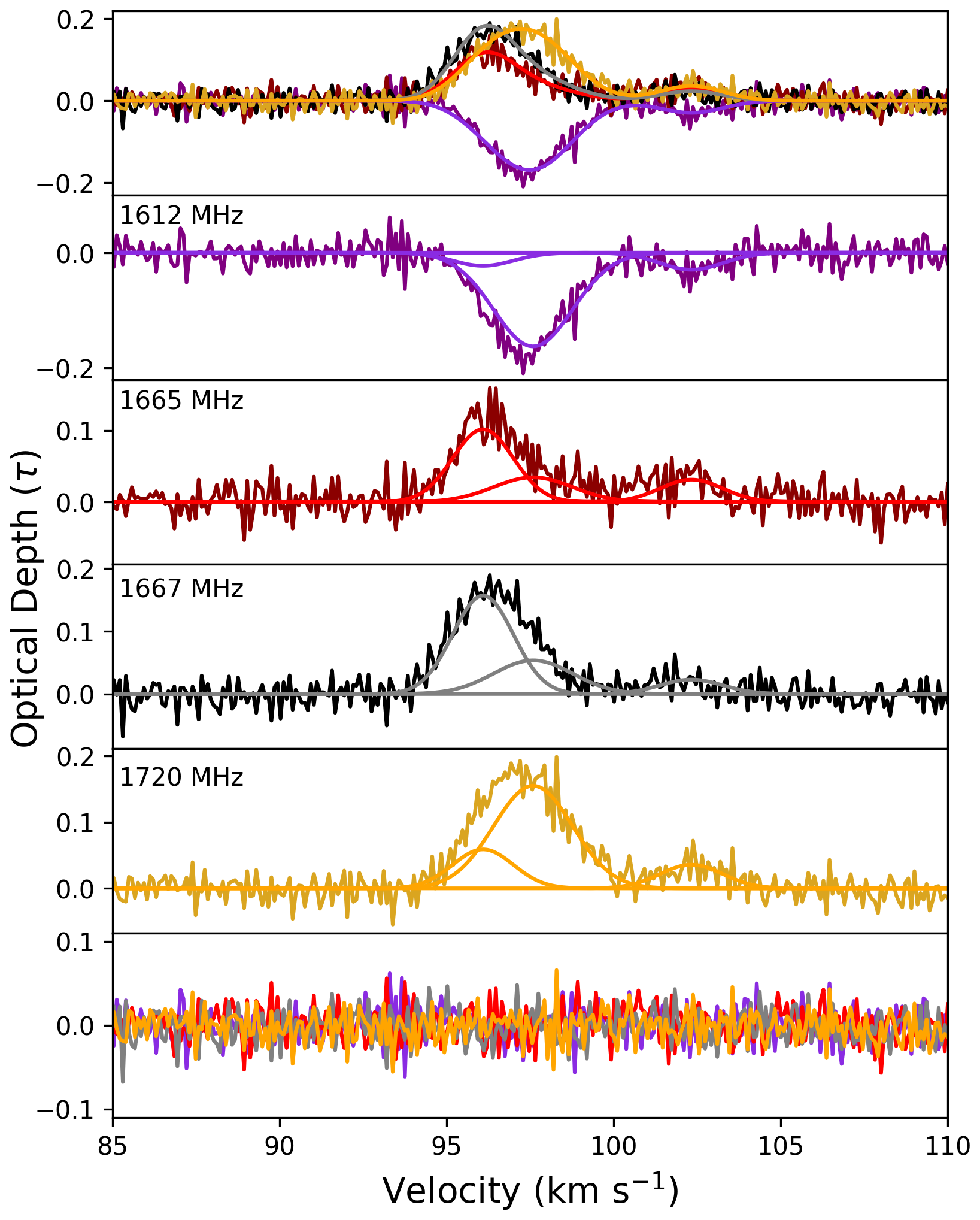}
    \caption{Same as Figure \ref{fig:G049MIMOSA} for a single sightline towards G024.471+0.492 (RA: 18$^{\text{h}}$ 34$^{\text{m}}$ 08$^{\text{s}}$, Dec: -7$^{\circ}$ 18' 07'').}
    \label{fig:G024MIMOSA}
\end{figure}

We present the full details of our OH fitting procedure in Appendix \ref{sec:appendixOHfit}, but we provide a brief overview here.
We first transform the OH spectra from brightness temperature spectra to optical depth spectra as $\tau_v = -\ln\left(T_{\text{b}}(v)/T_{\text{c}}\right)$, where $T_{\text{b}}(v)$ is the observed spectral brightness temperature (including the continuum) and $T_{\text{c}}$ is the continuum brightness temperature. To boost the S/N of the OH spectra, we select a subset of pixels containing all visible OH emission and the we spatially bin those pixels using Voronoi tessellation \citep{voronoi2003MNRAS.342..345C}. We simultaneously fit the four OH transitions from each binned pixel with $N$ Gaussians using \texttt{lmfit} \citep{lmfit}. To promote spatial coherence across the fits, the subset of OH pixels are binned into a series of `resolution cubes', inspired by \texttt{ROHSA}. Each resolution cube is sequentially fit, from coarsest to finest resolution, passing on their best fits as initial guesses to the corresponding pixels in the subsequent cube. The best fit models for each pixel are chosen with the Bayesian Information Criterion (BIC). Example fits are presented for G049 (Figure \ref{fig:G049MIMOSA}), G034 (Figure \ref{fig:G034MIMOSA}), and G024 (Figure \ref{fig:G024MIMOSA}). We then group and sort the fitted spectral features into `spatio-kinematic' components that map each feature, which are presented in Section \ref{sec:OHresults}.

\section{Results}
\label{sec:results}

\subsection{H RRL Kinematics \& 1.6 GHz Continuum Morphology}
\label{sec:RRL-cont-results}

\begin{figure*}[!htb]
    \centering
    \includegraphics[width=0.8\textwidth]{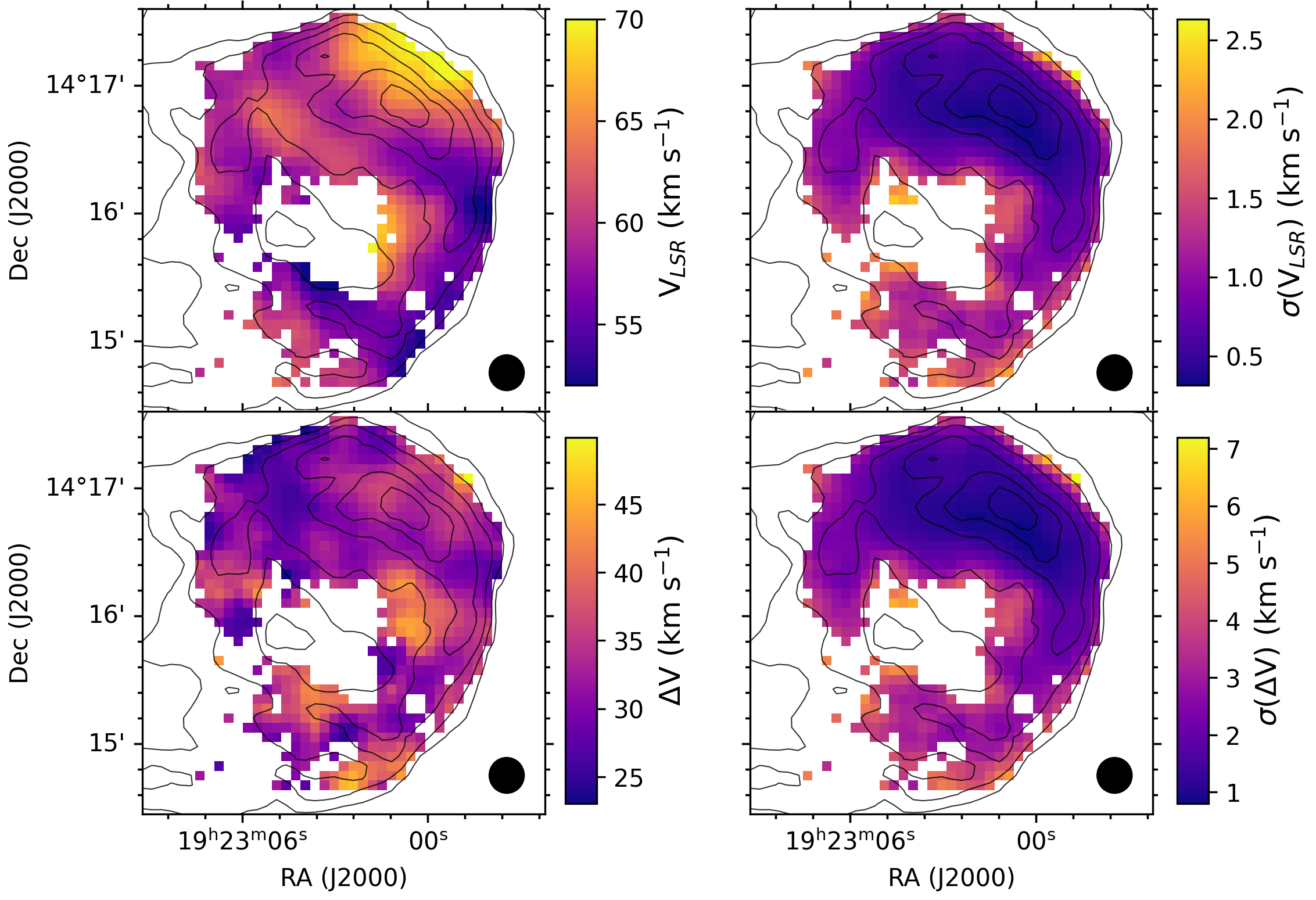}
    \caption{Maps of fitted \avghna RRL spectra from G049.205$-$0.343. Top: Line of sight velocity \VLSR from centroids of the Gaussian fits of the \avghna spectra and associated fit error $\sigma$(\VLSR). Bottom: Fitted Gaussian FWHM line width $\Delta$V and associated fit error $\sigma(\Delta$V$)$. The contours are 1\%, 5\%, 10\%, 20\%, 40\% 60\% and 80\% of the brightest radio continuum, $T_{\text{c}}$=1460 K. The synthesized beam size is represented by the filled black ellipse in the bottom right of each plot.}
    \label{fig:G049ionised}
\end{figure*}

\begin{figure*}[!htb]
    \centering
    \includegraphics[width=0.8\textwidth]{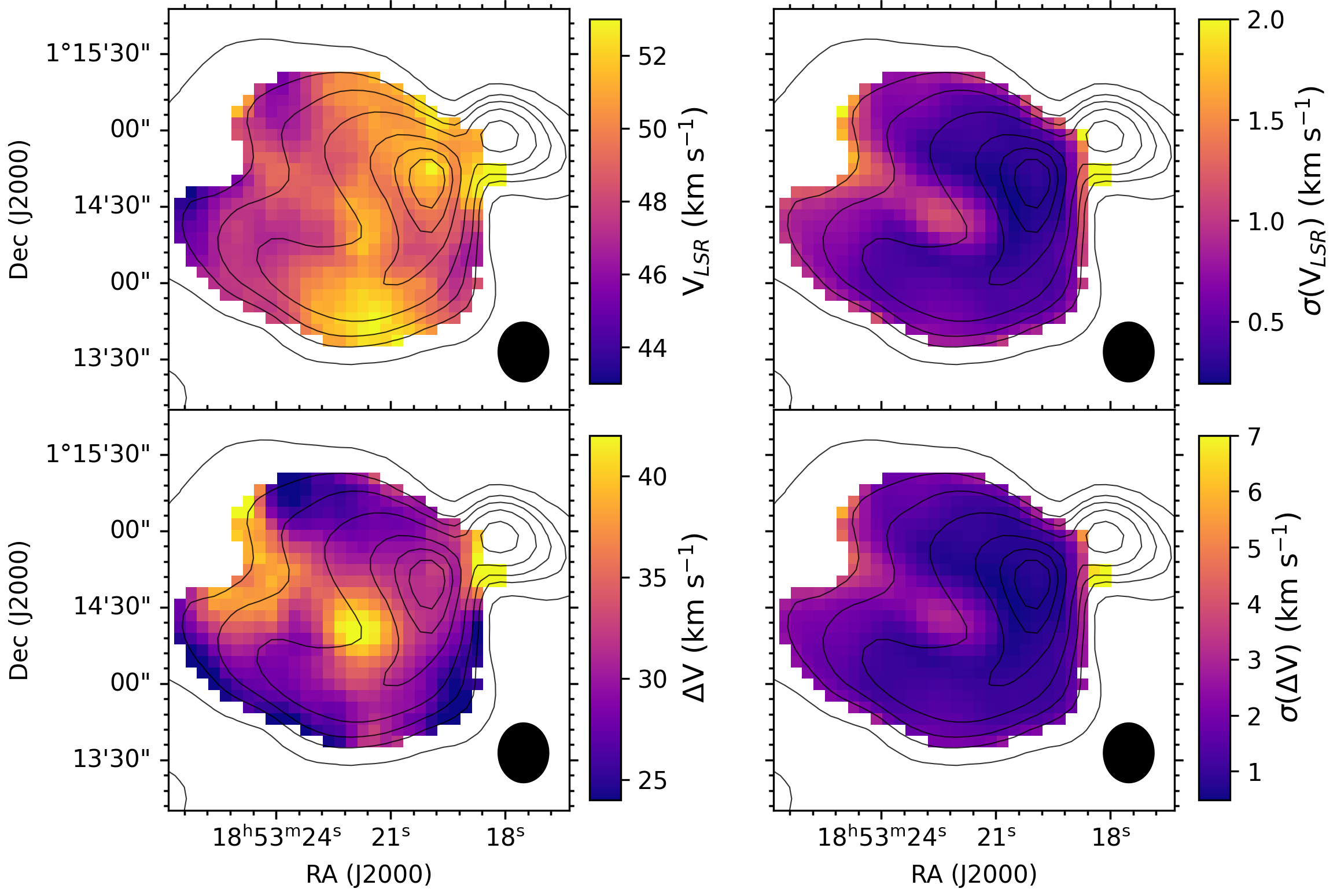}
    \caption{Maps of fitted \avghna RRL spectra from G034.256+0.145. The contours are 1\%, 5\%, 10\%, 20\%, 40\% 60\% and 80\% of brightest radio continuum, $T_{\text{c}}$=2310 K. Other panel details are the same as for Figure \ref{fig:G049ionised}.}
    \label{fig:G034ionised}
\end{figure*}

\begin{figure*}[!htb]
    \centering
    \includegraphics[width=0.8\textwidth]{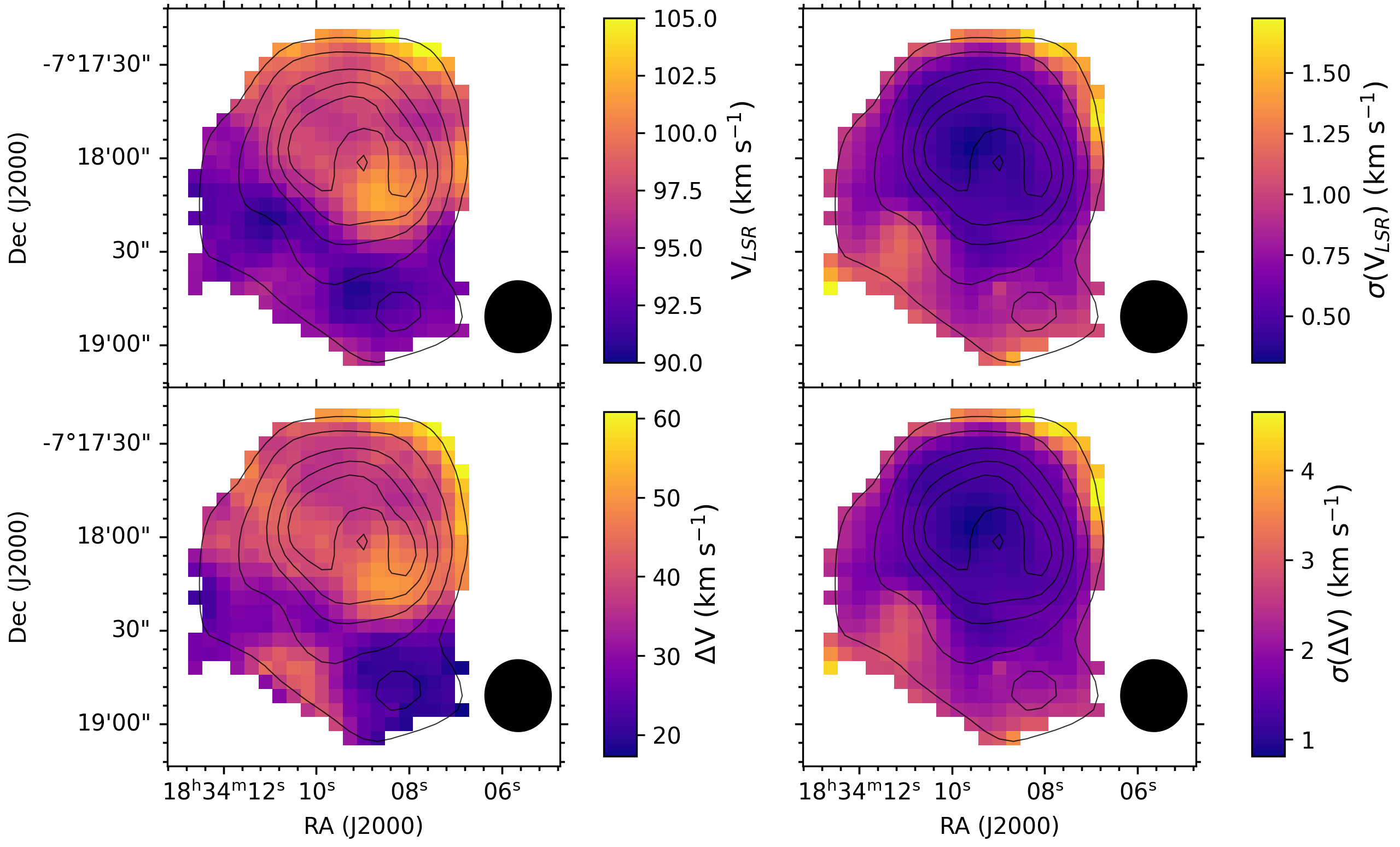}
    \caption{Maps of fitted \avghna RRL spectra from G024.471+0.492. The contours are 1\%, 5\%, 10\%, 20\%, 40\% 60\% and 80\% of brightest radio continuum, $T_{\text{c}}$=810 K. Other panel details are the same as for Figure \ref{fig:G049ionised}.} 
    \label{fig:G024ionised}
\end{figure*}

We investigate the kinematics and morphology of the ionized gas within the \HII regions in this work using spatially resolved maps of the \avghna RRL line-of-sight velocity and FWHM distributions.
We present the 1.6 GHz continuum and \avghna RRL kinematic maps of G049.205$-$0.343 (Figure \ref{fig:G049ionised}), G034.256+0.145 (Figure \ref{fig:G034ionised}), and G024.471+0.492 (Figure \ref{fig:G024ionised}). Each figure presents the line-of-sight velocity, V$_{\text{LSR}}$, and FWHM, $\Delta V$, of the single-component Gaussian fits and their associated fit uncertainties.

The 1.6 GHz continuum morphology of G049 shows a bright `head' to the north-west and an apparent outflow to the south-east, which is in agreement with G049's classification as a champagne flow type \HII region \citep{Carpenter1998AJ....116.1856C}. While the ionized gas velocity structure across the region is complex, the \avghna RRL kinematics of G049 show evidence for an overall velocity gradient that is consistent with its champagne flow morphology. The north-west is more redshifted and the south-east, tracing the outflow, is more blueshifted, suggesting that the outflow is flowing towards the observer. There is also complex structure in the FWHM distribution.

G034 Extended and G034 Compact can be seen distinctly in the 1.6 GHz continuum, with G034 Compact to the north-west of G034 Extended. G034 Extended has a similar brighter `head' to G049, but its continuum peak is offset to the north-west from the center of the bright `head'. G034 Compact is only partially resolved in these observations.

Similarly to G049, the \avghna RRL kinematics of G034 Extended show an overall velocity gradient from the west towards the north-east, with the west being slightly more redshifted and the north-east being more blueshifted. The velocity structure is inhomogeneous across the source and the FWHM distribution is centrally peaked. \citet{Balser2021ApJ...921..176B} made simple models of RRL spectra from \HII regions, where they found that \HII regions undergoing bulk rotation were characterized by a velocity gradient and centrally-peaked FWHM distributions. The slight velocity gradient and FWHM distribution of G034 Extended appear to match this characterization.

No RRLs are detected towards G034 Compact in our 1--2 GHz observations. For G034 Compact we measure a spectral index ($\alpha = \frac{d\log S}{d\log\nu}$, see Appendix \ref{sec:appendixalpha}) $\alpha  = 0.87 \pm 0.02$, indicating a high optical depth ($\sim \tau=1$), which is consistent with our non-detection of RRLs. The high frequency H53$\alpha$ RRL ($\sim$ 43 GHz) was detected towards G034 Compact in \citet{Sewilo2011} at V$_{\text{LSR}}$=47.2$\pm$1.2 km s$^{-1}$, indicating that our non-detection is indeed an optical depth effect.

The 1.6 GHz continuum morphology of G024 traces the ring shape described in \citet{Saha2024ApJ...970L..40S}, albeit at lower angular resolution. Our VLA observations do not have the spatial resolution to detect the individual UC and HC \HII regions identified in that work. The \avghna RRL kinematics of G024 do not show a smooth velocity gradient. Instead they appear to trace two distinct kinematic components with an abrupt jump of $\sim$10 km\,s$^{-1}$ between them, the north-west being more redshifted and the south-east section more blueshifted. A similar abrupt jump is also seen in the FWHM distribution between these two sections. The previously detected UC and HC \HII regions are in the north-west, but they have been blended together in the VLA beam.

Observations of optically thin radio continuum emission can be used with RRL spectra to measure physical properties of the ionized gas within an \HII region, such as electron temperature or electron density. We find, however, that the 1--2 GHz continuum from our sources is not optically thin. We fit the spectral indices, $\alpha$, of every pixel across our sources (see Appendix \ref{sec:appendixalpha}) and find that $\alpha > 0.05$ across all three sources, whereas an optically thin \HII region has $\alpha \sim -0.1$. Examples of fitted Spectral Energy Distributions (SEDs) are presented in Appendix \ref{sec:appendixalpha}. An alternative is to use the continuum optical depth alone, as measured from the SEDs. However, there is a caveat that the spatial filtering of the interferometer will tend to filter out more extended emission at higher frequencies---consequently, this decreases the SED slope and the gas appears more optically thin, and means that our measured optical depth is almost certainly an underestimate. We therefore do not attempt to use continuum optical depths directly to infer any physical properties of the ionized gas.

\subsection{OH Distribution \& Kinematics}
\label{sec:OHresults}

\begin{figure*}
    \centering
    \includegraphics[width=\textwidth]{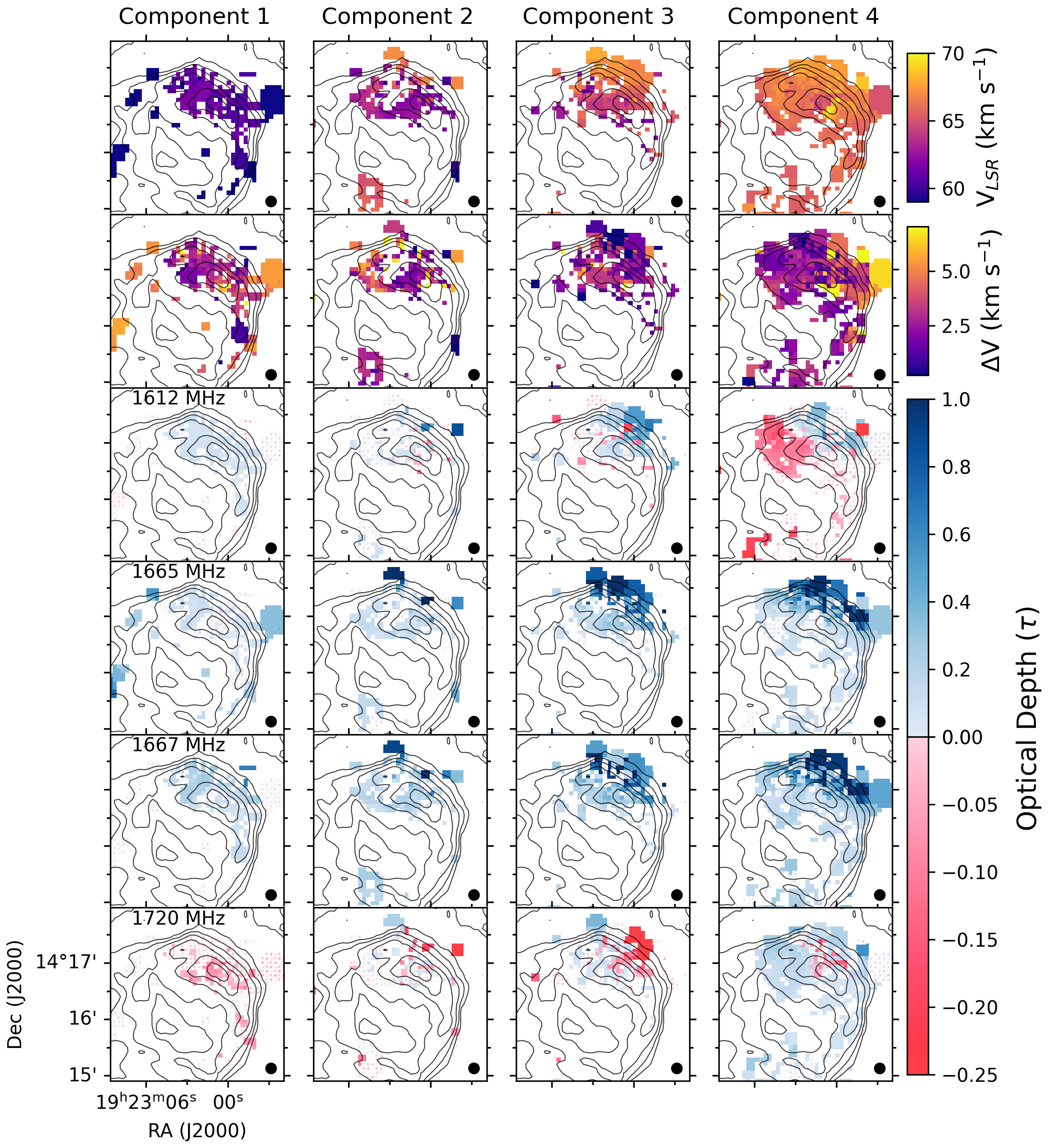}
    \caption{Maps of the fitted OH spatio-kinematic Components towards G049.205$-$0.343. Each column corresponds to a spatio-kinematic Component (1--4), ordered in increasing velocity. Each row shows a fit parameter: Top row: Line of sight velocity, V$_{\text{LSR}}$; second row: Gaussian FWHM line width, $\Delta$V; 3rd to 6th rows: optical depth amplitudes for the 1612-MHz, 1665-MHz, 1667-MHz and 1720-MHz lines. Negative optical depth corresponding to stimulated emission is in red, positive optical depth corresponding to absorption is in blue. The off-white boxes mark pixels where the fitted amplitudes have S/N$<$3 (see Section \ref{sec:results}). The contours are 1\%, 5\%, 10\%, 20\%, 40\%, 60\% and 80\% of the brightest radio continuum, $T_{\text{c}}$=1460 K. The synthesized beam size is represented by the filled black ellipses. Note, we do not plot the fit parameter errors as they are not a useful measurement of the true uncertainty on the models.}
    \label{fig:G049OH}
\end{figure*}

\begin{figure*}
    \centering
    \includegraphics[width=0.77\textwidth]{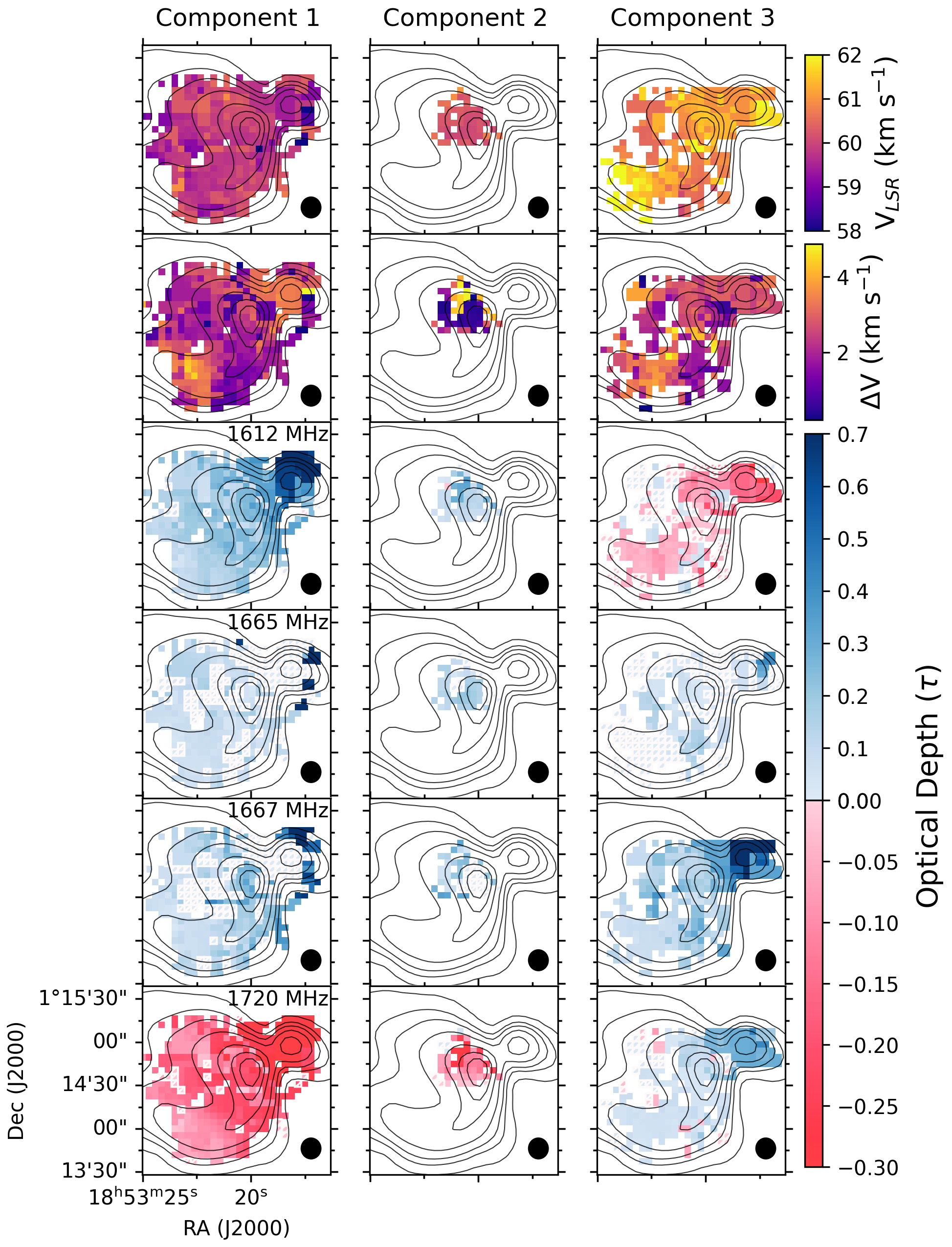}
    \caption{Maps of the fitted spatio-kinematic Components from OH optical depth spectra towards Galactic \HII region G034.256+0.145. Each column corresponds to a spatio-kinematic Component (1--3), ordered in increasing velocity. The contours are 1\%, 5\%, 10\%, 20\%, 40\%, 60\% and 80\% of the brightest radio continuum, $T_{\text{c}}$=2310 K. Other panel details are the same as for Figure \ref{fig:G049OH}.}
    \label{fig:G034OH}
\end{figure*}

\begin{figure*}[!htb]
    \centering
    \includegraphics[width=0.8\textwidth]{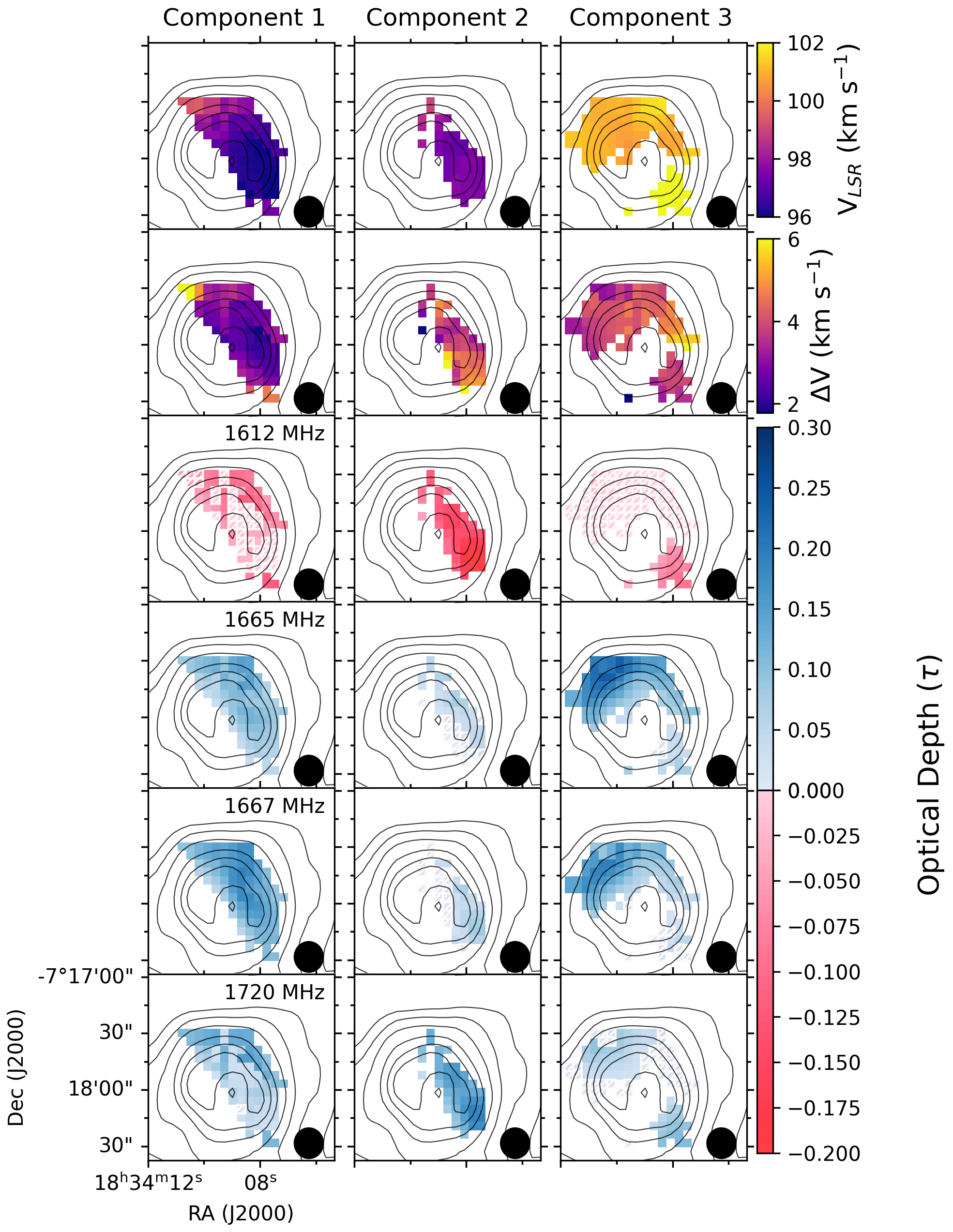}
    \caption{Maps of the fitted spatio-kinematic Components from OH optical depth spectra towards Galactic \HII region G024.471+0.492. Each column corresponds to a spatio-kinematic component (1--3), ordered in increasing velocity. The contours are 1\%, 5\%, 10\%, 20\%, 40\%, 60\% and 80\% of the brightest radio continuum, $T_{\text{b}}$=810 K. Other panel details are the same as for Figure \ref{fig:G049OH}. }
    \label{fig:G024OH}
\end{figure*}

OH is detected towards all three sources out to a continuum background brightness of $T_c \sim 250$ K, which corresponds to the expected sensitivity limit of the VLA observations. We detect the satellite-line flip towards two of our sources, G049.205$-$0.343 and G034.256+0.145. The flip is not present in G024.471+0.492, implying that the prior weak detection we identified in the (lower sensitivity) THOR survey data \citep{BeutherTHOR2016A&A...595A..32B} is spurious. We present maps of each fitted OH Gaussian spatio-kinematic component from G049 (Figure \ref{fig:G049OH}), G034 (Figure \ref{fig:G034OH}), and G024 (Figure \ref{fig:G024OH}). Each figure shows the fitted V$_{\text{LSR}}$, FWHM and optical depth of the four OH transitions, for all fitted velocity components.

The \citetalias{HafnerFlipPaper} hypothesis predicts that where there is an OH flip associated with an \HII region, the blueshifted feature with 1720-MHz inversion and 1612-MHz absorption traces dynamically-interacting molecular gas, and the redshifted feature with 1612-MHz inversion and 1720-MHz absorption traces the undisturbed gas of the parent cloud. The OH spectra towards G049 are best fit with four Gaussian components. The champagne flow morphology and multiple, inhomogeneous velocity components in this source evidence a level of complexity not captured in the simple two-component model of \citetalias{HafnerFlipPaper}, but nevertheless, an OH flip is clearly seen. We clearly identify the flip between the most blueshifted component (Component One), which shows 1720-MHz inversion, and the most redshifted component (Component Four), which mostly shows 1612-MHz inversion. The two central components (Components Two and Three) have significant inhomogeneity in their optical depths, with the north-west tending to show more gas with 1720-MHz inversion and the south-east tending to show more gas with 1612-MHz inversion. The line-of-sight velocity of Component Four overlaps with the velocity of the `68 km s$^{-1}$' parent molecular cloud of G049 \citep{Carpenter1998AJ....116.1856C, Bik2019A&A...624A..63B}, and the north-west of this component shows all four OH lines in absorption with high optical depth. Interestingly, the central components of G049 show similar velocity gradients to the \avghna RRL velocity distribution, with the north-east being redshifted compared to the south-west. Regions of apparently higher FWHM, seen where the continuum is weak, likely reflect the tendency of multiple spectral features to blend together when the S/N is poor. 

In G034 Extended we detect the OH flip between Components One and Two, which have 1720-MHz inversion, and the redshifted Component Three, which has 1612-MHz inversion. Component Two is only detected towards the center of G034 Extended where the continuum is the brightest. This component may be very blended with Component One in other sections of G034 Extended where the background continuum from the \HII region is weaker. Unlike G049, we find that the OH features are redshifted by $\sim$ 10 km s$^{-1}$ with respect to the \avghna RRL \VLSR distribution. In general, the OH spectra towards G034 Extended show a much ``cleaner'' flip than G049, with fewer spectral components, and less spectral and spatial inhomogeneity across the source.

In G034 Compact, the main lines were mostly masked out as they contained masers. We see the flip in Components One and Three where the satellite lines are stronger than in G034 Extended.

The OH spectra towards G024 do not contain an OH satellite-line flip, and instead are best fit with three components where 1612-MHz is always in stimulated emission and 1720-MHz is always in absorption. This is unexpected as the flip was apparently detected in the THOR observations. However, of the three sources, the detection in G024 was the most marginal---with the 1720\,MHz emission showing at $\sim3\sigma$ in only a single 1.5 km s$^{-1}$ channel. The higher S/N observations in this work make it clear that the detection in the THOR data was a false positive. Component One and Three have distinct morphologies, only partially overlapping along the line of sight. Component Two appears where there is a brighter continuum background and may otherwise blend with Component One where the background is weaker.

\section{Discussion}
\label{sec:disc}

The aim of this work is to map the OH satellite-line flip associated with Galactic \HII regions and to confirm and constrain the basic features of the \citetalias{HafnerFlipPaper} flip hypothesis. We now ask whether the observed multi-wavelength spectral features, kinematics and morphology of our sources are in agreement with the picture put forward by \citetalias{HafnerFlipPaper}.

\citetalias{HafnerFlipPaper} used literature spectra to demonstrate kinematic and spatial association between examples of the OH flip and cataloged \HII regions. The majority of their reported examples were single-pointing or single-dish observations, with association determined on the basis of on-sky position and RRL velocities from the literature. In this work we perform spatially-resolved interferometric mapping to explore the behavior of the flip across our sources, and to allow for more robust spatial and kinematic association between the OH flip and the \HII region. Appendix \ref{sec:appendix2Dspectra} shows grids of OH and RRL spectra across all three of our \HII regions. The OH spectral features fall within the FWHM of the corresponding \avghna RRL for every position across every source, providing strong evidence for spatial and kinematic association, as expected from the \citetalias{HafnerFlipPaper} hypothesis. Focusing on the two \HII regions with the flip, we find that in G049 the offset between the RRL fitted velocity centroids and closest OH feature is typically $\lesssim5$ km\,s$^{-1}$, with a tendency of the north-west portions of the ionized gas to be redshifted relative to the OH and the south-east to be blueshifted. In G034 the \avghna RRL V$_{\text{LSR}}$ are systematically blueshifted relative to the OH by $\sim$10 km s$^{-1}$ across the region. Since G034's RRL linewidths are generally large---as high as $\Delta V\sim40$ km\,s$^{-1}$---this does not suggest a significant kinematic discrepancy. However, the offset is interesting and will be discussed further below.  

The observed OH spectra towards G049 and G034 also agree with the \citetalias{HafnerFlipPaper} prediction that for an OH flip associated with an \HII region, the component with 1720-MHz stimulated emission is always blueshifted with respect to the component with 1612-MHz stimulated emission. While this behavior was suggested from the lower S/N THOR data prior to our observations, we demonstrate here that it is seen across the full spatial extent of the sources, and emerges despite the existence of complex, multi-component velocity structure. \citetalias{HafnerFlipPaper} state that the 1720-MHz stimulated emission component is blueshifted because it is tracing dynamically-interacting shocked gas that has been pushed towards the observer (noting that OH is only detected in front of the \HII region). While we cannot directly test here whether the molecular gas has been shocked, we may test whether the velocity difference between the two components corresponds to a realistic shock velocity. Numerical 1D models of \HII region expansion give a plausible range of velocities for molecular gas shocked by the expansion, $\sim$1--8 km s$^{-1}$ \citep{Hosokawa2005ApJ...623..917H}. To test this prediction we measure the velocity difference between the first and last velocity components of G049 and G034, as they most clearly show the OH flip. We find that this is $\sim 8$ km s$^{-1}$ and $\sim 3$ km s$^{-1}$ for G049 and G034, respectively, which is within the expected range.

\begin{figure*}
    \centering
    \includegraphics[width=0.32\textwidth]{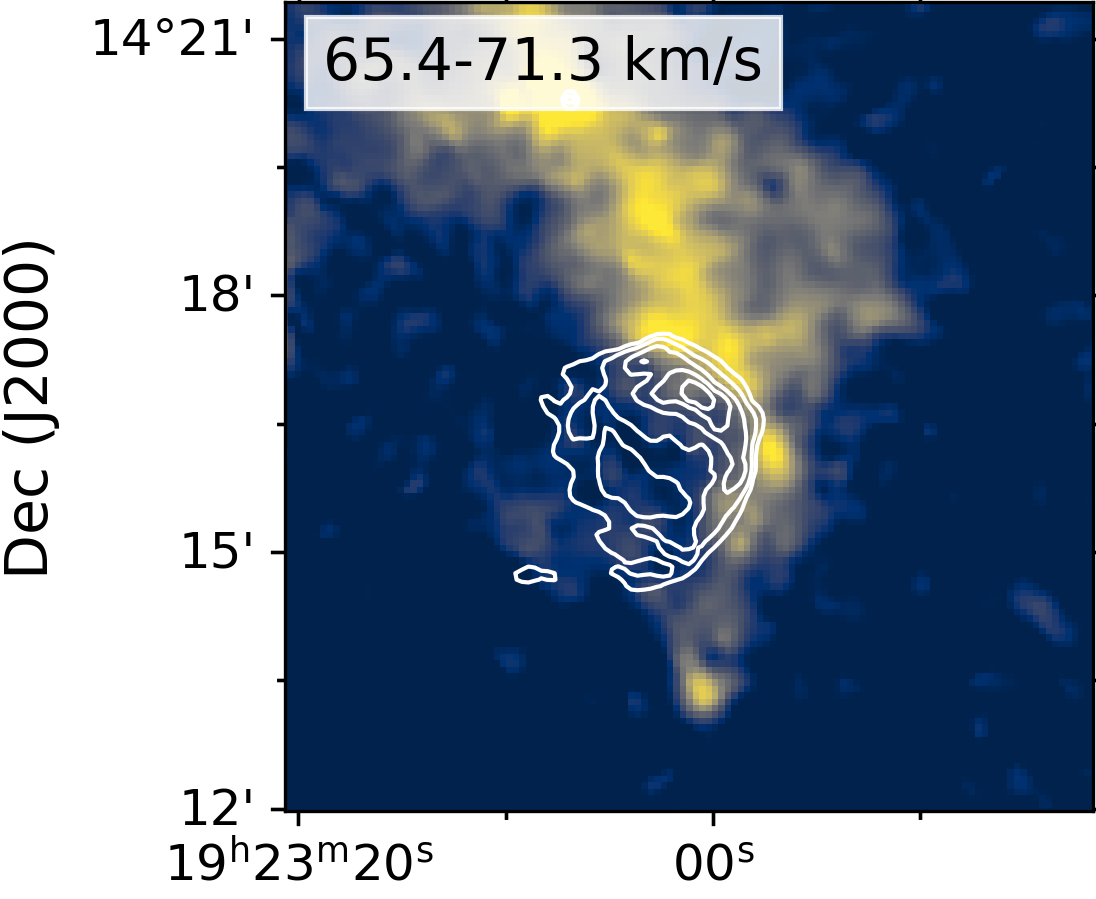}
    \includegraphics[width=0.32\textwidth]{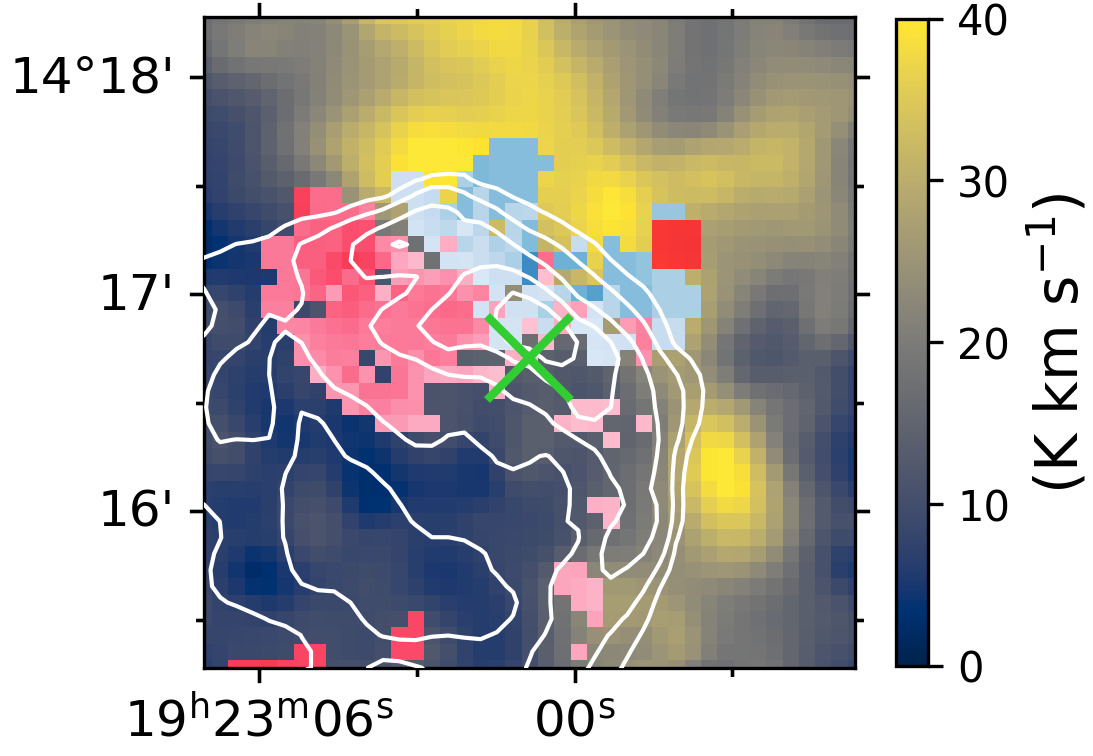}
    \includegraphics[width=0.315\textwidth]{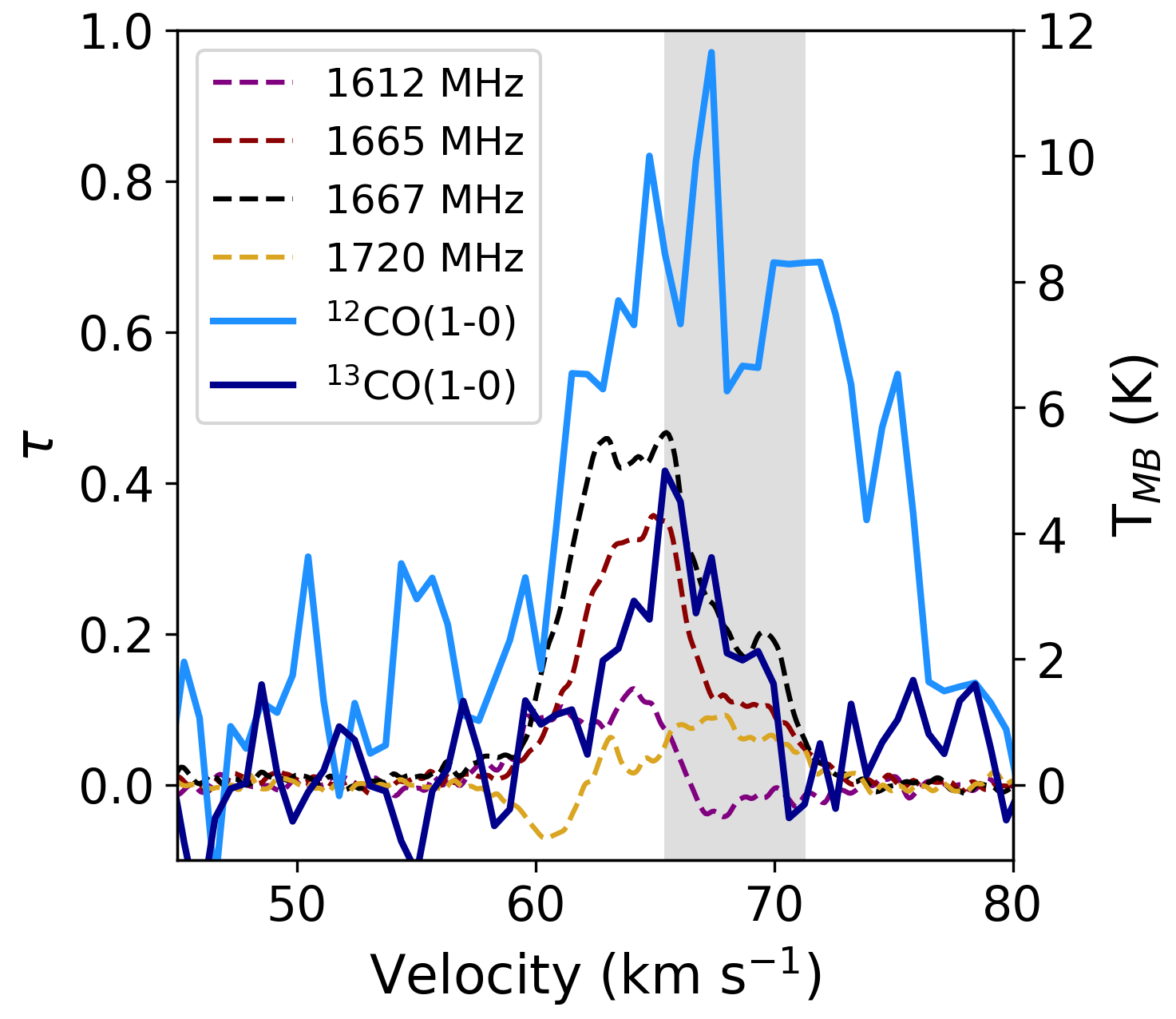}
    \includegraphics[width=0.3\textwidth]{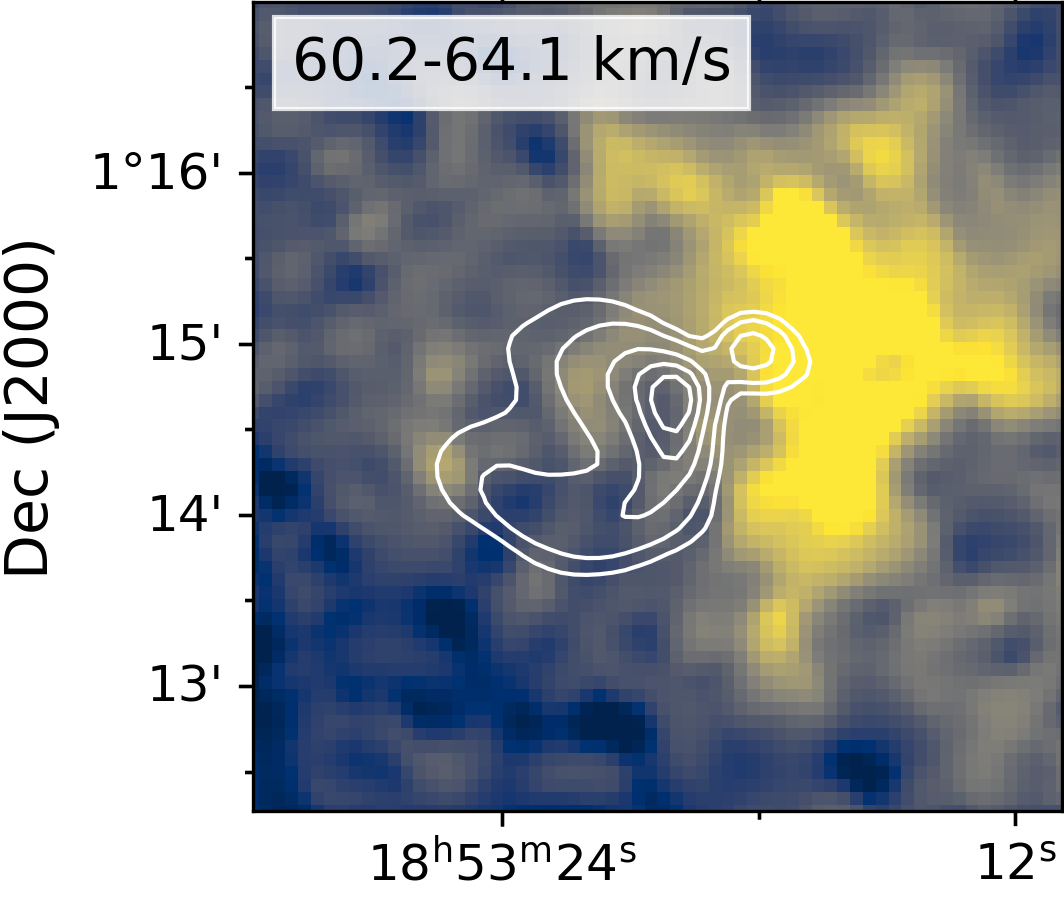}
    \includegraphics[width=0.32\textwidth]{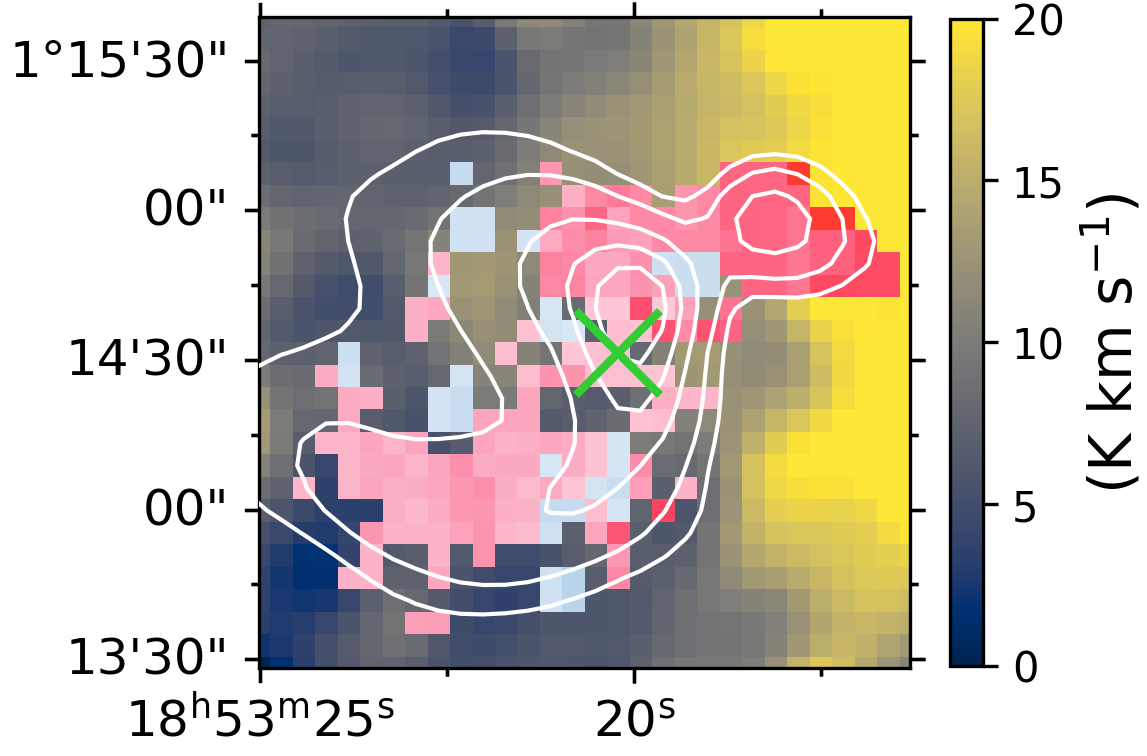}
    \includegraphics[width=0.315\textwidth]{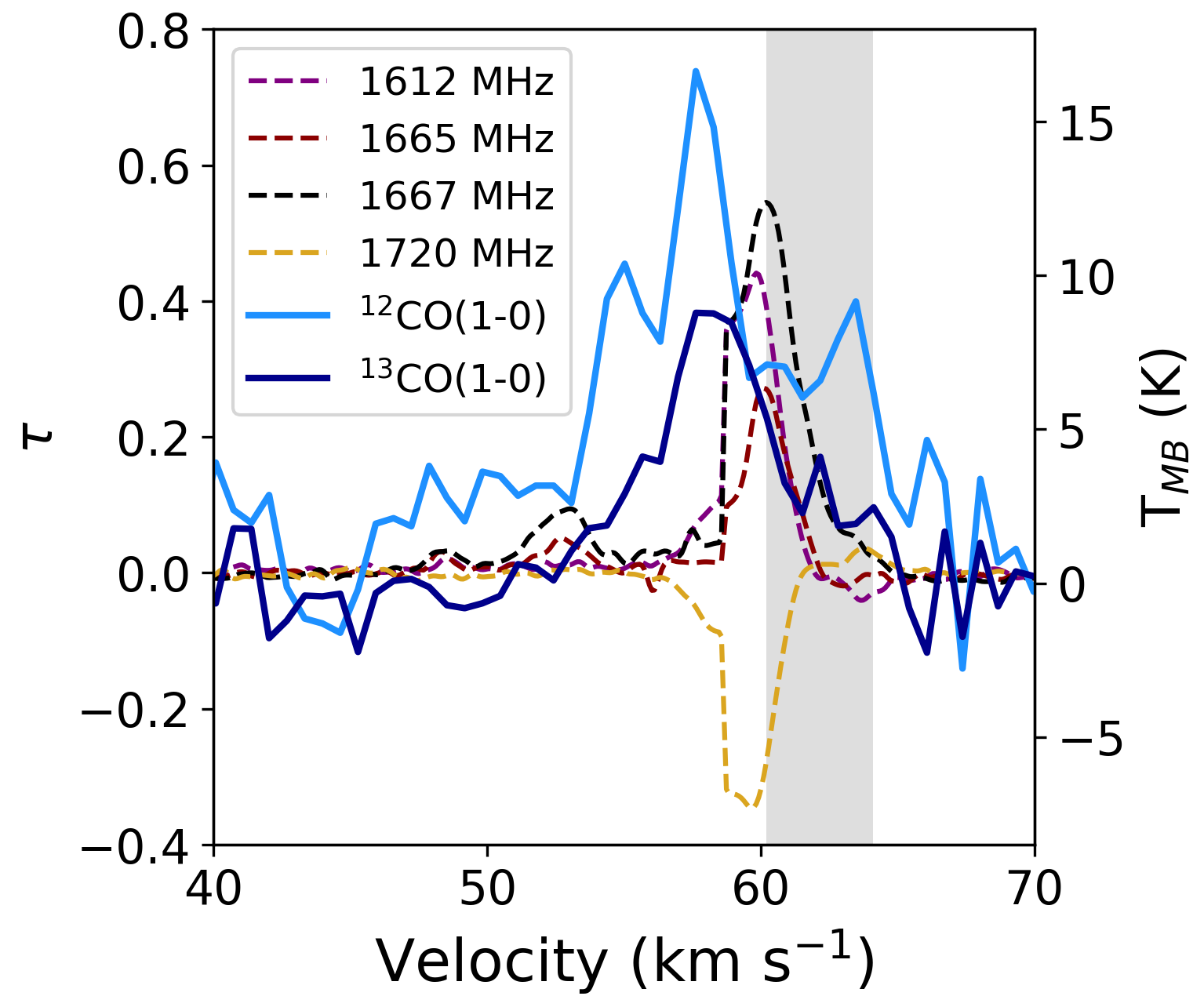}
    \includegraphics[width=0.3\textwidth]{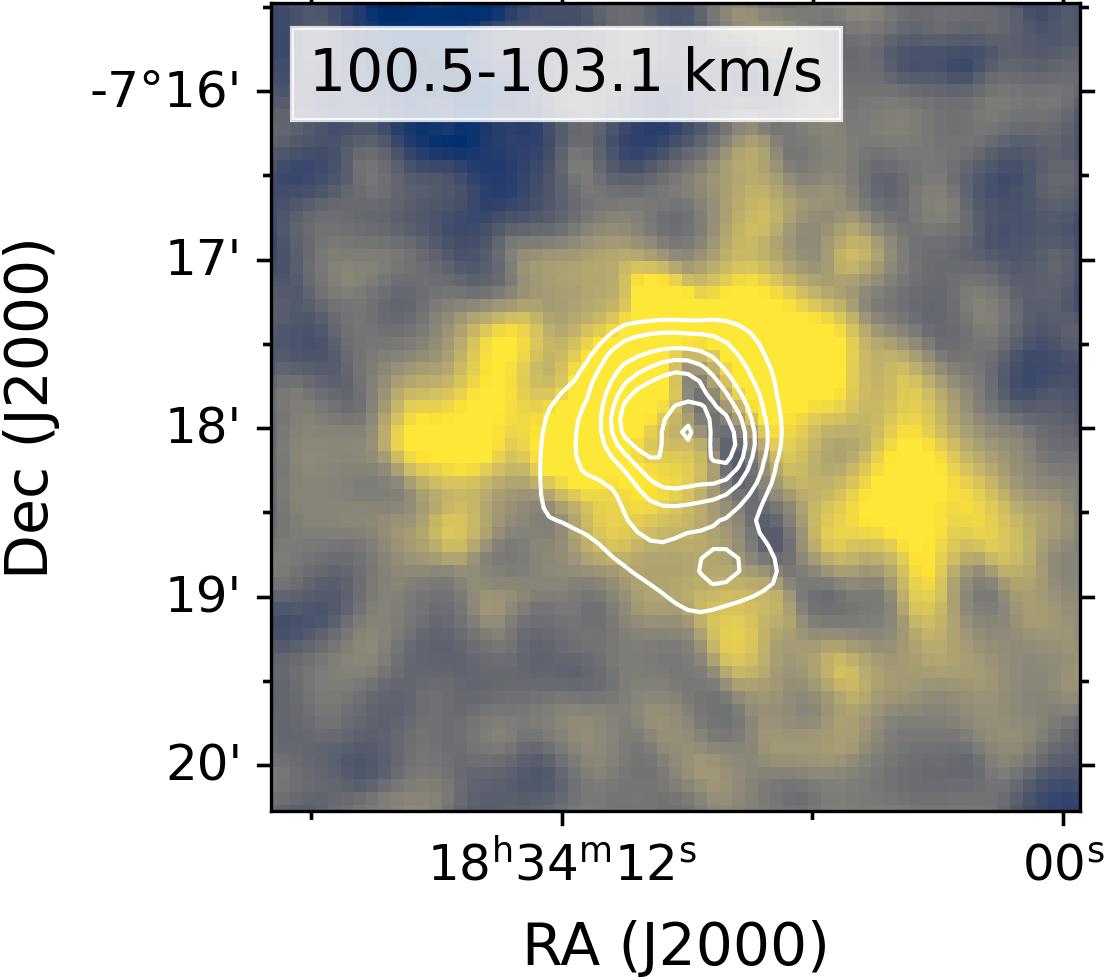}
    \includegraphics[width=0.32\textwidth]{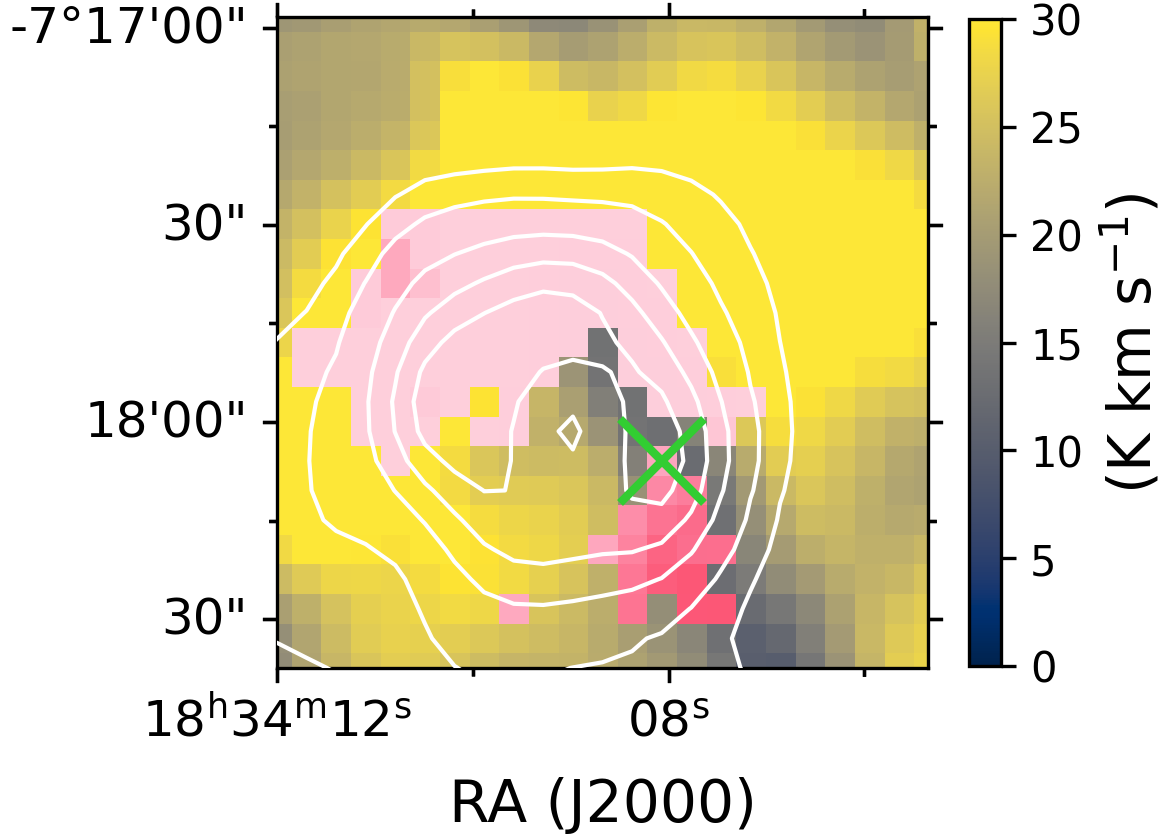}
    \includegraphics[width=0.315\textwidth]{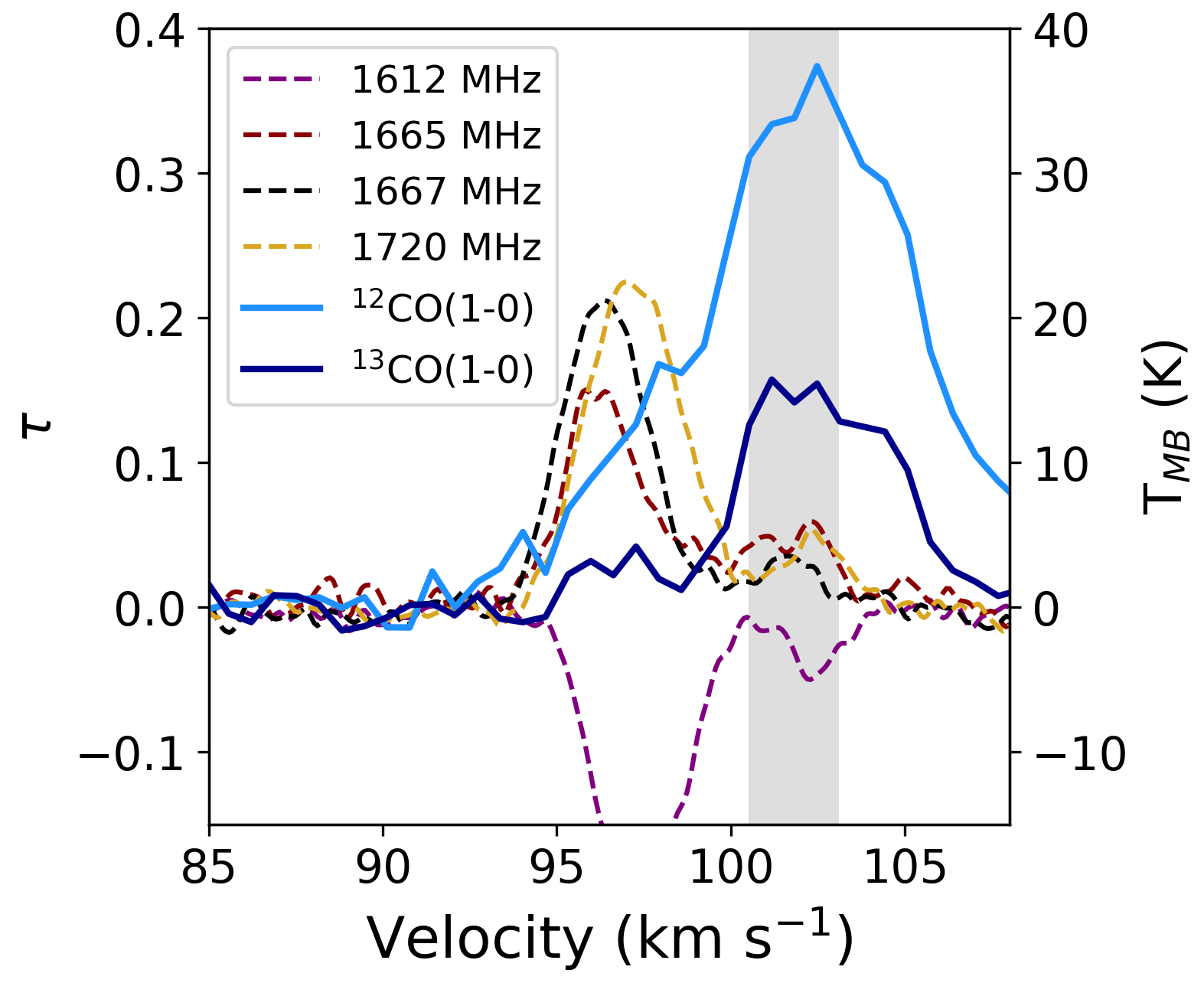}
    \caption{Comparison of CO and OH molecular gas towards Galactic \HII regions G049.205$-$0.343 (top row), G034.256+0.145 (middle row) and G024.471+0.492 (bottom row). Left column: $^{13}$CO(1-0) emission from FUGIN survey \citep{FUGIN}, integrated between 65.4--71.3 km s$^{-1}$ for G049, 60.2--64.1 km s$^{-1}$ for G034 and 100.5--103.1 km s$^{-1}$ for G024. Center column: Zoom in on the same integrated $^{13}$CO(1-0) emission, overlaid with 1612-MHz OH optical depths, where red shows inversion and blue shows absorption. The left and center column panels are overlaid with their respective 1--2 GHz continuum emission from this work. Right column: Example molecular gas spectra (from the green cross shown in the middle figure), $^{12}$CO(1-0) (dark blue) and $^{13}$CO(1-0) (light blue) from FUGIN survey and OH optical depth spectra are the dashed lines, 1612-MHz (purple), 1665-MHz (red), 1667-MHz (black) and 1720-MHz (gold). For the spectral comparisons, the OH data are smoothed to an effective beam size of 21'' to match the CO, and the gray box marks the range that the CO spectra has been integrated over.}
    \label{fig:CO}
\end{figure*}

We next investigate the \citetalias{HafnerFlipPaper} prediction that the spectral features with 1612-MHz inversion trace the parent cloud of the \HII region. In our interferometric observations, OH is only detected in front of the radio continuum from the \HII region and does not trace the full, extended parent cloud. Therefore we use $^{13}$CO(1--0) emission data from the FUGIN survey \citep{FUGIN} to trace the broader molecular environment beyond the \HII region continuum. We integrate the CO across the velocity range of the most redshifted OH component, which shows 1612-MHz inversion in all three sources, to see if that velocity component corresponds to a wider parent cloud as seen in CO.

The example CO and OH spectra towards all three \HII regions show that the redshifted velocity component we detect in OH is coincident with a broader/blended velocity component in both the $^{12}$CO and $^{13}$CO. Figure \ref{fig:CO} shows the integrated $^{13}$CO(1--0) emission (left column), a zoomed-in view overlaid with the 1612-MHz optical depths (middle column), and example $^{12}$CO(1--0), $^{13}$CO(1--0) and OH optical depth spectra (right column). Although there are not distinct components in CO with the same centroid velocity and line width as the redshifted OH components, integration over the velocity range of the 1612 component reveals extended molecular clouds, all of which are in spatial and kinematic agreement with their corresponding OH.

The distribution of these extended CO components is consistent with the hypothesis that they are the parent clouds of the \HII regions.
In Figure \ref{fig:CO}, the 1--2 GHz continuum from all three \HII regions has a clear spatial imprint on the extended molecular clouds. 
The parent cloud of G049 can be seen distinctly in $^{13}$CO(1--0) emission, wrapping around on the west side of the \HII region, with a similar morphology in the 8 $\mu$m emission (Figure \ref{fig:GLIMPSE}). This structure has been previously identified as the host cloud of G049 \citep{Carpenter1998AJ....116.1856C}. The known champagne flow exits the cloud to the south-east. In the vicinity of G034, the $^{13}$CO(1--0) emission shows that the molecular gas wraps around to the west of the \HII region, with the and 8 $\mu$m emission showing similar behavior. The CO emission is less distinct than in the G049 region, partially due to poor S/N, but has a similar morphology to G049's parent cloud. We also note that the UC\HII region G034 Compact remains apparently-embedded within the parent cloud, as expected for its earlier evolutionary stage. 

While G024 does not have an OH flip, it does have 1612-MHz inversion, and this sense of the inversion is predicted to trace the non-disrupted parent cloud of the \HII region in \citetalias{HafnerFlipPaper}. Therefore we include the $^{13}$CO(1--0) towards G024 here. We find that the CO gas kinematically coincident with its most redshifted OH component is spatially coincident with bubble-like structure in the 8 $\mu$m emission, strongly implying that a designation as the parent cloud is reasonable (bottom row of Figure \ref{fig:GLIMPSE}). 

Conversely, we are unable to reliably identify a component in CO corresponding to the blueshifted OH component with 1720-MHz inversion for any of our sources. While there is spatially-coincident CO emission in the relevant velocity ranges for all sources, the CO in these channels is generally not localized to the \HII regions, and includes messy and confused structures that extend well beyond the 1--2 GHz continuum. The `smoking gun' for gas that has been accelerated by the central source would be a distinct (yet possibly faint) blueshifted component localized to the \HII region \citepalias[][]{HafnerFlipPaper}. However, it is well within expectations that such gas would be difficult to distinguish clearly from the complex and blended CO emission of active star-forming regions within the Galactic Plane; indeed it is well-accepted that it is difficult to trace interacting gas in a confused tracer like CO \citep[e.g.,][]{LisztPety2012A&A...541A..58L, Penaloza2017MNRAS.465.2277P}. In the present case, the task is made particularly difficult by the variable S/N and coarser velocity resolution of the FUGIN data, which confound attempts to identify a velocity feature corresponding to the 1720-inverted gas via spectral decomposition. Future data with finer spectral resolution and high/uniform sensitivity would be useful to help connect what is observed in OH with the broader molecular cloud. However, this difficulty itself highlights the strength and usefulness of using OH to identify and study \HII region-molecular cloud interactions.

\citetalias{HafnerFlipPaper} suggest that the far-infrared radiation from warm dust associated with an \HII region pumps the 1612-MHz line inversion, but their excitation models show that as the molecular gas becomes colder and denser, neither satellite line is able to invert. 
The north-west section of G049 does not show the OH flip, instead it has very high OH optical depths in the most redshifted component, with the 1667-MHz line reaching $\tau \sim$1--1.2 in the area where the continuum contours abut the parent cloud. These are the highest OH optical depths we are aware of in the literature. Interestingly, in this region of high optical depth, all four OH lines are in absorption. Potentially this section of high optical depth is a colder and/or denser section of the parent molecular cloud where the required excitation conditions for the 1612-MHz inversion have not been met. The 8$\mu$m emission towards G049 also shows a notable decrease in emission in this area. 

Examination of the multiwavelength data for G034 allows us to suggest an explanation for the systematic $\sim10$ km\,s$^{-1}$ offset between its OH and \avghna RRL spectra. We suggest that this may be explained by a champagne flow morphology where the angle of the ionized gas outflow is close to the line-of-sight \citep[e.g.][]{Israel1978A&A....70..769I}. We may imagine a scenario where the \HII region of G034 has expanded within its molecular cloud, ionized gas has reached the front of the molecular cloud (with respect to the observer) and then outflowed beyond it.
The morphology, dust and gas distributions support this classification, with the general appearance of the CO and 8$\mu$m emission appearing very similar to G049. Similarly, there is a steeper gradient in the radio continuum of G034 towards the west, where it abuts the parent molecular cloud, and a potential opening towards the left, devoid of CO emission in the parent cloud component. The 8 $\mu$m morphology supports this picture, wrapping around the west of G034, which is where the outflow may originate.

Generally, the morphology and kinematics of G049 and G034 are in agreement with the testable predictions from \citetalias{HafnerFlipPaper}. G049's and G034's OH satellite-line flips have more components than presented in the simple model of \citetalias{HafnerFlipPaper}. This is not unexpected, however, because the \citetalias{HafnerFlipPaper} model is deliberately idealized, and also drew largely on examples of the flip from the literature, without performing detailed spectral decomposition. The spectral and spatial complexity in our resolved observations does not contradict the hypothesis; instead, it shows how the OH spectra can be used to trace complex structures within the molecular gas.

In G024, where no flip is detected, we suggest that the two major OH components are tracing separate molecular gas structures, as evidenced by their different morphologies. \citet{Saha2024ApJ...970L..40S} identified expanding molecular rings in HCO$^+$. The OH components identified in this work match the two more blueshifted components in the HCO$^+$ spectra. As OH is only tracing gas in front of an \HII region, with respect to the observer, the kinematics of the OH are in agreement with the expanding molecular gas ring picture. The distinct morphologies of the main components (1 and 3) suggest they may be tracing different rings or sections of the molecular ring.

\begin{table*}[]
\centering
\begin{tabular}{@{}cccc}
\hline
Source & Mass & Momentum & Kinetic Energy \\
& (M$_\odot$) & (M$_\odot$ km s$^{-1}$) & ($\times10^{47}$erg) \\
\hline
G049.205$-$0.343 & 433 & 3470 & 2.76 \\
G034.256+0.145 & 38.3 & 115 & 0.0343 \\ \hline
\end{tabular}
\caption{The mass, momentum and kinetic energy of the dynamically interacting molecular gas towards Galactic \HII regions G049.205$-$0.343 and G034.256+0.145. This gas is traced by the OH satellite-line flip spectral feature in the \citetalias{HafnerFlipPaper} hypothesis. See Section \ref{sec:disc} and Appendix \ref{sec:appendixMass} for details.}
\label{tab:mass}
\end{table*}

The OH flip is a potential tool for quantifying \HII feedback since it enables an estimate of the mass, momentum and kinetic energy of the dynamically interacting molecular gas. Here we demonstrate this procedure for G049 and G034. We select the first spatio-kinematic component to be representative components of dynamically interacting molecular gas as they show the highest fraction of pixels with 1720-MHz inversion. We estimate the mass of these components by estimating the column density of the OH gas and summing it over the area where OH is detected (see Appendix \ref{sec:appendixMass} for details). The relative momentum and kinetic energy of the interacting gas are $p=mv$ and $E=\frac{1}{2}mv^2$, where $v$ is the jump in velocity across the shock between the most redshifted and most blueshifted OH components. We present our obtained mass, momentum and kinetic energy for G049 and G034 in Table \ref{tab:mass}. These derived values are within the expected order of magnitude range for these types of systems \citep[e.g.,][]{Gendelev2012ApJ...745..158G, Xu2018A&A...609A..43X, Li2022RAA....22d5008L}. We use an assumed $T_{\text{ex}}$ and $X_{\text{OH}}$ for our estimates, which are sources of uncertainty for our values. Also, as we cannot observe the molecular gas behind the \HII region with our OH, and we only observe the line-of-sight component of the jump in velocity, our estimated parameters are lower limits. Nevertheless, these values demonstrate the potential utility of OH spectra, particularly the OH flip, in quantifying the impact of feedback from \HII regions on parent clouds.

\section{Conclusions}
\label{sec:conc}
In this work, we study the kinematics and morphology of three Galactic \HII regions with dedicated VLA observations of the four ground-state OH lines and multiple hydrogen radio recombination lines. We use these data to explore the basic features of the \citetalias{HafnerFlipPaper} hypothesis, which proposes that a characteristic spectral signature in the OH satellite lines---the OH flip---traces gas that is dynamically interacting with an expanding \HII region.

We unambiguously detect the OH flip towards two of our sources, G049.205$-$0.343 and G034.256+0.145. Using a novel Gaussian decomposition approach, we identify complex, blended spatio-kinematic components in the OH spectra towards all three sources. We do not detect the OH flip towards G024.471+0.492, which  shows only the 1612-MHz inversion. 

We explore the \citetalias{HafnerFlipPaper} hypothesis by comparing the kinematics and morphology of the observed 1.6 GHz continuum; OH, \avghna RRL, $^{12}$CO(1--0) and $^{13}$CO(1--0) spectra; and 8$\mu$m emission with a set of testable predictions from \citetalias{HafnerFlipPaper}. We find an agreement between several predictions of the \citetalias{HafnerFlipPaper} hypothesis and the multi-wavelength observations of G049 and G034. Namely, there is a strong spatial and kinematic association between the OH flip and the ionized gas of the \HII regions---the first time this has been demonstrated for resolved sources---and evidence from $^{13}$CO(1--0) emission that the expected OH component (the 1612-MHz inversion) traces the parent cloud. While G024 does not show the OH flip, it is important to note that this does not disprove the hypothesis that the flip (when present) is tracing dynamically-interacting and undisturbed molecular gas. Instead we find that the OH in G024 is tracing a previously identified expanding molecular ring. We also estimate the mass, momentum and kinetic energy of the dynamically interacting molecular gas. These values are within the expected order of magnitude ranges for \HII region-molecular cloud systems. Taken together, the results from all three \HII regions demonstrate the ability of OH to trace \HII region-molecular cloud interactions, with G049's and G034's observations providing direct support for the OH flip's ability to disentangle dynamically interacting and non-interacting gas, as proposed by \citetalias{HafnerFlipPaper}. 

Further work is now required to expand on the initial findings provided here. For example, \citetalias{HafnerFlipPaper} suggests that there should be an increase in gas temperature and number density in the blueshifted component (1720-MHz inversion) compared to the redshifted component, reflecting the fact that the former has been shocked by the expanding \HII region. Dedicated excitation modeling of the fitted OH spectra from our sources would allow us to test this hypothesis, by constraining the physical parameters of each OH component, and comparing those with the expected shock-generated environmental conditions. 

While the results we present here are promising, a larger sample of sources is required to draw more robust conclusions. The Tracing OH and Recombination lines from molecular Clouds and \HII regions (TORCH) Survey (VLA/25A-022, PI Cappellazzo) will observe 13 fields containing Galactic \HII regions with a larger range of morphologies and evolutionary stages.
These observations also include higher frequency observations of G049, G034 and G024, which will allow for spatially resolved maps of ionized gas physical properties. Other recent and upcoming sensitive OH surveys (e.g. SPLASH, \citealt{DawsonSPLASH2022MNRAS.512.3345D}; GASKAP-OH \citealt{DickeyGASKAP2013PASA...30....3D,Dawson2024IAUS..380..486D}) and RRL surveys \citep[SHRDS][]{WengerSHRDS2019ApJS..240...24W} can be used to identify more examples of the flip in the Southern Sky and study \HII region-molecular cloud interactions.

\begin{acknowledgments}
The National Radio Astronomy Observatory is a facility of the National Science Foundation operated under cooperative agreement by Associated Universities, Inc. This publication makes use of data from FUGIN, FOREST Unbiased Galactic plane Imaging survey with the Nobeyama 45-m telescope, a legacy project in the Nobeyama 45-m radio telescope.
T.V.W. is supported by a National Science Foundation Astronomy and Astrophysics Postdoctoral Fellowship under award AST-2202340.
M. R. Rugel is a Jansky Fellow of the National Radio Astronomy Observatory.
We thank the referee, Erik Rosolowsky, for his helpful comments that improved the manuscript.
\end{acknowledgments}

\appendix

\section{VLA Correlator Set-up}
\label{sec:appendixA}

We observed the three Galactic \HII regions in this work with the VLA in L-band (1--2 GHz) using the correlator setup listed in Table \ref{tab:corr}. For each spectral window, window number, central frequency, bandwidth, number of channels, channel width, spectral line and its rest frequency are listed. We observed 38 H RRLs, the four 18-cm OH transitions, the \HI 21-cm line and continuum. The ``$*$'' indicates spectral windows that were dominated by RFI ($>$50\% of channels) or completely flagged and hence are unusable. 

\begin{longtable}{ccccccc}
\caption{Spectral window and continuum window configuration. Note: ``$*$'' indicates spectral windows that were dominated by RFI ($>$50\% of channels) or completely flagged and hence are unusable.}
\label{tab:corr}
\\
\hline
Window & $\nu_{\text{center}}$ & Bandwidth & Channels & Channel Width & Line & $\nu_{\text{L,rest}}$ \\ 
(Number) & (MHz) & (MHz) & & (kHz) & (name) & (MHz) \\
\hline
\endhead
0 & 1818.2323 & 2 & 512 & 3.906 & H153$\alpha$ & 1818.24591 \\
1 & 1854.2349 & 2 & 512 & 3.906 & H152$\alpha$ & 1854.25027 \\
2 & 1891.1945 & 2 & 512 & 3.906 & H151$\alpha$ & 1891.21153 \\
3 & 1929.1429 & 2 & 512 & 3.906 & H150$\alpha$ & 1929.16170 \\
4$*$ & 1968.1135 & 2 & 512 & 3.906 & H149$\alpha$ & 1968.13408 \\
5 & 1720.5219 & 2 & 4096 & 0.488 & OH & 1720.52990 \\
6 & 1667.3534 & 2 & 4096 & 0.488 & OH & 1667.35900 \\
7 & 1420.4111 & 4 & 2048 & 1.953 & HI & 1420.40580 \\
8$*$ & 2017.0721 & 128 & 64 & 2000 & Continuum & - \\
9 & 1889.0721 & 128 & 64 & 2000 & Continuum & - \\
10 & 1761.0721 & 128 & 64 & 2000 & Continuum & - \\
11$*$ & 1633.0721 & 128 & 64 & 2000 & Continuum & - \\
12$*$ & 1505.0721 & 128 & 64 & 2000 & Continuum & - \\
13 & 1377.0721 & 128 & 64 & 2000 & Continuum & - \\
14$*$ & 1249.0721 & 128 & 64 & 2000 & Continuum & - \\
15$*$ & 1121.0721 & 128 & 64 & 2000 & Continuum & - \\
16 & 1400.0701 & 2 & 512 & 3.906 & H167$\alpha$ & 1399.36771 \\
17 & 1424.7380 & 2 & 512 & 3.906 & H166$\alpha$ & 1424.73359 \\
18 & 1450.7195 & 2 & 512 & 3.906 & H165$\alpha$ & 1450.71626 \\
19 & 1477.3366 & 2 & 512 & 3.906 & H164$\alpha$ & 1477.33457 \\
20 & 1504.6088 & 2 & 512 & 3.906 & H163$\alpha$ & 1504.60810 \\
21$*$ & 1532.5566 & 2 & 512 & 3.906 & H162$\alpha$ & 1532.55712 \\
22$*$ & 1561.2008 & 2 & 512 & 3.906 & H161$\alpha$ & 1561.20269 \\
23$*$ & 1590.5634 & 2 & 512 & 3.906 & H160$\alpha$ & 1590.56662 \\
24$*$ & 1620.6670 & 2 & 512 & 3.906 & H159$\alpha$ & 1620.67158 \\
25 & 1651.0701 & 2 & 512 & 3.906 & H158$\alpha$ & 1651.54111 \\
26$*$ & 1684.0701 & 2 & 512 & 3.906 & H157$\alpha$ & 1683.19962 \\
27 & 1715.6635 & 2 & 512 & 3.906 & H156$\alpha$ & 1715.67248 \\
28 & 1748.9756 & 2 & 512 & 3.906 & H155$\alpha$ & 1748.98605 \\
29 & 1784.0701 & 2 & 512 & 3.906 & H154$\alpha$ & 1783.16770 \\
30$*$ & 1281.1863 & 2 & 512 & 3.906 & H172$\alpha$ & 1281.17526 \\
31$*$ & 1303.7277 & 2 & 512 & 3.906 & H171$\alpha$ & 1303.71768 \\
32 & 1326.8010 & 2 & 512 & 3.906 & H170$\alpha$ & 1326.79206 \\
33 & 1350.4221 & 2 & 512 & 3.906 & H169$\alpha$ & 1350.41420 \\
34 & 1374.6072 & 2 & 512 & 3.906 & H168$\alpha$ & 1374.60043 \\
35 & 1665.3963 & 2 & 4096 & 0.488 & OH & 1665.40180 \\
36 & 1612.2279 & 2 & 4096 & 0.488 & OH & 1612.23090 \\
37 & 1420.4111 & 4 & 2048 & 1.953 & HI & 1420.40580 \\
38 & 1888.0721 & 128 & 64 & 2000 & Continuum &  \\
39 & 1760.0721 & 128 & 64 & 2000 & Continuum &  \\
40$*$ & 1632.0721 & 128 & 64 & 2000 & Continuum &  \\
41$*$ & 1504.0721 & 128 & 64 & 2000 & Continuum &  \\
42 & 1376.0721 & 128 & 64 & 2000 & Continuum &  \\
43$*$ & 1248.0721 & 128 & 64 & 2000 & Continuum &  \\
44$*$ & 1120.0721 & 128 & 64 & 2000 & Continuum &  \\
45$*$ & 992.0721 & 128 & 64 & 2000 & Continuum &  \\
46 & 1013.7917 & 1 & 512 & 1.953 & H186$\alpha$ & 1013.76730 \\
47 & 1030.2747 & 1 & 512 & 1.953 & H185$\alpha$ & 1030.25116 \\
48 & 1047.1172 & 1 & 512 & 1.953 & H184$\alpha$ & 1047.09434 \\
49 & 1064.3288 & 1 & 512 & 1.953 & H183$\alpha$ & 1064.30668 \\
50 & 1081.9195 & 1 & 512 & 1.953 & H182$\alpha$ & 1081.89835 \\ 
51 & 1099.9003 & 1 & 512 & 1.953 & H181$\alpha$ & 1099.87985 \\
52 & 1118.2816 & 1 & 512 & 1.953 & H180$\alpha$ & 1118.26206 \\
53 & 1137.0749 & 1 & 512 & 1.953 & H179$\alpha$ & 1137.05618 \\
54 & 1156.2916 & 1 & 512 & 1.953 & H178$\alpha$ & 1156.27383 \\
55$*$ & 1175.9429 & 1 & 256 & 3.906 & H177$\alpha$ & 1175.92701 \\
56$*$ & 1196.0431 & 1 & 256 & 3.906 & H176$\alpha$ & 1196.02811 \\
57$*$ & 1216.6040 & 2 & 512 & 3.906 & H175$\alpha$ & 1216.58997 \\
58$*$ & 1237.6389 & 2 & 512 & 3.906 & H174$\alpha$ & 1237.62588 \\
59$*$ & 1259.1616 & 2 & 512 & 3.906 & H173$\alpha$ & 1259.14957 \\ \hline
\end{longtable}

\section{OH Gaussian Decomposition}
\label{sec:appendixOHfit}

In this section we detail our approach for spatially coherent Gaussian decomposition of OH spectra. First, the OH spectra are transformed into optical depth $\tau_\nu$ spectra using:
\begin{equation}
    \label{eqn:OHtau}
    \tau_v = -\ln\left(\frac{T_{\text{b}}(v)}{T_{\text{c}}} \right),
\end{equation}
where $T_{\text{b}(v)}$ is the brightness temperature of the OH spectra (which still include the background continuum) and $T_{\text{c}}$ is the observed background continuum brightness temperature, measured as the mean of the line free channels for each OH transition. Equation \ref{eqn:OHtau} assumes either that the OH emission from the broader molecular cloud is smooth and therefore filtered out by the radio interferometric observations, or that the excitation temperature of that emission is much smaller than the brightness temperature of the continuum emission. For OH, the latter is a reasonable assumption as our sources have $T_{\text{c}} \sim 1000$K and generally $T_{\text{ex}} \sim 5$K \citep{HafnerGNOMES2023}.

The OH spectra from each pixel are fit with model $M_N$, which has $N$ Gaussian components, the $i$th of which has parameters $\theta_i = [V_{\text{LSR},i}, \Delta V_i, \tau_{1612,i}, \tau_{1665,i}, \tau_{1667,i}, \tau_{1720,i}]$. Where $V_{\text{LSR}}$ is the line-of-sight velocity centroid, $\Delta V$ is the FWHM line width, and $\tau_{\nu}$ is the optical depth of the $\nu$ transition. The fitting of each spectrum is performed using \texttt{lmfit: Non-Linear Least-Squares Minimization and Curve-Fitting for Python} \citep{lmfit2023zndo...7887568N}.

The S/N of the OH spectra are low, so we use Voronoi tessellation to spatially bin the data to increase the S/N. First, from visual inspection of the data cubes, a subset of $S \times S$ pixels around each \HII region is chosen that contain all visible OH emission and absorption; $64\times64$ pixels for G049, $32\times32$ pixels for G034 and $16\times16$ pixels for G024. These subsets of pixels are spatially binned with the \texttt{VorBin} package \citep{voronoi2003MNRAS.342..345C, voronoisoftware2012ascl.soft11006C} to achieve a constant continuum S/N for each bin. The aim is to bin the pixels that have a lower continuum background to increase their signal to noise, while leaving those with a bright continuum background as single pixels to maximize the spatial resolution. We adopt a S/N threshold of 30 as a reasonable compromise between S/N and angular resolution. The resulting binned subset still spans $S \times S$ pixels, but pixels containing spectra with lower S/N are replaced by the average spectrum of the original pixel and its neighboring pixels within a given Voronoi bin to achieve the desired S/N. We refer to this binned subset as $S^* \times S^*$ hereafter.

\begin{figure}
    \centering
    \includegraphics[width=\linewidth]{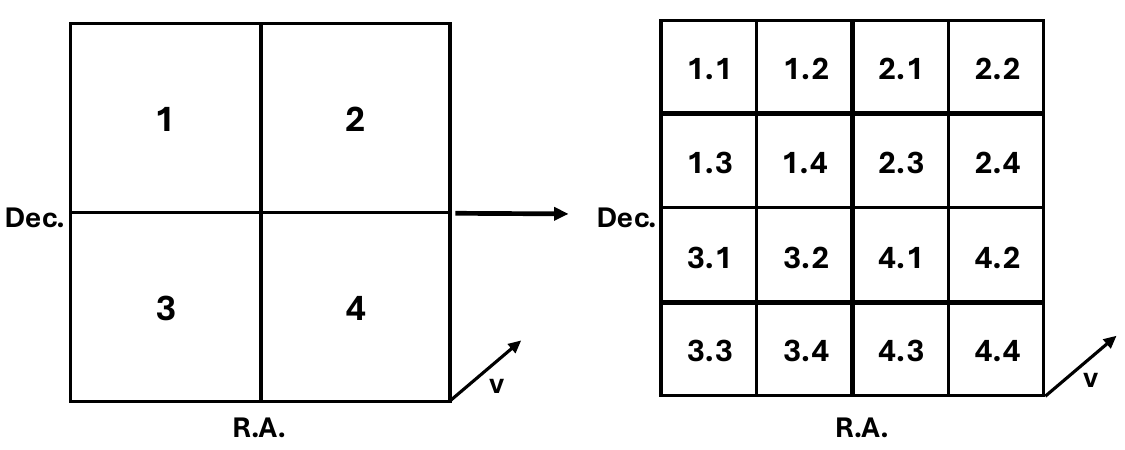}
    \caption{A diagram demonstrating the how the sequentially `resolution cube' fitting process works. In this example, the fit parameters from the $2\times2$ pixel cube (left) are passed as initial guesses to the $4\times4$ pixel cube, which is then fit.}
    \label{fig:rescube}
\end{figure}

In order to promote spatial coherence, the binned pixels within the $S^* \times S^*$ subset (that has been through Voronoi tessellation) are then averaged together to form a series of `resolution cubes' with different spatial grids, starting with a coarse 1$\times$1 grid, then 2$\times$2 and so on until reaching the original $S^* \times S^*$ grid. Each resolution cube is fit sequentially, from coarse grid to fine grid. At each stage, the best fit parameters of the current resolution cube are passed on as initial guess parameters of the next finest resolution cube. 

The first iteration starts by fitting the 1$\times$1 resolution cube using $N$ Gaussian components where we specify $N$. The initial guess parameters are tuned by visual inspection of the fitted Gaussian components and fit residuals, then these parameters are manually input into the fitting algorithm. For all subsequent iterations, the initial guess parameters for each pixel are automatically determined from the best fit of the corresponding pixel in the previous resolution cube (see Figure \ref{fig:rescube}). After fitting every pixel in a resolution cube, the S/N of the best fits for each pixel are checked. In order for a fit to be retained and passed onto the next resolution cube, at least one Gaussian component in at least one transition must have S/N $>$ 3 (where $\text{S/N}=\frac{\tau_{\text{peak}}}{\text{rms}}$). Otherwise the best fits from the previous resolution cube are used. This process of fitting $N$ Gaussian components to each pixel, evaluating the fits and then passing to the next resolution cube is repeated until the original $S^* \times S^*$ resolution cube is reached. To retain only the most reliable fits in the final models, a final cull is performed to remove pixels where no fit Gaussian component exceeded a S/N of 3. Due to the Voronoi tessellation process, some binned pixels spanned more than one pixel in a coarser resolution cube. In that case, the pixel would receive multiple initial guesses and only the best fit of that Voronoi binned pixel was retained. 

This process of fitting each resolution cube (from 1$\times$1 to $S^* \times S^*$) is repeated with one to seven Gaussian components. We then use the Bayesian Information Criterion to identify the best model for each pixel. The BIC penalizes more complicated models to prevent overfitting, and is commonly used to evaluate the best fit model for Gaussian decomposition of astrophysical spectra \citep[e.g.,][]{Koch2021, Avery10.1093/mnras/stab780}. The BIC is defined as:
\begin{equation}
    \text{BIC} = \chi^2 + d\ln(n_{\text{data}})
\end{equation}
where $d$ is the number of free parameters and $n_{\text{data}}$ is the number of data points.
To choose the best model, the equation $\text{BIC}_N - \text{BIC}_{N+1} < -10$ is evaluated from smaller to larger number of Gaussian components until the BIC stops decreasing by 10, which indicates a stronger preference for model $N$, following \citet{Kass1995doi:10.1080/01621459.1995.10476572}.

In order to understand the kinematics of the molecular gas and identify Gaussian components that correspond to either sense of the OH flip, individual spectral features need to be identified and tracked across each \HII region. For each source we sort and group the fitted Gaussian components into `spatio-kinematic Components' that map each feature across the sources, which are presented in Section \ref{sec:OHresults}.

If there is a velocity gradient in the group of pixels that are averaged together in a Voronoi bin, the resulting averaged spectrum may have a wider line width and lower amplitude than the individual pixels. In this work, the maximum number of pixels we bin is four and the pixel-to-pixel velocity gradient is $\sim$1 km/s, so the worst-case-scenario could be a $\sim$4--5 km/s velocity gradient that is missing. Based on the typical line widths that we fit from individual pixels ($\sim$2 km s$^{-1}$), in this worst-case, the line width would increase by a factor of $\sim$1.5 and the optical depth would be multiplied by a factor of $\sim$0.6. This work does not rely on exact optical depth measurements, and the estimated column density uses the area under the spectral line. In Voronoi-binned spectra, a wider line with shallower amplitude will still have the same area and produce the same column density value.

The final OH spectra that were fitted for G034 and G024 do not have any binned pixels and we bin very few pixels from G049. Therefore any missed velocity gradients will not have a significant effect on the final results of this work.

\section{Spectral Index Fits}
\label{sec:appendixalpha}

\begin{figure*} 
    \centering
    \includegraphics[width=0.35\linewidth]{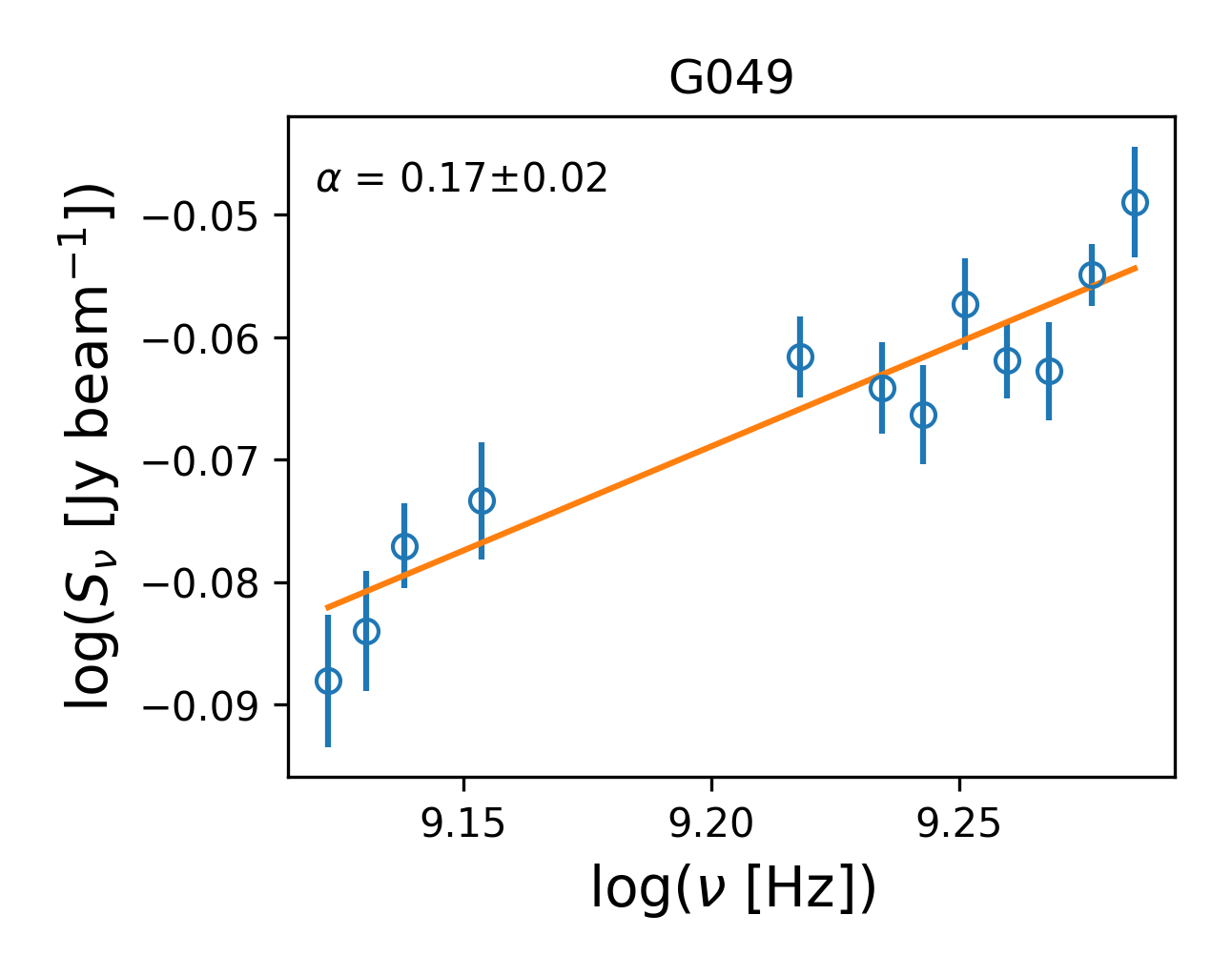}
    \includegraphics[width=0.35\linewidth]{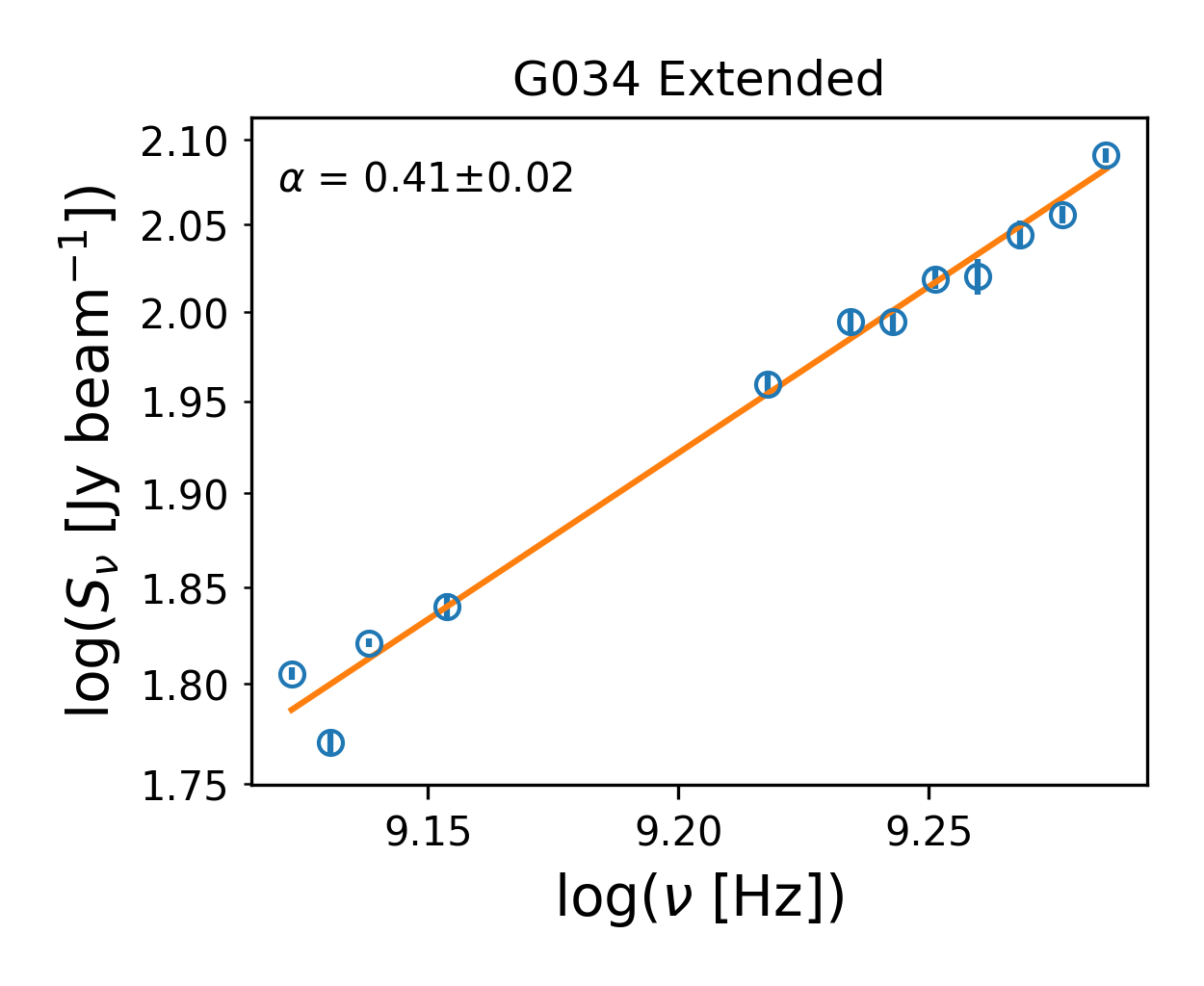}
    \includegraphics[width=0.35\linewidth]{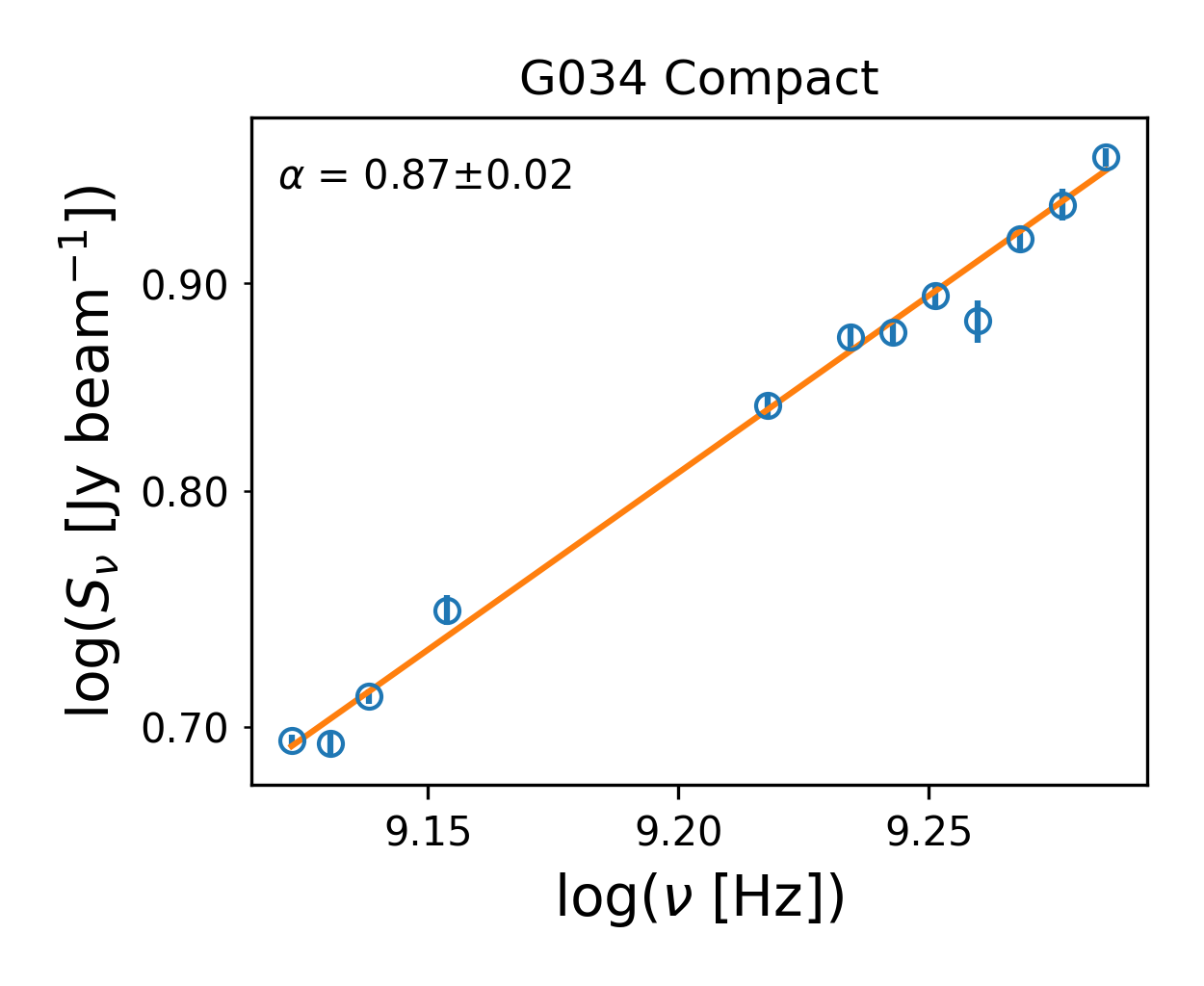}
    \includegraphics[width=0.35\linewidth]{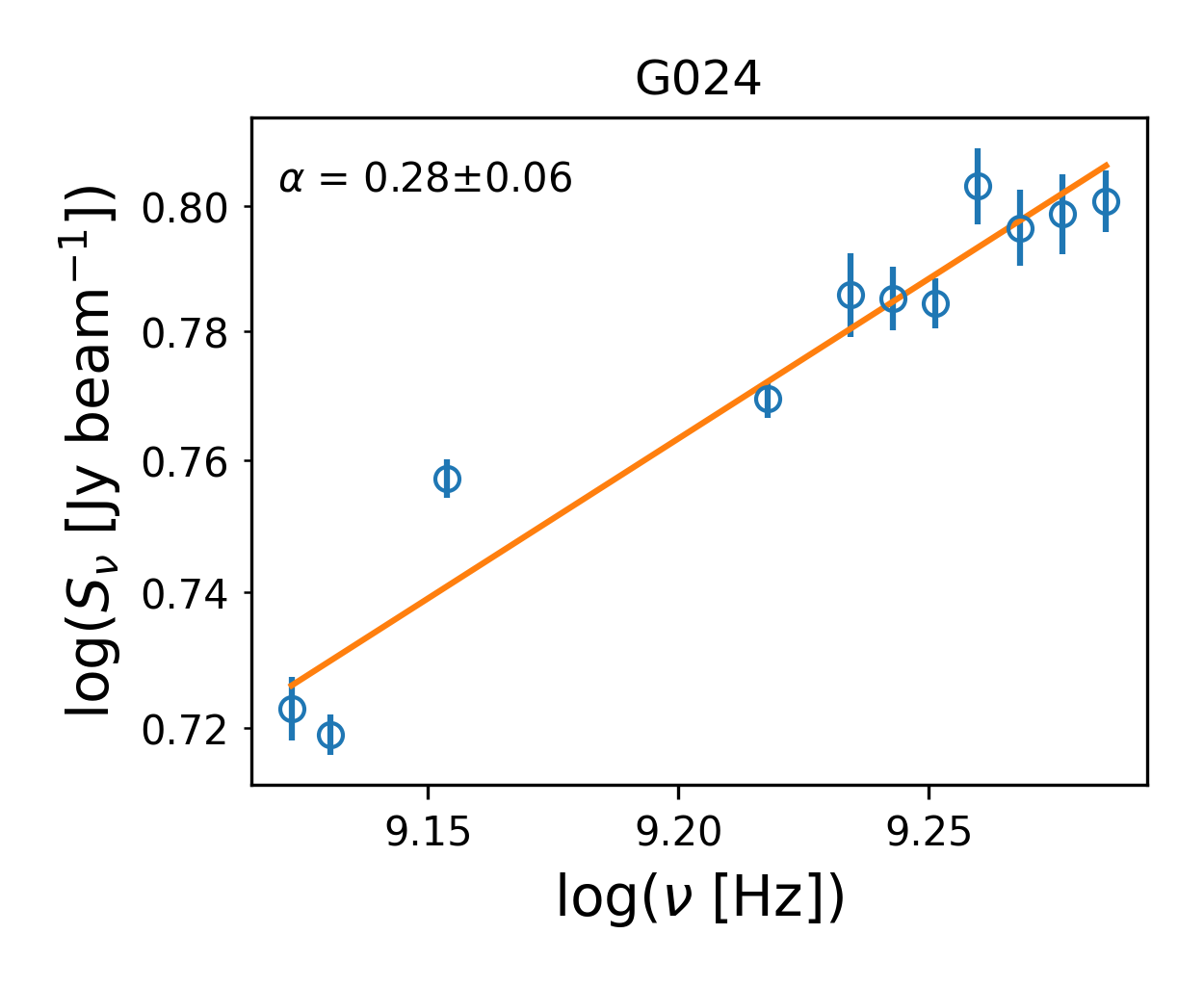}
    \caption{Examples of fitted radio SEDs from central sightlines in our Galactic \HII regions. Top left: G049 (RA: $19^{\text{h}}23^{\text{m}}01\text{s}$, Dec: $+14^\circ16'47''$). Top right: Extended G034 (RA: $18^{\text{h}}53^{\text{m}}20.2\text{s}$, Dec: $+1^\circ4'40''$). Bottom left: Compact G034 (RA: $18^{\text{h}}53^{\text{m}}18\text{s}$, Dec: $+1^\circ14'59''$). Bottom right: G024 (RA: $18^{\text{h}}34^{\text{m}}0.93\text{s}$, Dec: $-7^\circ17'52''$). The blue data points are measurements of the radio continuum taken from line free channels of each H RRL spectral window, and the orange line is the linear fit, where the gradient of the line is the spectral index $\alpha$. See Section \ref{sec:RRL-cont-results} for details.}
    \label{fig:alphafits}
\end{figure*}

The spectral index is defined as $\alpha = \frac{d\log S}{d\log\nu}$, where $S$ is the radio continuum intensity in Jy/beam and $\nu$ is the frequency in Hz. We estimate $\alpha$ for each source by fitting a straight line to each pixel's SED in log-log space using \texttt{curve\_fit} from \texttt{scipy.optimize} \citep{scipy}.
Examples of fitted SEDs used to derive $\alpha$ for each \HII region are presented in Figure \ref{fig:alphafits}. The SEDs show slopes with $\alpha > 0.05$, indicating that the gas is not optically thin. See Section \ref{sec:RRL-cont-results} for details.

\section{Spectra Grid}
\label{sec:appendix2Dspectra}

\begin{figure*}[htb]
    \centering
    \includegraphics[width=\linewidth]{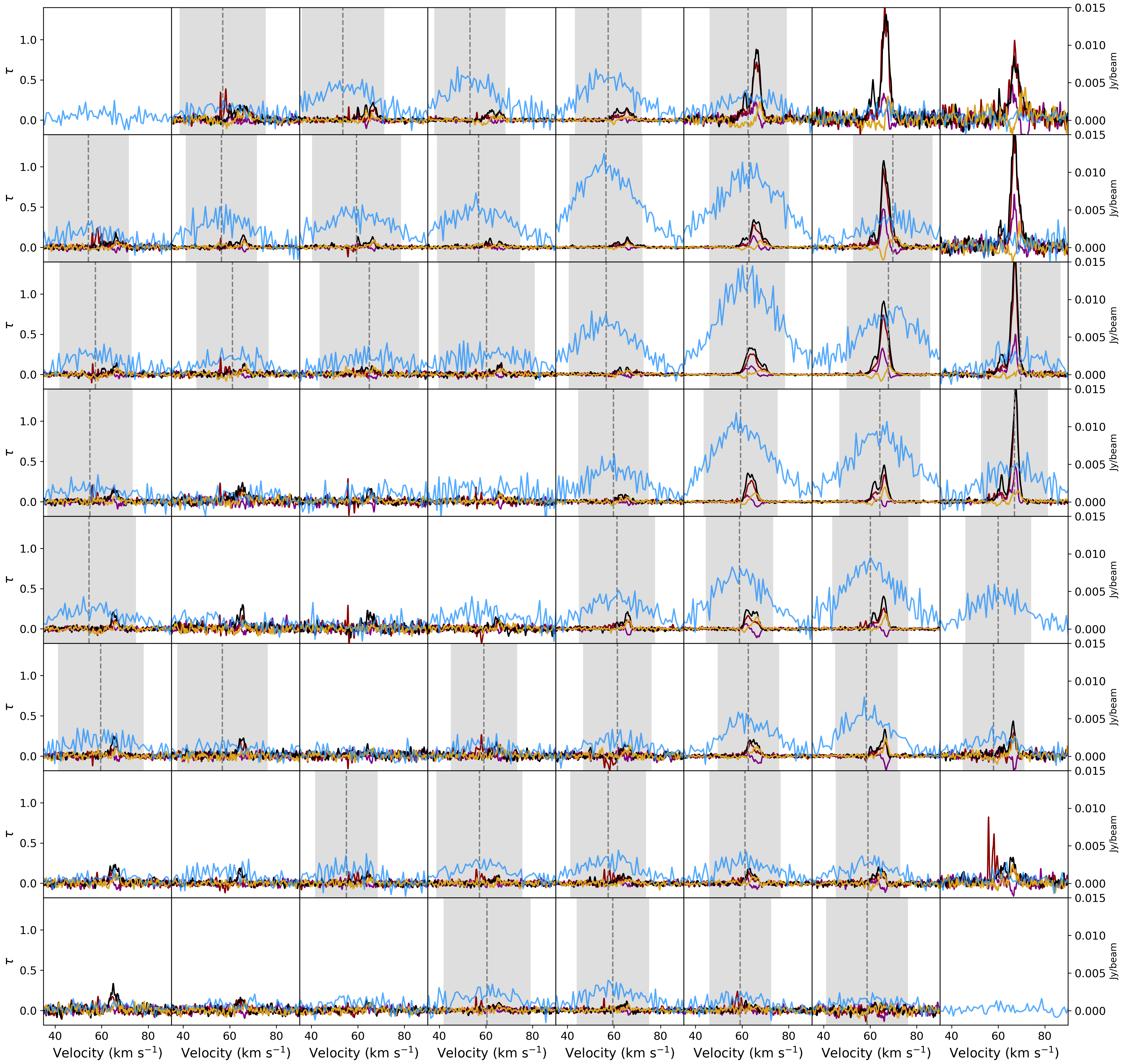}
    \caption{OH optical depth spectra and \avghna RRL spectra from Galactic \HII region G049.205$-$0.343, skipping every 4 pixels. Purple is the 1612-MHz, red is the 1665-MHz, black is the 1667-MHz and gold is the 1720-MHz OH transitions. The \avghna RRL spectrum is shown in blue. In each panel, the vertical dashed line marks the \VLSR and the gray box marks the span of the FWHM of the fitted \avghna spectrum in that position.}
    \label{fig:G049grid}
\end{figure*}

\begin{figure*}
    \centering
    \includegraphics[width=\linewidth]{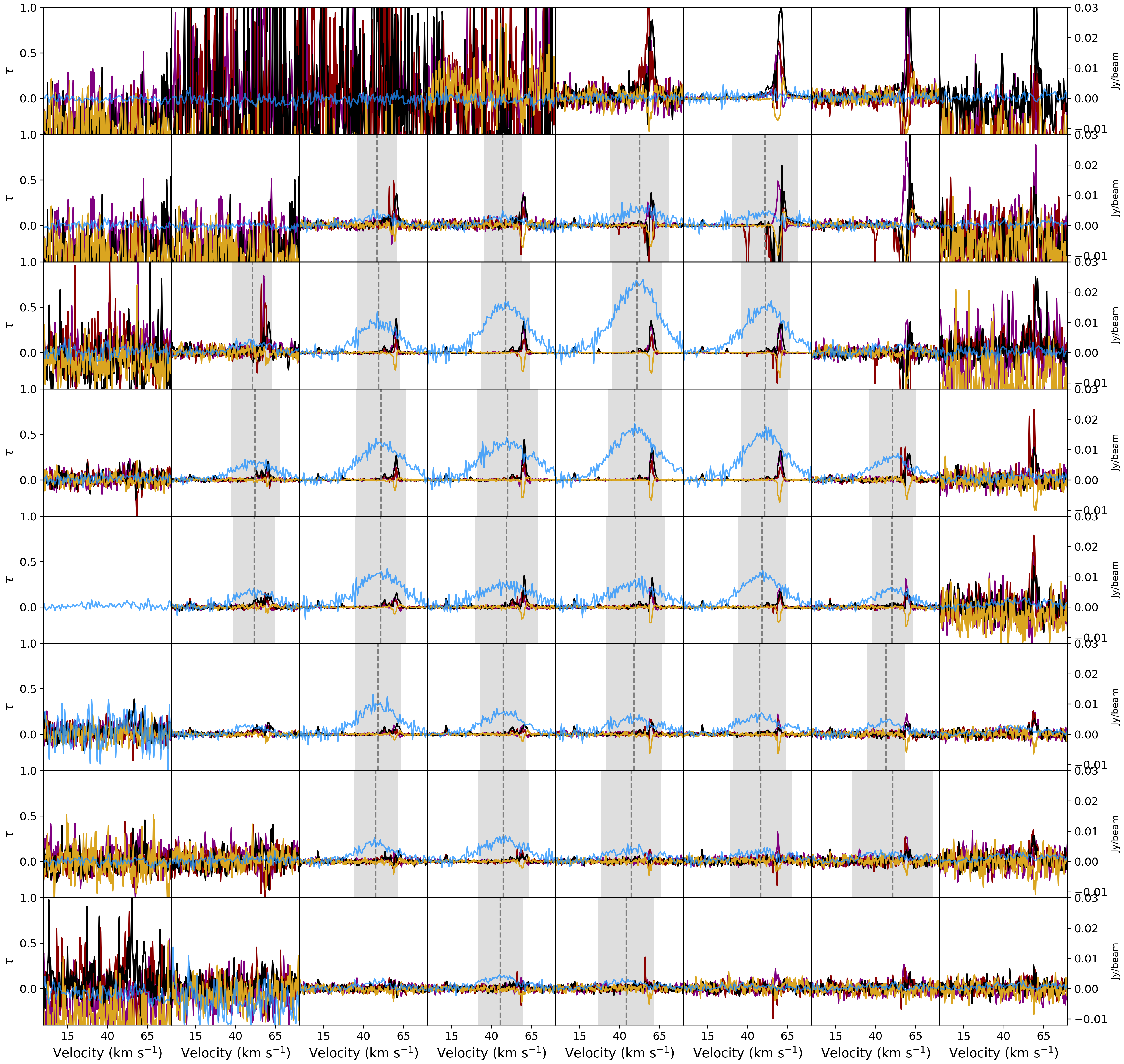}
    \caption{Same as Figure \ref{fig:G049grid} for G034.256+0.145.}
    \label{fig:G034grid}
\end{figure*}

\begin{figure*}
    \centering
    \includegraphics[width=\linewidth]{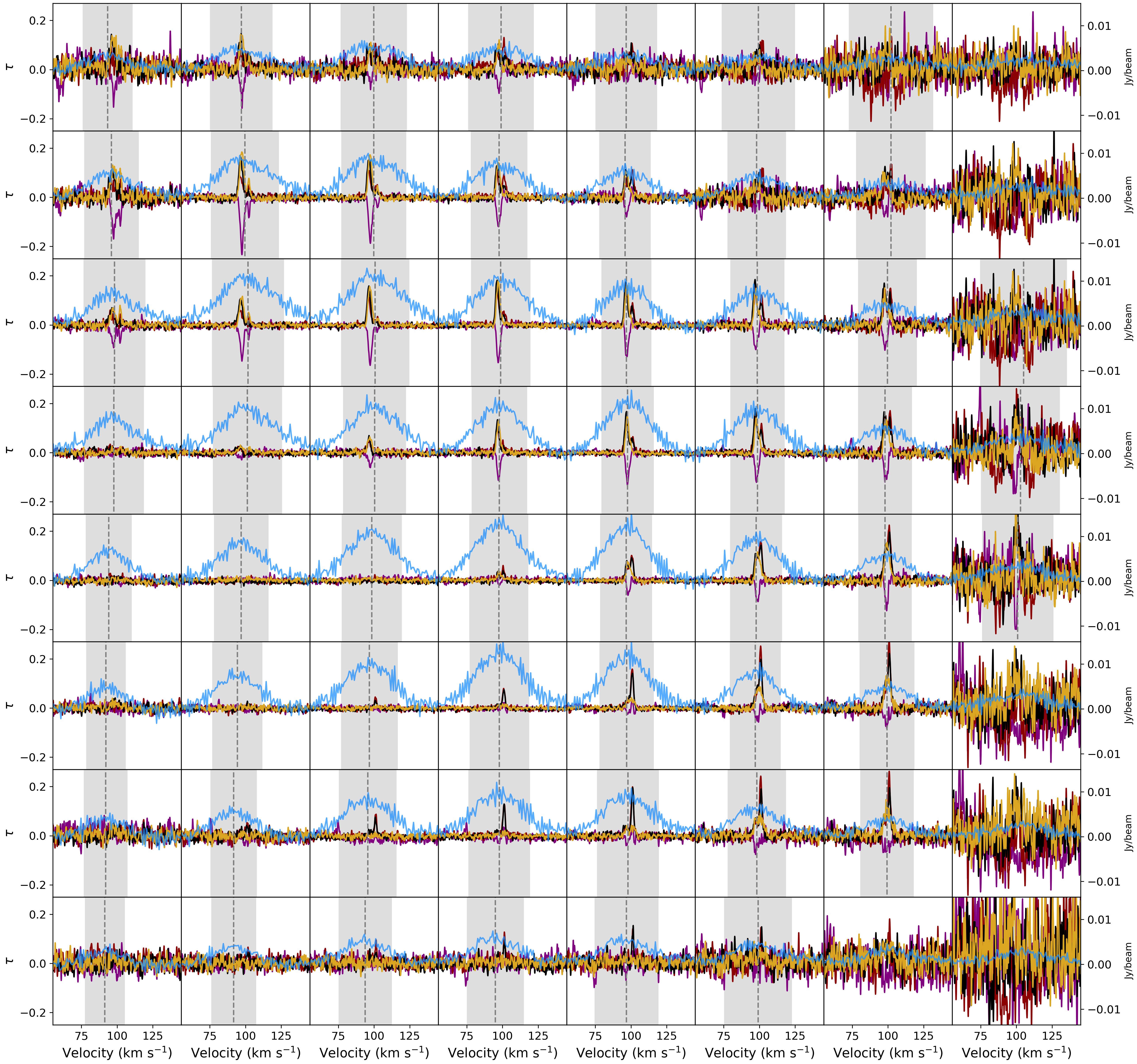}
    \caption{Same as Figure \ref{fig:G049grid} for G024.491+0.492, skipping every 2 pixels.}
    \label{fig:G024grid}
\end{figure*}

We present the OH optical depth spectra and \avghna RRL spectra observed with the VLA in this work towards G049 (Figure \ref{fig:G049grid}), G034 (Figure \ref{fig:G034grid}), and G024 (Figure \ref{fig:G024grid}). In these plots we skip every 4 pixels across the $S^* \times S^*$ grid that was chosen for fitting (see Appendix \ref{sec:appendixOHfit}). Each panel is overlaid with the fitted \avghna RRL V$\text{LSR}$ and the fitted \avghna RRL FWHM is represented as a shaded box. The OH features for all three \HII regions lie within the FWHM of the \avghna RRL, indicating a strong kinematic association. The \avghna RRL \VLSR overlaps with the OH spectral features for G049 and G024 across those sources. G034, however, has a $\sim$10 km s$^{-1}$ offset between the \avghna RRL \VLSR and the OH features. We suggest that this offset is due to G034 having an ionized outflow (see Section \ref{sec:disc}). 

\section{Molecular Gas Mass}
\label{sec:appendixMass}

We estimate the mass of the dynamically interacting molecular gas by first estimating the column density of the observed OH gas:
\begin{equation}
    \frac{N_\text{OH}}{[10^{14} \text{ cm}^{-2}]} = C \ \tau_0 \ \frac{T_\text{ex}}{\text{[K]}} \ \frac{\Delta v}{\text{[km/s]}},
\end{equation}
where $\Delta v$ is the Gaussian line width, $T_\text{ex}$ is the excitation temperature, $\tau_0$ is the optical depth at the line center, and $C$ is 4.23 for the 1667 transition \citep[e.g.,][]{DawsonSPLASH2022MNRAS.512.3345D}.
As the main lines are less affected by anomalous excitation than the satellites in diffuse gas, and 1667-MHz is generally the strongest main line, we use $\tau_\text{1667 MHz}$ to estimate $N(\text{OH})$ \citep{Li2018, DawsonSPLASH2014MNRAS.439.1596D, HafnerGNOMES2023}. $T_\text{ex}$ cannot be determined from interferometric observations alone, we therefore assume $T_{\text{ex,1667}}=4$ K as it is the mode value in the literature \citep{Li2018, HafnerGNOMES2023}.

$N(\text{OH})$ is converted to $N(\text{H}_2)$ using the $N(\text{OH})/N(\text{H}_2)$ abundance ratio, $X_{\text{OH}} \sim 1.0 \times 10^{-7}$ \citep{Liszt2002A&A...391..693L}. We estimate the mass of the gas within each pixel as $m_{\text{pixel}}= 2.8 m_{\text{H}} \, d_{\text{H\,\textsc{ii}}}^2 \, l^2 \, N(\text{H}_2)$ where $d_{\text{H\,\textsc{ii}}}$ is the distance to the \HII region (in Table \ref{tab:sources}), $l$ is the pixel size and $m_{\text{H}}$ is the mass of hydrogen, 1.67$\times 10^{-24}$ g. The total mass of the interacting molecular gas is found by summing all pixel masses together.

\bibliography{paper}{}

\begin{thebibliography}{}
\expandafter\ifx\csname natexlab\endcsname\relax\def\natexlab#1{#1}\fi
\providecommand{\url}[1]{\href{#1}{#1}}
\providecommand{\dodoi}[1]{doi:~\href{http://doi.org/#1}{\nolinkurl{#1}}}
\providecommand{\doeprint}[1]{\href{http://ascl.net/#1}{\nolinkurl{http://ascl.net/#1}}}
\providecommand{\doarXiv}[1]{\href{https://arxiv.org/abs/#1}{\nolinkurl{https://arxiv.org/abs/#1}}}

\bibitem[{C.~R. Avery {et~al.}(2021)Avery, Wuyts, Förster Schreiber, Villforth, Bertemes, Chang, Hamer, Toshikawa, \& Zhang}]{Avery10.1093/mnras/stab780}
Avery, C.~R., Wuyts, S., Förster Schreiber, N.~M., {et~al.} 2021, \bibinfo{title}{{Incidence, scaling relations and physical conditions of ionized gas outflows in MaNGA},} Monthly Notices of the Royal Astronomical Society, 503, 5134, \dodoi{10.1093/mnras/stab780}

\bibitem[{D.~S. {Balser} \& T.~V. {Wenger}(2024){Balser} \& {Wenger}}]{BalserWenger2024ApJ...964...47B}
{Balser}, D.~S., \& {Wenger}, T.~V. 2024, \bibinfo{title}{{The Metallicity{\textendash}Electron Temperature Relationship in H II Regions},} \apj, 964, 47, \dodoi{10.3847/1538-4357/ad2458}

\bibitem[{D.~S. {Balser} {et~al.}(2021){Balser}, {Wenger}, {Anderson}, {Armentrout}, {Bania}, {Dawson}, \& {Dickey}}]{Balser2021ApJ...921..176B}
{Balser}, D.~S., {Wenger}, T.~V., {Anderson}, L.~D., {et~al.} 2021, \bibinfo{title}{{Discovery of a New Population of Galactic H II Regions with Ionized Gas Velocity Gradients},} \apj, 921, 176, \dodoi{10.3847/1538-4357/ac1db3}

\bibitem[{A.~T. {Barnes} {et~al.}(2021){Barnes}, {Glover}, {Kreckel}, {Ostriker}, {Bigiel}, {Belfiore}, {Be{\v{s}}li{\'c}}, {Blanc}, {Chevance}, {Dale}, {Egorov}, {Eibensteiner}, {Emsellem}, {Grasha}, {Groves}, {Klessen}, {Kruijssen}, {Leroy}, {Longmore}, {Lopez}, {McElroy}, {Meidt}, {Murphy}, {Rosolowsky}, {Saito}, {Santoro}, {Schinnerer}, {Schruba}, {Sun}, {Watkins}, \& {Williams}}]{Barnes2021MNRAS.508.5362B}
{Barnes}, A.~T., {Glover}, S.~C.~O., {Kreckel}, K., {et~al.} 2021, \bibinfo{title}{{Comparing the pre-SNe feedback and environmental pressures for 6000 H II regions across 19 nearby spiral galaxies},} \mnras, 508, 5362, \dodoi{10.1093/mnras/stab2958}

\bibitem[{H. {Beuther} {et~al.}(2016){Beuther}, {Bihr}, {Rugel}, {Johnston}, {Wang}, {Walter}, {Brunthaler}, {Walsh}, {Ott}, {Stil}, {Henning}, {Schierhuber}, {Kainulainen}, {Heyer}, {Goldsmith}, {Anderson}, {Longmore}, {Klessen}, {Glover}, {Urquhart}, {Plume}, {Ragan}, {Schneider}, {McClure-Griffiths}, {Menten}, {Smith}, {Roy}, {Shanahan}, {Nguyen-Luong}, \& {Bigiel}}]{BeutherTHOR2016A&A...595A..32B}
{Beuther}, H., {Bihr}, S., {Rugel}, M., {et~al.} 2016, \bibinfo{title}{{The HI/OH/Recombination line survey of the inner Milky Way (THOR). Survey overview and data release 1},} \aap, 595, A32, \dodoi{10.1051/0004-6361/201629143}

\bibitem[{A. {Bik} {et~al.}(2019){Bik}, {Henning}, {Wu}, {Zhang}, {Brandner}, {Pasquali}, \& {Stolte}}]{Bik2019A&A...624A..63B}
{Bik}, A., {Henning}, T., {Wu}, S.~W., {et~al.} 2019, \bibinfo{title}{{Near-infrared spectroscopy of the massive stellar population of W51: evidence for multi-seeded star formation},} \aap, 624, A63, \dodoi{10.1051/0004-6361/201935061}

\bibitem[{D.~S. {Briggs}(1995){Briggs}}]{Briggs1995AAS...18711202B}
{Briggs}, D.~S. 1995, in American Astronomical Society Meeting Abstracts, Vol. 187, American Astronomical Society Meeting Abstracts, 112.02

\bibitem[{M. {Cappellari} \& Y. {Copin}(2003){Cappellari} \& {Copin}}]{voronoi2003MNRAS.342..345C}
{Cappellari}, M., \& {Copin}, Y. 2003, \bibinfo{title}{{Adaptive spatial binning of integral-field spectroscopic data using Voronoi tessellations},} \mnras, 342, 345, \dodoi{10.1046/j.1365-8711.2003.06541.x}

\bibitem[{M. {Cappellari} \& Y. {Copin}(2012){Cappellari} \& {Copin}}]{voronoisoftware2012ascl.soft11006C}
{Cappellari}, M., \& {Copin}, Y. 2012, \bibinfo{title}{{VorBin: Voronoi binning method},}, Astrophysics Source Code Library, record ascl:1211.006

\bibitem[{J.~M. {Carpenter} \& D.~B. {Sanders}(1998){Carpenter} \& {Sanders}}]{Carpenter1998AJ....116.1856C}
{Carpenter}, J.~M., \& {Sanders}, D.~B. 1998, \bibinfo{title}{{The W51 Giant Molecular Cloud},} \aj, 116, 1856, \dodoi{10.1086/300534}

\bibitem[{M. {Chevance} {et~al.}(2022){Chevance}, {Kruijssen}, {Krumholz}, {Groves}, {Keller}, {Hughes}, {Glover}, {Henshaw}, {Herrera}, {Kim}, {Leroy}, {Pety}, {Razza}, {Rosolowsky}, {Schinnerer}, {Schruba}, {Barnes}, {Bigiel}, {Blanc}, {Dale}, {Emsellem}, {Faesi}, {Grasha}, {Klessen}, {Kreckel}, {Liu}, {Longmore}, {Meidt}, {Querejeta}, {Saito}, {Sun}, \& {Usero}}]{Chevance2022MNRAS.509..272C}
{Chevance}, M., {Kruijssen}, J.~M.~D., {Krumholz}, M.~R., {et~al.} 2022, \bibinfo{title}{{Pre-supernova feedback mechanisms drive the destruction of molecular clouds in nearby star-forming disc galaxies},} \mnras, 509, 272, \dodoi{10.1093/mnras/stab2938}

\bibitem[{E. {Churchwell} {et~al.}(2009){Churchwell}, {Babler}, {Meade}, {Whitney}, {Benjamin}, {Indebetouw}, {Cyganowski}, {Robitaille}, {Povich}, {Watson}, \& {Bracker}}]{GLIMPSE2009PASP..121..213C}
{Churchwell}, E., {Babler}, B.~L., {Meade}, M.~R., {et~al.} 2009, \bibinfo{title}{{The Spitzer/GLIMPSE Surveys: A New View of the Milky Way},} \pasp, 121, 213, \dodoi{10.1086/597811}

\bibitem[{T. Cornwell(2008)Cornwell}]{multiscaleCornwell}
Cornwell, T. 2008, \bibinfo{title}{Multiscale CLEAN Deconvolution of Radio Synthesis Images,} Selected Topics in Signal Processing, IEEE Journal of, 2, 793 , \dodoi{10.1109/JSTSP.2008.2006388}

\bibitem[{J.~R. {Dawson} {et~al.}(2024){Dawson}, {Breen}, \& {Gaskap-Oh Team}}]{Dawson2024IAUS..380..486D}
{Dawson}, J.~R., {Breen}, S.~L., \& {Gaskap-Oh Team}. 2024, in IAU Symposium, Vol. 380, Cosmic Masers: Proper Motion Toward the Next-Generation Large Projects, ed. T.~{Hirota}, H.~{Imai}, K.~{Menten}, \& Y.~{Pihlstr{\"o}m}, 486--490, \dodoi{10.1017/S1743921323002405}

\bibitem[{J.~R. {Dawson} {et~al.}(2014){Dawson}, {Walsh}, {Jones}, {Breen}, {Cunningham}, {Lowe}, {Jones}, {Purcell}, {Caswell}, {Carretti}, {McClure-Griffiths}, {Ellingsen}, {Green}, {G{\'o}mez}, {Krishnan}, {Dickey}, {Imai}, {Gibson}, {Hennebelle}, {Lo}, {Hayakawa}, {Fukui}, \& {Mizuno}}]{DawsonSPLASH2014MNRAS.439.1596D}
{Dawson}, J.~R., {Walsh}, A.~J., {Jones}, P.~A., {et~al.} 2014, \bibinfo{title}{{SPLASH: the Southern Parkes Large-Area Survey in Hydroxyl - first science from the pilot region},} \mnras, 439, 1596, \dodoi{10.1093/mnras/stu032}

\bibitem[{J.~R. {Dawson} {et~al.}(2022){Dawson}, {Jones}, {Purcell}, {Walsh}, {Breen}, {Brown}, {Carretti}, {Cunningham}, {Dickey}, {Ellingsen}, {Gibson}, {G{\'o}mez}, {Green}, {Imai}, {Krishnan}, {Lo}, {Lowe}, {Marquarding}, \& {McClure-Griffiths}}]{DawsonSPLASH2022MNRAS.512.3345D}
{Dawson}, J.~R., {Jones}, P.~A., {Purcell}, C., {et~al.} 2022, \bibinfo{title}{{SPLASH: the Southern Parkes Large-Area Survey in Hydroxyl - data description and release},} \mnras, 512, 3345, \dodoi{10.1093/mnras/stac636}

\bibitem[{J.~M. {Dickey} {et~al.}(2013){Dickey}, {McClure-Griffiths}, {Gibson}, {G{\'o}mez}, {Imai}, {Jones}, {Stanimirovi{\'c}}, {Van Loon}, {Walsh}, {Alberdi}, {Anglada}, {Uscanga}, {Arce}, {Bailey}, {Begum}, {Wakker}, {Bekhti}, {Kalberla}, {Winkel}, {Bekki}, {For}, {Staveley-Smith}, {Westmeier}, {Burton}, {Cunningham}, {Dawson}, {Ellingsen}, {Diamond}, {Green}, {Hill}, {Koribalski}, {McConnell}, {Rathborne}, {Voronkov}, {Douglas}, {English}, {Ford}, {Lockman}, {Foster}, {Gomez}, {Green}, {Bland-Hawthorn}, {Gulyaev}, {Hoare}, {Joncas}, {Kang}, {Kerton}, {Koo}, {Leahy}, {Lo}, {Migenes}, {Nakashima}, {Zhang}, {Nidever}, {Peek}, {Tafoya}, {Tian}, \& {Wu}}]{DickeyGASKAP2013PASA...30....3D}
{Dickey}, J.~M., {McClure-Griffiths}, N., {Gibson}, S.~J., {et~al.} 2013, \bibinfo{title}{{GASKAP-The Galactic ASKAP Survey},} \pasa, 30, e003, \dodoi{10.1017/pasa.2012.003}

\bibitem[{Y. {Ebisawa} {et~al.}(2020){Ebisawa}, {Sakai}, {Menten}, {Oya}, \& {Yamamoto}}]{Ebisawa2020ApJ...904..136E}
{Ebisawa}, Y., {Sakai}, N., {Menten}, K.~M., {Oya}, Y., \& {Yamamoto}, S. 2020, \bibinfo{title}{{Temperature Structure of the Pipe Nebula Studied by the Intensity Anomaly of the OH 18 cm Transition},} \apj, 904, 136, \dodoi{10.3847/1538-4357/abc16f}

\bibitem[{Y. {Ebisawa} {et~al.}(2019){Ebisawa}, {Sakai}, {Menten}, \& {Yamamoto}}]{Ebisawa2019ApJ...871...89E}
{Ebisawa}, Y., {Sakai}, N., {Menten}, K.~M., \& {Yamamoto}, S. 2019, \bibinfo{title}{{The Effect of Far-infrared Radiation on the Hyperfine Anomaly of the OH 18 cm Transition},} \apj, 871, 89, \dodoi{10.3847/1538-4357/aaf72b}

\bibitem[{M. {Elitzur}(1976){Elitzur}}]{Elitzur17201976ApJ...203..124E}
{Elitzur}, M. 1976, \bibinfo{title}{{Inversion of the OH 1720-MHz Line},} \apj, 203, 124, \dodoi{10.1086/154054}

\bibitem[{M. {Elitzur} {et~al.}(1976){Elitzur}, {Goldreich}, \& {Scoville}}]{Elitzur16121976ApJ...205..384E}
{Elitzur}, M., {Goldreich}, P., \& {Scoville}, N. 1976, \bibinfo{title}{{OH-IR stars. II. A model for the 1612 MHz masers.},} \apj, 205, 384, \dodoi{10.1086/154289}

\bibitem[{C. {Federrath} \& R.~S. {Klessen}(2013){Federrath} \& {Klessen}}]{Federrath2013ApJ...763...51F}
{Federrath}, C., \& {Klessen}, R.~S. 2013, \bibinfo{title}{{On the Star Formation Efficiency of Turbulent Magnetized Clouds},} \apj, 763, 51, \dodoi{10.1088/0004-637X/763/1/51}

\bibitem[{L. {Gendelev} \& M.~R. {Krumholz}(2012){Gendelev} \& {Krumholz}}]{Gendelev2012ApJ...745..158G}
{Gendelev}, L., \& {Krumholz}, M.~R. 2012, \bibinfo{title}{{Evolution of Blister-type H II Regions in a Magnetized Medium},} \apj, 745, 158, \dodoi{10.1088/0004-637X/745/2/158}

\bibitem[{M.~Y. {Grudi{\'c}} {et~al.}(2022){Grudi{\'c}}, {Guszejnov}, {Offner}, {Rosen}, {Raju}, {Faucher-Gigu{\`e}re}, \& {Hopkins}}]{Grudic2022MNRAS.512..216G}
{Grudi{\'c}}, M.~Y., {Guszejnov}, D., {Offner}, S. S.~R., {et~al.} 2022, \bibinfo{title}{{The dynamics and outcome of star formation with jets, radiation, winds, and supernovae in concert},} \mnras, 512, 216, \dodoi{10.1093/mnras/stac526}

\bibitem[{J. {Guibert} {et~al.}(1978){Guibert}, {Rieu}, \& {Elitzur}}]{Guibert1978A&A....66..395G}
{Guibert}, J., {Rieu}, N.~Q., \& {Elitzur}, M. 1978, \bibinfo{title}{{OH excitation in interstellar clouds.},} \aap, 66, 395

\bibitem[{A. Hafner {et~al.}(2020)Hafner, Dawson, \& Wardle}]{HafnerFlipPaper}
Hafner, A., Dawson, J.~R., \& Wardle, M. 2020, \bibinfo{title}{{The hydroxyl satellite-line ‘flip’ as a tracer of expanding H ii regions},} Monthly Notices of the Royal Astronomical Society, 497, 4066, \dodoi{10.1093/mnras/staa2234}

\bibitem[{A. {Hafner} {et~al.}(2021){Hafner}, {Dawson}, \& {Wardle}}]{HafnerAMOEBA2021}
{Hafner}, A., {Dawson}, J.~R., \& {Wardle}, M. 2021, \bibinfo{title}{{Amoeba: Automated Molecular Excitation Bayesian Line-fitting Algorithm},} \apj, 923, 261, \dodoi{10.3847/1538-4357/ac2f42}

\bibitem[{A. {Hafner} {et~al.}(2023){Hafner}, {Dawson}, {Nguyen}, {Heiles}, {Wardle}, {Lee}, {Murray}, {Thompson}, \& {Stanimirovi{\'c}}}]{HafnerGNOMES2023}
{Hafner}, A., {Dawson}, J.~R., {Nguyen}, H., {et~al.} 2023, \bibinfo{title}{{GNOMES II: Analysis of the Galactic diffuse molecular ISM in all four ground state hydroxyl transitions using AMOEBA},} \pasa, 40, e015, \dodoi{10.1017/pasa.2023.8}

\bibitem[{T. {Hosokawa} \& S.-i. {Inutsuka}(2005){Hosokawa} \& {Inutsuka}}]{Hosokawa2005ApJ...623..917H}
{Hosokawa}, T., \& {Inutsuka}, S.-i. 2005, \bibinfo{title}{{Dynamical Expansion of Ionization and Dissociation Fronts around a Massive Star. I. A Mode of Triggered Star Formation},} \apj, 623, 917, \dodoi{10.1086/428648}

\bibitem[{F.~P. {Israel}(1978){Israel}}]{Israel1978A&A....70..769I}
{Israel}, F.~P. 1978, \bibinfo{title}{{H II regions and CO clouds: the blister model.},} \aap, 70, 769

\bibitem[{M. {Kang} {et~al.}(2010){Kang}, {Bieging}, {Kulesa}, {Lee}, {Choi}, \& {Peters}}]{Kang2010ApJS..190...58K}
{Kang}, M., {Bieging}, J.~H., {Kulesa}, C.~A., {et~al.} 2010, \bibinfo{title}{{A CO Line and Infrared Continuum Study of the Active Star-forming Complex W51},} \apjs, 190, 58, \dodoi{10.1088/0067-0049/190/1/58}

\bibitem[{R.~E. Kass \& A.~E. Raftery(1995)Kass \& Raftery}]{Kass1995doi:10.1080/01621459.1995.10476572}
Kass, R.~E., \& Raftery, A.~E. 1995, \bibinfo{title}{Bayes Factors,} Journal of the American Statistical Association, 90, 773, \dodoi{10.1080/01621459.1995.10476572}

\bibitem[{B.~W. {Keller} {et~al.}(2022){Keller}, {Kruijssen}, \& {Chevance}}]{Keller2022MNRAS.514.5355K}
{Keller}, B.~W., {Kruijssen}, J.~M.~D., \& {Chevance}, M. 2022, \bibinfo{title}{{Empirically motivated early feedback: momentum input by stellar feedback in galaxy simulations inferred through observations},} \mnras, 514, 5355, \dodoi{10.1093/mnras/stac1607}

\bibitem[{S. {Khan} {et~al.}(2022){Khan}, {Pandian}, {Lal}, {Rugel}, {Brunthaler}, {Menten}, {Wyrowski}, {Medina}, {Dzib}, \& {Nguyen}}]{Khan2022A&A...664A.140K}
{Khan}, S., {Pandian}, J.~D., {Lal}, D.~V., {et~al.} 2022, \bibinfo{title}{{A multiwavelength study of the W33 Main ultracompact HII region},} \aap, 664, A140, \dodoi{10.1051/0004-6361/202140914}

\bibitem[{J.-G. {Kim} {et~al.}(2018){Kim}, {Kim}, \& {Ostriker}}]{Kim2018ApJ...859...68K}
{Kim}, J.-G., {Kim}, W.-T., \& {Ostriker}, E.~C. 2018, \bibinfo{title}{{Modeling UV Radiation Feedback from Massive Stars. II. Dispersal of Star-forming Giant Molecular Clouds by Photoionization and Radiation Pressure},} \apj, 859, 68, \dodoi{10.3847/1538-4357/aabe27}

\bibitem[{E.~W. {Koch} {et~al.}(2021){Koch}, {Rosolowsky}, {Leroy}, {Chastenet}, {Chiang}, {Dalcanton}, {Kepley}, {Sandstrom}, {Schruba}, {Stanimirovi{\'c}}, {Utomo}, \& {Williams}}]{Koch2021}
{Koch}, E.~W., {Rosolowsky}, E.~W., {Leroy}, A.~K., {et~al.} 2021, \bibinfo{title}{{A lack of constraints on the cold opaque H I mass: H I spectra in M31 and M33 prefer multicomponent models over a single cold opaque component},} \mnras, 504, 1801, \dodoi{10.1093/mnras/stab981}

\bibitem[{B.-C. {Koo}(1999){Koo}}]{Koo1999ApJ...518..760K}
{Koo}, B.-C. 1999, \bibinfo{title}{{CO Observations of the W51B H II Region Complex},} \apj, 518, 760, \dodoi{10.1086/307316}

\bibitem[{J.~M.~D. {Kruijssen} {et~al.}(2019){Kruijssen}, {Schruba}, {Chevance}, {Longmore}, {Hygate}, {Haydon}, {McLeod}, {Dalcanton}, {Tacconi}, \& {van Dishoeck}}]{Kruijssen2019Natur.569..519K}
{Kruijssen}, J.~M.~D., {Schruba}, A., {Chevance}, M., {et~al.} 2019, \bibinfo{title}{{Fast and inefficient star formation due to short-lived molecular clouds and rapid feedback},} \nat, 569, 519, \dodoi{10.1038/s41586-019-1194-3}

\bibitem[{M.~R. {Krumholz}(2014){Krumholz}}]{Krumholz2014PhR...539...49K}
{Krumholz}, M.~R. 2014, \bibinfo{title}{{The big problems in star formation: The star formation rate, stellar clustering, and the initial mass function},} \physrep, 539, 49, \dodoi{10.1016/j.physrep.2014.02.001}

\bibitem[{C.-X. {Li} {et~al.}(2022){Li}, {Wang}, {Ma}, {Zhang}, {Li}, \& {Zheng}}]{Li2022RAA....22d5008L}
{Li}, C.-X., {Wang}, H.-C., {Ma}, Y.-H., {et~al.} 2022, \bibinfo{title}{{Molecular Clouds Associated with H II Regions and Candidates within l = 106.{\textdegree}65 to 109.{\textdegree}50 and b = -1.{\textdegree}85 to 0.{\textdegree}95},} Research in Astronomy and Astrophysics, 22, 045008, \dodoi{10.1088/1674-4527/ac52a0}

\bibitem[{D. {Li} {et~al.}(2018){Li}, {Tang}, {Nguyen}, {Dawson}, {Heiles}, {Xu}, {Pan}, {Goldsmith}, {Gibson}, {Murray}, {Robishaw}, {McClure-Griffiths}, {Dickey}, {Pineda}, {Stanimirovi{\'c}}, {Bronfman}, {Troland}, \& {PRIMO Collaboration}}]{Li2018}
{Li}, D., {Tang}, N., {Nguyen}, H., {et~al.} 2018, \bibinfo{title}{{Where is OH and Does It Trace the Dark Molecular Gas (DMG)?},} \apjs, 235, 1, \dodoi{10.3847/1538-4365/aaa762}

\bibitem[{H. {Liszt} \& R. {Lucas}(2002){Liszt} \& {Lucas}}]{Liszt2002A&A...391..693L}
{Liszt}, H., \& {Lucas}, R. 2002, \bibinfo{title}{{Comparative chemistry of diffuse clouds. IV: CH},} \aap, 391, 693, \dodoi{10.1051/0004-6361:20020849}

\bibitem[{H.~S. {Liszt} \& J. {Pety}(2012){Liszt} \& {Pety}}]{LisztPety2012A&A...541A..58L}
{Liszt}, H.~S., \& {Pety}, J. 2012, \bibinfo{title}{{Imaging diffuse clouds: bright and dark gas mapped in CO},} \aap, 541, A58, \dodoi{10.1051/0004-6361/201218771}

\bibitem[{A. {Marchal} {et~al.}(2019){Marchal}, {Miville-Desch{\^e}nes}, {Orieux}, {Gac}, {Soussen}, {Lesot}, {d'Allonnes}, \& {Salom{\'e}}}]{ROHSA2019A&A...626A.101M}
{Marchal}, A., {Miville-Desch{\^e}nes}, M.-A., {Orieux}, F., {et~al.} 2019, \bibinfo{title}{{ROHSA: Regularized Optimization for Hyper-Spectral Analysis. Application to phase separation of 21 cm data},} \aap, 626, A101, \dodoi{10.1051/0004-6361/201935335}

\bibitem[{J.~P. {McMullin} {et~al.}(2007){McMullin}, {Waters}, {Schiebel}, {Young}, \& {Golap}}]{CASA}
{McMullin}, J.~P., {Waters}, B., {Schiebel}, D., {Young}, W., \& {Golap}, K. 2007, in Astronomical Society of the Pacific Conference Series, Vol. 376, Astronomical Data Analysis Software and Systems XVI, ed. R.~A. {Shaw}, F.~{Hill}, \& D.~J. {Bell}, 127

\bibitem[{G. {Mellema} {et~al.}(2006){Mellema}, {Arthur}, {Henney}, {Iliev}, \& {Shapiro}}]{Mellema2006ApJ...647..397M}
{Mellema}, G., {Arthur}, S.~J., {Henney}, W.~J., {Iliev}, I.~T., \& {Shapiro}, P.~R. 2006, \bibinfo{title}{{Dynamical H II Region Evolution in Turbulent Molecular Clouds},} \apj, 647, 397, \dodoi{10.1086/505294}

\bibitem[{B. {Mookerjea} {et~al.}(2007){Mookerjea}, {Casper}, {Mundy}, \& {Looney}}]{Mookerjea2007ApJ...659..447M}
{Mookerjea}, B., {Casper}, E., {Mundy}, L.~G., \& {Looney}, L.~W. 2007, \bibinfo{title}{{Kinematics and Chemistry of the Hot Molecular Core in G34.26+0.15 at High Resolution},} \apj, 659, 447, \dodoi{10.1086/512095}

\bibitem[{M. {Newville} {et~al.}(2014){Newville}, {Stensitzki}, {Allen}, \& {Ingargiola}}]{lmfit}
{Newville}, M., {Stensitzki}, T., {Allen}, D.~B., \& {Ingargiola}, A. 2014, \bibinfo{title}{{LMFIT: Non-Linear Least-Square Minimization and Curve-Fitting for Python},}, 0.8.0, Zenodo Zenodo, \dodoi{10.5281/zenodo.11813}

\bibitem[{M. {Newville} {et~al.}(2023){Newville}, {Otten}, {Nelson}, {Stensitzki}, {Ingargiola}, {Allan}, {Fox}, {Carter}, {Micha{\l}}, {Osborn}, {Pustakhod}, {Lneuhaus}, {Weigand}, {Aristov}, {Glenn}, {Deil}, {Mgunyho}, {Mark}, {Hansen}, {Pasquevich}, {Foks}, {Zobrist}, {Frost}, {Stuermer}, {Azelcer}, {Polloreno}, {Persaud}, {Hedegaard Nielsen}, {Pompili}, \& {Caldwell}}]{lmfit2023zndo...7887568N}
{Newville}, M., {Otten}, R., {Nelson}, A., {et~al.} 2023, \bibinfo{title}{{lmfit/lmfit-py: 1.2.1},}, 1.2.1 Zenodo, \dodoi{10.5281/zenodo.7887568}

\bibitem[{G.~M. {Olivier} {et~al.}(2021){Olivier}, {Lopez}, {Rosen}, {Nayak}, {Reiter}, {Krumholz}, \& {Bolatto}}]{Olivier2021ApJ...908...68O}
{Olivier}, G.~M., {Lopez}, L.~A., {Rosen}, A.~L., {et~al.} 2021, \bibinfo{title}{{Evolution of Stellar Feedback in H II Regions},} \apj, 908, 68, \dodoi{10.3847/1538-4357/abd24a}

\bibitem[{C.~H.~M. {Pabst} {et~al.}(2024){Pabst}, {Goicoechea}, {Cuadrado}, {Salas}, {Tielens}, \& {Marcelino}}]{Pabst2024arXiv240417963P}
{Pabst}, C.~H.~M., {Goicoechea}, J.~R., {Cuadrado}, S., {et~al.} 2024, \bibinfo{title}{{Multiline observations of hydrogen, helium, and carbon radio-recombination lines toward Orion A: A detailed dynamical study and direct determination of physical conditions},} arXiv e-prints, arXiv:2404.17963, \dodoi{10.48550/arXiv.2404.17963}

\bibitem[{N. {Panwar} {et~al.}(2020){Panwar}, {Sharma}, {Ojha}, {Baug}, {Dewangan}, {Bhatt}, \& {Pandey}}]{Panwar2020ApJ...905...61P}
{Panwar}, N., {Sharma}, S., {Ojha}, D.~K., {et~al.} 2020, \bibinfo{title}{{Star Formation and Evolution of Blister-type H II Region Sh2-112},} \apj, 905, 61, \dodoi{10.3847/1538-4357/abc42e}

\bibitem[{D. {Pathak} {et~al.}(2025){Pathak}, {Leroy}, {Thompson}, {Lopez}, {Barnes}, {Dale}, {Blackstone}, {Glover}, {Menon}, {Sutter}, {Williams}, {Baron}, {Belfiore}, {Bigiel}, {Bolatto}, {Boquien}, {Chandar}, {Chevance}, {Chown}, {Grasha}, {Groves}, {Klessen}, {Kreckel}, {Li}, {M{\'e}ndez-Delgado}, {Rosolowsky}, {Sandstrom}, {Sarbadhicary}, {Sun}, \& {{\'U}beda}}]{Pathak2025ApJ...982..140P}
{Pathak}, D., {Leroy}, A.~K., {Thompson}, T.~A., {et~al.} 2025, \bibinfo{title}{{Linking Stellar Populations to H II Regions across Nearby Galaxies. II. Infrared Reprocessed and UV Direct Radiation Pressure in H II Regions},} \apj, 982, 140, \dodoi{10.3847/1538-4357/adb484}

\bibitem[{C.~H. {Pe{\~n}aloza} {et~al.}(2017){Pe{\~n}aloza}, {Clark}, {Glover}, {Shetty}, \& {Klessen}}]{Penaloza2017MNRAS.465.2277P}
{Pe{\~n}aloza}, C.~H., {Clark}, P.~C., {Glover}, S. C.~O., {Shetty}, R., \& {Klessen}, R.~S. 2017, \bibinfo{title}{{Using CO line ratios to trace the physical properties of molecular clouds},} \mnras, 465, 2277, \dodoi{10.1093/mnras/stw2892}

\bibitem[{D. {Rahner} {et~al.}(2017){Rahner}, {Pellegrini}, {Glover}, \& {Klessen}}]{Rahner2017MNRAS.470.4453R}
{Rahner}, D., {Pellegrini}, E.~W., {Glover}, S. C.~O., \& {Klessen}, R.~S. 2017, \bibinfo{title}{{Winds and radiation in unison: a new semi-analytic feedback model for cloud dissolution},} \mnras, 470, 4453, \dodoi{10.1093/mnras/stx1532}

\bibitem[{D. {Rahner} {et~al.}(2018){Rahner}, {Pellegrini}, {Glover}, \& {Klessen}}]{Rahner2018MNRAS.473L..11R}
{Rahner}, D., {Pellegrini}, E.~W., {Glover}, S. C.~O., \& {Klessen}, R.~S. 2018, \bibinfo{title}{{Forming clusters within clusters: how 30 Doradus recollapsed and gave birth again},} \mnras, 473, L11, \dodoi{10.1093/mnrasl/slx149}

\bibitem[{U. {Rau} \& T.~J. {Cornwell}(2011){Rau} \& {Cornwell}}]{MSMFS2011A&A...532A..71R}
{Rau}, U., \& {Cornwell}, T.~J. 2011, \bibinfo{title}{{A multi-scale multi-frequency deconvolution algorithm for synthesis imaging in radio interferometry},} \aap, 532, A71, \dodoi{10.1051/0004-6361/201117104}

\bibitem[{M.~J. {Reid} \& P.~T.~P. {Ho}(1985){Reid} \& {Ho}}]{Reid1985ApJ...288L..17R}
{Reid}, M.~J., \& {Ho}, P.~T.~P. 1985, \bibinfo{title}{{G 34.3+0.2 : a ``cometary'' HII region.},} \apjl, 288, L17, \dodoi{10.1086/184412}

\bibitem[{L.~E. {Rowland} {et~al.}(2024){Rowland}, {McLeod}, {Fattahi}, {Belfiore}, {Cresci}, {Hunt}, {Krumholz}, {Kumari}, {Marasco}, \& {Venturi}}]{Rowland2024A&A...685A..46R}
{Rowland}, L.~E., {McLeod}, A.~F., {Fattahi}, A., {et~al.} 2024, \bibinfo{title}{{Pre-supernova stellar feedback in nearby starburst dwarf galaxies},} \aap, 685, A46, \dodoi{10.1051/0004-6361/202348029}

\bibitem[{M.~R. {Rugel} {et~al.}(2018){Rugel}, {Beuther}, {Bihr}, {Wang}, {Ott}, {Brunthaler}, {Walsh}, {Glover}, {Goldsmith}, {Anderson}, {Schneider}, {Menten}, {Ragan}, {Urquhart}, {Klessen}, {Soler}, {Roy}, {Kainulainen}, {Henning}, {Bigiel}, {Smith}, {Wyrowski}, \& {Longmore}}]{RugelTHOR2018A&A...618A.159R}
{Rugel}, M.~R., {Beuther}, H., {Bihr}, S., {et~al.} 2018, \bibinfo{title}{{OH absorption in the first quadrant of the Milky Way as seen by THOR},} \aap, 618, A159, \dodoi{10.1051/0004-6361/201731872}

\bibitem[{M.~R. {Rugel} {et~al.}(2019){Rugel}, {Rahner}, {Beuther}, {Pellegrini}, {Wang}, {Soler}, {Ott}, {Brunthaler}, {Anderson}, {Mottram}, {Henning}, {Goldsmith}, {Heyer}, {Klessen}, {Bihr}, {Menten}, {Smith}, {Urquhart}, {Ragan}, {Glover}, {McClure-Griffiths}, {Bigiel}, \& {Roy}}]{RugelW49A2019A&A...622A..48R}
{Rugel}, M.~R., {Rahner}, D., {Beuther}, H., {et~al.} 2019, \bibinfo{title}{{Feedback in W49A diagnosed with radio recombination lines and models},} \aap, 622, A48, \dodoi{10.1051/0004-6361/201834068}

\bibitem[{A. {Saha} {et~al.}(2024){Saha}, {Tej}, {Liu}, {Liu}, {Garay}, {Goldsmith}, {Lee}, {He}, {Juvela}, {Bronfman}, {Baug}, {V{\'a}zquez-Semadeni}, {Sanhueza}, {Li}, {Chibueze}, {Bhadari}, {Dewangan}, {Das}, {Xu}, {Issac}, {Hwang}, \& {T{\'o}th}}]{Saha2024ApJ...970L..40S}
{Saha}, A., {Tej}, A., {Liu}, H.-L., {et~al.} 2024, \bibinfo{title}{{Direct Observational Evidence of Multi-epoch Massive Star Formation in G24.47+0.49},} \apjl, 970, L40, \dodoi{10.3847/2041-8213/ad6144}

\bibitem[{M. {Sewiło} {et~al.}(2011){Sewiło}, {Churchwell}, {Kurtz}, {Goss}, \& {Hofner}}]{Sewilo2011}
{Sewiło}, M., {Churchwell}, E., {Kurtz}, S., {Goss}, W.~M., \& {Hofner}, P. 2011, \bibinfo{title}{{A Very Large Array Study of Ultracompact and Hypercompact H II Regions from 0.7 to 3.6 cm},} \apjs, 194, 44, \dodoi{10.1088/0067-0049/194/2/44}

\bibitem[{W.~S. {Tan} {et~al.}(2023){Tan}, {Araya}, {Rigg}, {Hofner}, {Kurtz}, {Linz}, \& {Rosero}}]{Tan2023}
{Tan}, W.~S., {Araya}, E.~D., {Rigg}, C., {et~al.} 2023, \bibinfo{title}{{Excited Hydroxyl Outflow in the High-mass Star-forming Region G34.26 + 0.15},} \apj, 953, 90, \dodoi{10.3847/1538-4357/acde7b}

\bibitem[{T. {Umemoto} {et~al.}(2017){Umemoto}, {Minamidani}, {Kuno}, {Fujita}, {Matsuo}, {Nishimura}, {Torii}, {Tosaki}, {Kohno}, {Kuriki}, {Tsuda}, {Hirota}, {Ohashi}, {Yamagishi}, {Handa}, {Nakanishi}, {Omodaka}, {Koide}, {Matsumoto}, {Onishi}, {Tokuda}, {Seta}, {Kobayashi}, {Tachihara}, {Sano}, {Hattori}, {Onodera}, {Oasa}, {Kamegai}, {Tsuboi}, {Sofue}, {Higuchi}, {Chibueze}, {Mizuno}, {Honma}, {Muller}, {Inoue}, {Morokuma-Matsui}, {Shinnaga}, {Ozawa}, {Takahashi}, {Yoshiike}, {Costes}, \& {Kuwahara}}]{FUGIN}
{Umemoto}, T., {Minamidani}, T., {Kuno}, N., {et~al.} 2017, \bibinfo{title}{{FOREST unbiased Galactic plane imaging survey with the Nobeyama 45 m telescope (FUGIN). I. Project overview and initial results},} \pasj, 69, 78, \dodoi{10.1093/pasj/psx061}

\bibitem[{H.~J. {van Langevelde} {et~al.}(1995){van Langevelde}, {van Dishoeck}, {Sevenster}, \& {Israel}}]{vanLangevelde1995}
{van Langevelde}, H.~J., {van Dishoeck}, E.~F., {Sevenster}, M.~N., \& {Israel}, F.~P. 1995, \bibinfo{title}{{Anomalously Excited OH and Competition between Maser Transitions toward Centaurus A},} \apjl, 448, L123, \dodoi{10.1086/309613}

\bibitem[{J.~S. {Vink} {et~al.}(2000){Vink}, {de Koter}, \& {Lamers}}]{VinkOBWind2000A&A...362..295V}
{Vink}, J.~S., {de Koter}, A., \& {Lamers}, H.~J.~G.~L.~M. 2000, \bibinfo{title}{{New theoretical mass-loss rates of O and B stars},} \aap, 362, 295, \dodoi{10.48550/arXiv.astro-ph/0008183}

\bibitem[{P. Virtanen {et~al.}(2020)Virtanen, Gommers, Oliphant, Haberland, Reddy, Cournapeau, Burovski, Peterson, Weckesser, Bright, {van der Walt}, Brett, Wilson, Millman, Mayorov, Nelson, Jones, Kern, Larson, Carey, Polat, Feng, Moore, {VanderPlas}, Laxalde, Perktold, Cimrman, Henriksen, Quintero, Harris, Archibald, Ribeiro, Pedregosa, {van Mulbregt}, \& {SciPy 1.0 Contributors}}]{scipy}
Virtanen, P., Gommers, R., Oliphant, T.~E., {et~al.} 2020, \bibinfo{title}{{{SciPy} 1.0: Fundamental Algorithms for Scientific Computing in Python},} Nature Methods, 17, 261, \dodoi{10.1038/s41592-019-0686-2}

\bibitem[{S.~K. {Walch} {et~al.}(2012){Walch}, {Whitworth}, {Bisbas}, {W{\"u}nsch}, \& {Hubber}}]{WalchDispersal2012MNRAS.427..625W}
{Walch}, S.~K., {Whitworth}, A.~P., {Bisbas}, T., {W{\"u}nsch}, R., \& {Hubber}, D. 2012, \bibinfo{title}{{Dispersal of molecular clouds by ionizing radiation},} \mnras, 427, 625, \dodoi{10.1111/j.1365-2966.2012.21767.x}

\bibitem[{Y. {Wang} {et~al.}(2020){Wang}, {Beuther}, {Rugel}, {Soler}, {Stil}, {Ott}, {Bihr}, {McClure-Griffiths}, {Anderson}, {Klessen}, {Goldsmith}, {Roy}, {Glover}, {Urquhart}, {Heyer}, {Linz}, {Smith}, {Bigiel}, {Dempsey}, \& {Henning}}]{WangTHOR2020A&A...634A..83W}
{Wang}, Y., {Beuther}, H., {Rugel}, M.~R., {et~al.} 2020, \bibinfo{title}{{The HI/OH/Recombination line survey of the inner Milky Way (THOR): data release 2 and H I overview},} \aap, 634, A83, \dodoi{10.1051/0004-6361/201937095}

\bibitem[{T.~V. {Wenger}(2018){Wenger}}]{WISP2018ascl.soft12001W}
{Wenger}, T.~V. 2018, \bibinfo{title}{{WISP: Wenger Interferometry Software Package},}, Astrophysics Source Code Library, record ascl:1812.001 \doeprint{1812.001}

\bibitem[{T.~V. {Wenger} {et~al.}(2019{\natexlab{a}}){Wenger}, {Balser}, {Anderson}, \& {Bania}}]{WengerMetallicity2019ApJ...887..114W}
{Wenger}, T.~V., {Balser}, D.~S., {Anderson}, L.~D., \& {Bania}, T.~M. 2019{\natexlab{a}}, \bibinfo{title}{{Metallicity Structure in the Milky Way Disk Revealed by Galactic H II Regions},} \apj, 887, 114, \dodoi{10.3847/1538-4357/ab53d3}

\bibitem[{T.~V. {Wenger} {et~al.}(2019{\natexlab{b}}){Wenger}, {Dickey}, {Jordan}, {Balser}, {Armentrout}, {Anderson}, {Bania}, {Dawson}, {McClure-Griffiths}, \& {Shea}}]{WengerSHRDS2019ApJS..240...24W}
{Wenger}, T.~V., {Dickey}, J.~M., {Jordan}, C.~H., {et~al.} 2019{\natexlab{b}}, \bibinfo{title}{{The Southern H II Region Discovery Survey. I. The Bright Catalog},} \apjs, 240, 24, \dodoi{10.3847/1538-4365/aaf8ba}

\bibitem[{Y.~W. {Wu} {et~al.}(2014){Wu}, {Sato}, {Reid}, {Moscadelli}, {Zhang}, {Xu}, {Brunthaler}, {Menten}, {Dame}, \& {Zheng}}]{Wu2014}
{Wu}, Y.~W., {Sato}, M., {Reid}, M.~J., {et~al.} 2014, \bibinfo{title}{{Trigonometric parallaxes of star-forming regions in the Sagittarius spiral arm},} \aap, 566, A17, \dodoi{10.1051/0004-6361/201322765}

\bibitem[{J.-L. {Xu} {et~al.}(2018){Xu}, {Xu}, {Zhang}, {Liu}, {Yu}, {Ning}, \& {Ju}}]{Xu2018A&A...609A..43X}
{Xu}, J.-L., {Xu}, Y., {Zhang}, C.-P., {et~al.} 2018, \bibinfo{title}{{Gas kinematics and star formation in the filamentary molecular cloud G47.06+0.26},} \aap, 609, A43, \dodoi{10.1051/0004-6361/201629189}

\bibitem[{F.-Y. {Zhu} {et~al.}(2023){Zhu}, {Wang}, {Yan}, {Zhu}, \& {Li}}]{ZhuM172023MNRAS.522..503Z}
{Zhu}, F.-Y., {Wang}, J., {Yan}, Y., {Zhu}, Q.-F., \& {Li}, J. 2023, \bibinfo{title}{{Spatial distributions and kinematics of shocked and ionized gas in M17},} \mnras, 522, 503, \dodoi{10.1093/mnras/stad996}

\end{thebibliography}
\bibliographystyle{aasjournalv7}

\end{document}